\documentclass[a4paper,11pt]{article}

\def\H{\mathcal{H}}

\def\n{{\hat{\boldsymbol{n}}}}
\def\p{{\boldsymbol{p}}}
\def\q{{\boldsymbol{q}}}

\def\x{{\boldsymbol{x}}}

\def\kone{\boldsymbol{k}_1}
\def\ktwo{\boldsymbol{k}_2}
\def\kthree{\boldsymbol{k}_3}
\def\kfour{\boldsymbol{k}_4}
\def\qone{\boldsymbol{q}_1}
\def\qtwo{\boldsymbol{q}_2}

\def\qonetwo{\boldsymbol{q}_{12}}
\def\qoneint{\int_{\boldsymbol{q}_1}}

\def\x{\boldsymbol{x}}
\def\Pz{P_\zeta}
\newcommand{\bk}[1]{\boldsymbol{k}_{#1}}

\newcommand{\af}{\hspace{0.1cm}}

\newcommand{\fnl}[1]{f_\text{NL}^{\text{#1}}}

\def\geqone{g_{\textrm{NL}}^{\dot{\sigma}^4}}
\def\geqtwo{g_{\textrm{NL}}^{\dot{\sigma}^2(\partial \sigma)^2}}
\def\geqthree{g_{\textrm{NL}}^{ (\partial \sigma)^4}}
\def\taunl{\tau_{\text{NL}}^{\textrm{loc}}}
\def\gnl{g_{\text{NL}}^{\textrm{loc}}}
\def\kmax{k_{\text{max}}}
\def\kmin{k_{\text{min}}}
\def\Mpc{\textrm{ Mpc}}

\def\zk{(z,\boldsymbol{k})}

\def\kmin{k_\mathrm{min}}
\def\kmax{k_\mathrm{max}}

\def\kb{k_\mathrm{B}}
\def\kJ{k_\mathrm{J}}

\def\beq{\begin{equation}}
\def\eeq{\end{equation}}

\def\kb{k_\mathrm{B}}
\def\Gc{\Gamma_\mathrm{C}}
\def\Gr{\Gamma_\mathrm{R}}

\def\Tg{T_\gamma}
\def\Tgb{\bar{T}_\gamma}

\def\Tgas{T_{\mathrm{gas}}}
\def\Tgasdot{\dot{T}_\mathrm{gas}}
\def\Tgasb{\bar{T}_\mathrm{gas}}

\def\nH{n_\mathrm{H}}
\def\np{n_p}
\def\nHdot{\dot{n}_\mathrm{H}}
\def\nHb{\bar{n}_\mathrm{H}}

\def\nHI{n_\mathrm{HI}}

\def\np{n_p}

\def\xe{x_e}
\def\xedot{\dot{x}_e}
\def\xeb{\bar{x}_e}
\def\xHe{x_\mathrm{He}}
\def\xHeb{\bar{x}_\mathrm{He}}

\def\dT{\delta_\mathrm{T}}
\def\dTup{\delta^\mathrm{T}}
\def\dH{\delta_\mathrm{H}}
\def\db{\delta_b}

\def\dv{\delta_v}
\def\dxe{\delta_{x_e}}
\def\dx{\delta^x}
\def\dxTup{\delta^{a}}
\def\dTcmtilde{\delta \tilde{T}_{21}}
\def\dTcm{\delta T_{21}}
\def\DT{\Delta^\mathrm{T}}
\def\Dx{\Delta^x}
\def\DxT{\Delta^{a}}
\def\SxT{\Sigma^{a}}
\def\ST{\Sigma^{\mathrm{T}}}
\def\Sx{\Sigma^x}

\def\dTdot{\dot{\delta}_\mathrm{T}}
\def\dbdot{\dot{\delta}_b}

\def\dxedot{\dot{\delta}_{x_e}}
\def\dxdot{\dot{\delta}^x}
\def\dTupdot{\dot{\delta}^\mathrm{T}}
\def\diracd{\delta_\mathrm{D}}

\def\Cx{\mathcal{C}^x}
\def\CT{\mathcal{C}^\mathrm{T}}
\def\CxT{\mathcal{C}^{a}}

\def\sigmaT{\sigma_\mathrm{T}}
\def\me{m_e}
\def\mp{m_p}

\def\tx{(t,\boldsymbol{x})}
\def\zx{(z,\boldsymbol{x})}
\def\order{\mathcal{O}}
\def\Cp{C_\mathrm{P}}
\def\alphaB{\alpha_\mathrm{B}}
\def\alphaBbkg{\bar{\alpha}_\mathrm{B}}
\def\betaB{\beta_\mathrm{B}}

\def\zx{(z,\boldsymbol{x})}

\def\Eonetwo{E_{21}}

\def\Ltwoone{\Lambda_{2s\to 1s}}
\def\lalpha{\lambda_\alpha}

\def\Ts{T_s}
\def\Tstar{T_*}
\def\Yc{Y_{\mathrm{c}}}
\def\Tcm{T_{21}}
\def\Tcmb{\bar{T}_{21}}
\def\Tcmtilde{\tilde{T}_{21}}
\def\taucm{\tau_{21}}
\def\lcm{\lambda_{21}}

\def\pp{\partial_\parallel}
\def\vp{v_\parallel}
\def\T{\mathcal{T}}
\def\Tbitj{\mathcal{T}_{b^i\mathrm{T}^j}}
\def\Tb{\mathcal{T}_b}
\def\TT{\mathcal{T}_\mathrm{T}}
\def\TbT{\mathcal{T}_{b\mathrm{T}}}
\def\Tbb{\mathcal{T}_{bb}}
\def\TTT{\mathcal{T}_{\mathrm{TT}}}
\def\Tbbb{\mathcal{T}_{bbb}}
\def\TTTT{\mathcal{T}_{\mathrm{TTT}}}
\def\TbTT{\mathcal{T}_{b\mathrm{TT}}}
\def\TbbT{\mathcal{T}_{bb\mathrm{T}}}

\def\camb{\texttt{CAMB}}

\def\p{\boldsymbol{p}}

\def\n{{\boldsymbol{n}}}

\pdfoutput=1 

\usepackage{jcappub} 
                     
\usepackage{mathtools, cases}
\usepackage[T1]{fontenc} 
\usepackage{float}
\usepackage{xcolor}

\title{\boldmath The Dark Ages' 21-cm Trispectrum}

\author[a,b]{Thomas Fl\"oss,}
\author[a]{Tim de Wild,}
\author[a]{P. Daniel Meerburg,}
\author[b]{L\'eon V.E. Koopmans}

\affiliation[a]{Van Swinderen Institute, University of Groningen, Nijenborgh 4, 9747 AG Groningen, The Netherlands}
\affiliation[b]{Kapteyn Astronomical Institute, University of Groningen, P.O.Box 800, 9700 AV Groningen, The Netherlands}

\emailAdd{t.s.floss@rug.nl}

\abstract{We investigate tomography of 21-cm brightness temperature fluctuations during the Dark Ages as a probe for constraining primordial non-Gaussianity. We expand the 21-cm brightness temperature up to cubic order in perturbation theory and improve previous models of the signal by including the effect of the free electron fraction. Using modified standard perturbation theory methods that include baryonic pressure effects we derive an improved secondary bispectrum and for the first time derive the secondary trispectrum of 21-cm brightness temperature fluctuations. We then forecast the amount of information available from the Dark Ages to constrain primordial non-Gaussianity, including the imprints of massive particle exchange during inflation and we determine how much signal is lost due to secondary non-Gaussianity. We find that although secondary non-Gaussianity swamps the primordial signal, primordial non-Gaussianity can still be extracted with signal-to-noise ratios that surpass current and future CMB experiments by several orders of magnitude, depending on the experimental setup. Furthermore, we conclude that for the bi- and trispectra of massive particle exchange marginalizing over other primordial shapes affects signal-to-noise ratios more severely than secondary shapes. Baryonic pressure effects turn out to have a negligible impact on our forecasts, even at scales close to the Jeans scale. The results of this work reinforce the prospects of 21-cm brightness temperature fluctuations from the Dark Ages as the ultimate probe for primordial non-Gaussianity.}

\begin{document}
\maketitle
\newpage
\section{Introduction}
Over the last few decades, inflation has become the leading paradigm for describing the early universe. Even the simplest models of inflation accurately predict the nearly Gaussian initial conditions of the universe and match the precise observations of the Cosmic Microwave Background (CMB) by the Planck satellite \cite{Planck2018Inflation,Planck2018PNG}. Slight deviations from these Gaussian initial conditions of the early universe, colloquially known as primordial non-Gaussianity (pnG), may be used to further constrain the vast space of inflationary theories \cite{Meerburg2019}. Currently, bounds on the size of pnG by CMB and Large Scale Structure (LSS) observations are consistent with a purely Gaussian distribution of fluctuations, and therefore unable to favour a particular theory of inflation. However, upcoming CMB and LSS experiments are forecast to improve on these bounds and might find a statistically significant deviation from Gaussianity.\\
\\
The size of pnG can be probed through the higher statistical moments of the distribution of initial conditions. For a purely Gaussian distribution all information is then contained in the two-point correlation function of the primordial density field. Generally, the size of pnG is determined by the interactions of the inflaton, the scalar field driving the supposed inflationary expansion. However, even in the absence of such \emph{direct} interactions, one expects \emph{indirect} interaction through gravity (i.e. through coupling to the metric). This gravitational interaction sets the minimum amount of pnG that is guaranteed to be present in the initial conditions sourced by inflation and is commonly referred to as the \emph{gravitational floor}. In the simplest theories such as single-field slow-roll inflation, this is the only source of pnG \cite{Maldacena2002,Creminelli:2003iq}, whereas more complicated models (i.e. inflaton self-interactions or multiple fields) could generate considerably larger amounts of non-Gaussianity. Another powerful probe of inflation is the famous Maldacena consistency condition \cite{Maldacena2002,Creminelli2004}, which predicts the vanishing of the squeezed bispectrum for all single-field models of inflation. Hence a non-zero measurement of the squeezed bispectrum would rule out all single-field inflationary models and point toward multi-field dynamics.\\
\\
Besides being a powerful probe of the physics driving the inflationary expansion, pnG can also contain signatures of additional physics at play during the inflationary era. Massive (spinning) extra fields could have been present, leaving their imprint on the higher order statistics of the initial conditions \cite{Arkani-Hamed2015,Lee:2016vti,Baumann2017}. More specifically, heavy fields induce a characteristic oscillatory shape in the squeezed and collapsed limit of the three- and four-point correlation function of fluctuations respectively. The amplitude and frequency of this oscillation is directly related to the mass, in principle allowing one to probe the particle spectrum during inflation. Such extra fields are naturally present in string theory in which one might ultimately aim to embed the theory of inflation as well. Hence, we are motivated to study the existence of these extra fields and use inflation as a cosmological particle collider \cite{Arkani-Hamed2015}. Although such oscillatory behaviour is a clean probe of the particle spectrum, its amplitude is unfortunately severely suppressed proportional to the mass, making it challenging for CMB and LSS experiments to ever measure such an effect.\\
\\
Clearly, in order to advance precision cosmology to measure pnG, new probes and experiments are necessary. In this work we will study the use of 21-cm fluctuations during the cosmic Dark Ages as the ultimate probe of pnG. The Dark Ages refer to the epoch between recombination ($z\approx1100$), when neutral hydrogen is formed and the CMB photons are released to free stream, and the formation of the first luminous objects (i.e stars) at $z<30$. Soon after the CMB photons have been released, they redshift out of the visible wavelengths and the universe becomes truly dark. During this time, a neutral hydrogen gas permeates the universe while occasionally scattering with CMB photons, sometimes exciting the hyper-fine state of the electrons in the neutral hydrogen atoms. When the electron relaxes to its ground state, a photon with a wavelength of 21 centimeter is released. Once the hydrogen gas has cooled sufficiently and its temperature is decoupled from the CMB temperature, this 21-cm signal can be observed in absorption or emission to the background of CMB photons. Higher density regions containing more hydrogen have a brighter 21-cm signal, thereby tracing the matter density field which in turn traces the primordial fluctuations seeded by inflation. In this way, the 21-cm signal during $30 \leq z \leq 100$ can be used as a probe of pnG \cite{Cooray2006,Pillepich2006}, containing an amount of information that is estimated to be several magnitudes more than that of the CMB and LSS, making it the ultimate probe of the early universe and possibly opening up the cosmological collider \cite{Meerburg2016}. However, several comments are in order. First, the small amplitude of the 21-cm signal makes it hard to measure even the global (mean) brightness temperature, let alone tiny fluctuations around it, making it a serious challenge from a technological and experimental point of view. Secondly, Earth's ionosphere is opaque to the red-shifted 21-cm signal emitted at $z > 30$. Hence, probing this era will require an observatory in space or even on the far side of the moon, where also radio frequency interference (RFI) is minimized \cite{Silk2020}. Thirdly, although the 21-cm signal is expected to be a rather pristine tracer of the primordial initial conditions compared to the galaxy distributions that are the target of LSS experiments, it is nevertheless affected by non-linearity (i.e. gravity and peculiar velocity). Both the non-linear dependence of the 21-cm brightness temperature on the initial conditions as well as gravitational effects therefore leave their imprint on the distribution of fluctuations, inducing secondary non-Gaussianities that obscure the sought-after primordial non-Gaussianity by several orders of magnitude \cite{Pillepich2006,Munoz2015}. A detailed understanding of the 21-cm signal as well as these non-linear effects is warranted in order to reliably and accurately extract the primordial contribution and learn about inflation. \\

Secondary non-Gaussianity of the 21-cm brightness temperature during the Dark Ages was addressed previously in \cite{Munoz2015} for the case of the bispectrum, where it was found to contribute significantly to the observed bispectrum of temperature fluctuations, introducing it as a nuisance that should be marginalized over. In this work we will improve on some of the assumptions and simplifications made in \cite{Munoz2015}, in order to more accurately model the 21-cm signal during the Dark Ages. Furthermore, it was recently found that observational sensitivity to the primordial trispectrum can receive an enhanced scaling with respect to the smallest observable scale $k_{\textrm{max}}$ (or $\ell_{\textrm{max}}$ for CMB surveys) \cite{Kalaja2020}. We will see that this also affects primordial trispectra sourced by massive extra fields, possibly making it of prime observational interest to the cosmological collider. Extracting the primordial trispectrum will require an accurate modelling of the secondary trispectrum, which we derive here for the first time. In an effort to motivate experiments targeted at measuring the 21-cm signal from the Dark Ages we will determine the total information content of the Dark Ages that can be used to constrain primordial non-Gaussianity. Finally, we forecast the sensitivity of a simple experimental setup, that might be realized in the future.\\

This paper is organized as follows. In Section 2 we review the physics of the 21-cm signal during the Dark Ages. Section 3 will cover the non-Gaussian contributions to the statistics of 21-cm fluctuations. Then, in Section 4, we present Fisher forecasts on the amount of information available to constrain non-Gaussianity from the Dark Ages, as well as the sensitivity of more realistic experimental scenarios. We summarize our conclusions and outlook for future research in Section 5.

\subsection*{Conventions}
\label{sec:conventions}
In this work we denote spatial vectors in boldface, $\bk{}$. The magnitude of a vector is denoted as $k \equiv |\bk{}|$. Hats denote unit vectors $\hat{\x} \equiv \x/x$. We denote $k_1 + k_j \equiv k_{ij}$ and $\bk{i} + \bk{j} \equiv \bk{ij}$. Note that $k_{ij} \neq |\bk{ij}|$. We also define Mandelstam-like variables $\boldsymbol{s} = \bk{1} + \bk{2}$, $\boldsymbol{t} = \bk{1} + \bk{4}$ and $\boldsymbol{u} = \bk{1} + \bk{3}$.
For correlation functions we often adopt the primed notation to hide usual prefactors and Dirac delta functions:
\begin{eqnarray}
    \langle \delta(\bk{1}) \delta(\bk{2})\cdots \delta(\bk{n}) \rangle = (2\pi)^3 \delta_D(\bk{1}+\bk{2}+\cdots+\bk{n}) \langle \delta(\bk{1}) \delta(\bk{2})\cdots \delta(\bk{n}) \rangle'
\end{eqnarray}
Momentum integrals are sometimes written in the following condensed notation for readability:
\begin{eqnarray}
\int_{\bk{i}} = \int \frac{d^3\bk{i}}{(2\pi)^3}
\end{eqnarray}
For numerical computations we consider flat $\Lambda$CDM cosmology with parameters from \emph{Planck} \cite{Planck2018Parameters}, see Table \ref{tab:cosmology}.

\begin{table}[h!]
    \centering
    \begin{tabular}{c  c  c}
     & $\boldsymbol{\Lambda}$\textbf{CDM} \textbf{parameters} &\\
        H$_0 = 67.66$ &  $\Omega_\mathrm{b} h^2 = 0.02242$ & $\sum$m$_\nu = 0.06$\\
        $\Omega_\mathrm{k} = 0$ &  $\Omega_\mathrm{c} h^2 = 0.11993$  & $\tau = 0.0561$\\
         $n_s = 0.9665$& $A_\mathrm{s} = 2.1056\times 10^{-9}$  & $r = 0$\\
    \end{tabular}
    \caption{Best-fit Planck parameters (specifically, Table 2 of Ref.~\cite{Planck2018Parameters} with $TT$, $TE$, $EE+$low$E+$lensing+BAO) used in our numerical computations.}
    \label{tab:cosmology}
\end{table}

\section{21-cm fluctuations during the Dark Ages}
In this section we review the physics that gives rise to the 21-cm brightness fluctuations during the Dark Ages. First we will restrict the analysis to the background (or global) temperature. Subsequently we will discuss how 21-cm brightness temperature fluctuations arise due to fluctuations in the underlying density fields. In particular, the goal will be to derive how 21-cm fluctuations trace those of the baryonic density and velocity field. 

\subsection{Global 21-cm signal}
We define the ratio of the abundance of Hydrogen in the hyperfine singlet ($F=0$) and triplet state ($F=1$) via the spin temperature $\Ts$ as follows \cite{Furlanetto:2006jb}:
\begin{equation}
    \frac{n_1}{n_0}\equiv \frac{g_1}{g_0}\;e^{-\Tstar/\Ts},
\end{equation}
where $\Tstar\equiv E_{10}/\kb$ is the energy gap between the two hyperfine states and the degeneracies of the levels are $g_{0,1}=1,3$. During the Dark Ages, the spin temperature is determined by two processes: collision and radiative transitions between the two states. These processes are described by the rates $C_{ij}$ and $R_{ij}$, respectively. Specifically, $R_{10}$ encodes spontaneous and stimulated emission whereas $R_{01}$ encodes absorption. The rates $C_{01}$ and $C_{10}$ describe upward and downward transition between the hyperfine states due to collisions between Hydrogen atoms. We are interested in the Dark Ages, before the first luminous objects are formed, and therefore do not take transitions induced by Lyman-$\alpha$ photons into account (known as the Wouthuysen-Field effect) \cite{Furlanetto:2006jb}. 

During the Dark Ages, the steady state approximation is very accurate, since $C_{ij}+R_{ij}\gg H$ at all times \cite{Lewis2007b}, and the abundances are related via:
\begin{equation}
    n_0(C_{01}+R_{01})=n_1(C_{10}+R_{10}). 
\end{equation}
Then, in the limit $\Tstar\ll \Tgas,\Tg$, also valid at all times of interest \cite{Munoz2015}, we can write the spin temperature as:
\begin{equation}
    T_s=\frac{\Tg+\Yc \Tgas}{1+\Yc},
\end{equation}
where $\Yc\equiv \Tstar C_{10}/\Tgas A_{10}$ and $A_{10}$ the spontaneous decay rate. From the definition of $\Yc$, we find that atomic collisions drive $\Ts\to \Tgas$, while radiative interactions drive $\Ts\to \Tg$. 

In accordance with the standard convention in the literature, we define the brightness temperature of 21-cm radiation as \cite{Lewis2007b,Ali-Haimoud2014,Munoz2015}:
\begin{equation}
    \Tcm =\frac{\Ts-\Tg}{1+z}\;\taucm,
\end{equation}
where the optical depth for the 21-cm transtion, $\taucm$, is given by:
\begin{equation}
    \taucm = \frac{3\kb}{32\pi}\frac{\Tstar}{\Ts}\nHI\lcm^3\frac{A_{10}}{H+\pp\vp},
\end{equation}
with wavelength $\lcm =hc/E_{10}$. Note that $\taucm$ depends implicitly on the electron fraction via $\nHI\equiv \nH(1-\xe)$. During the Dark Ages, however, $\xe\sim 10^{-4}$, so that we may approximate $\nH=\nHI$ and the optical depth $\taucm$ becomes independent of the free electron fraction. We denote by $\pp\vp$ the gradient of the component of the peculiar velocity along the line-of-sight ($\vp$). We define the dimensionless velocity gradient:\footnote{In literature, this is sometimes defined without a minus sign (e.g. \cite{Ali-Haimoud2014}).}
\begin{equation}
    \dv\equiv -\frac{\pp\vp}{H}.
\end{equation}
Then, we may write the 21-cm brightness temperature in the following way:
\begin{equation}\label{eq:T21}
    \Tcm(\Ts,\nH,\dv)=\frac{3\kb}{32\pi}\frac{\Tstar}{\Ts}\nH\lcm^3\frac{A_{10}}{H ( 1- \dv)}\frac{\Ts-\Tg}{1+z}.
\end{equation}
Notice that the brightness temperature depends explicitly on $\nH$ and $\dv$ and implicitly on $\Tgas$ via $\Ts$, which in turn depends on $\nH$, making the temperature a non-linear tracer of the underlying density field. For later convenience, we define $\Tcmtilde$ as the brightness temperature in which the dependence on the velocity gradient term $\dv$ is factored out:
\begin{equation}
    \Tcmtilde(\Ts,\nH)\equiv \Tcm(\Ts,\nH,\dv)\times (1-\dv). 
\end{equation}

\subsection{Fluctuations in the 21-cm signal}
\label{subsec:21cmfluctuations}
The final expression for the 21-cm brightness temperature in equation \eqref{eq:T21} shows that it depends on the hydrogen density $\nH$, velocity gradient $\dv$ and gas temperature $\Tgas$ and the photon temperature $T_\gamma$. Therefore, fluctuations in the brightness ($\dTcm$) are in principle sourced by those in the hydrogen density, velocity gradient, gas and photon temperature. However, following \cite{Munoz2015,Ali-Haimoud2014}, we will restrict the analysis to sub-horizon scales. On these scales, we assume the photon temperature to be uniform, since their high sound speed ($c_s=c/\sqrt{3}$) results in photon fluctuations being suppressed relative to those in the other fields. In this section, we will find out how 21-cm brightness fluctuations trace those in the hydrogen density, gas temperature and velocity gradient. 

\subsubsection{21-cm brightness temperature}
We start by considering fluctuations in the density and gas temperature, and only include the velocity gradient fluctuations $\dv$ once we transform to momentum space.\footnote{Since spatial operators such as the gradient transform to simple products in momentum space, it is convenient to consider $\dv$ only once we have moved to momentum space.} That is, we effectively expand $\Tcmtilde$ in terms of fluctuations in $\nH$ and $\Tgas$. We define fluctuations in the latter by:
\begin{equation}
    \nH\tx=\nHb(t)\Big[1+\delta_b\tx\Big],\quad\quad\quad \Tgas\tx\equiv\Tgasb(t)
    \Big[1+\dT\tx\Big],
\end{equation}
where we have used that $\dH=\db$ up to negligible corrections of order $\order(\me/\mp)$ \cite{Ali-Haimoud2014}. 

\begin{figure}
    \centering
    \includegraphics[scale=0.95]{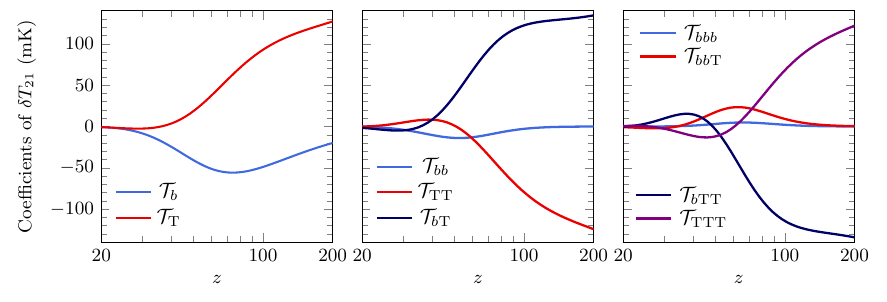}
    \caption{Coefficients $\T$ coupling brightness temperature fluctuations to the gas temperature and baryon density fluctuations. Left, middle and right panel show the coefficients and linear, quadratic and cubic order, respectively.}
    \label{fig:Tijk}
\end{figure}

Taylor expanding $\dTcm$ around $\dT=\db=0$ up to cubic order in fluctuations, we find:
\begin{align}\label{eq:T21expansion}
    \dTcmtilde\tx&=\TT\dT+\Tb\db\nonumber\\
    &+\TTT\dT^2+\TbT\db\dT+\Tbb\db^2\nonumber\\
    &+\TTTT\dT^3+\TbTT\db\dT^2+\TbbT\db^2\dT+\Tbbb\db^3,
\end{align}
where the coefficients $\Tbitj(t)$ depend only on time. We have numerically solved for the coefficients $\T$ following \cite{Ali-Haimoud2014} and show them in Figure \ref{fig:Tijk}.  

We can qualitatively understand the behavior of the coefficients $\T$ in the low and high-redshift limits as follows. 
\begin{itemize}
    \item \textit{Efficient Collisional Coupling.}---At $z\gtrsim 100$, collisions between Hydrogen atoms efficiently couple $\Ts=\Tgas$. The brightness temperature, in the absence of the velocity gradient term, then depends on $\nH$ and $\Tgas$ as:
    \begin{equation}
        \Tcm\propto \nH\bigg(1-\frac{\Tg}{\Tgas}\bigg). 
    \end{equation}
    Since $\Tcm\propto \nH$, we find that $\Tb\to \Tcmb$ and:
    \begin{equation}
        \Tbb\simeq\Tbbb\simeq\TbbT\to 0. 
    \end{equation}
    Note also that $\Tcm$ is proportional to the difference $\Tgas-\Tg$ in this regime, which asymptotes to zero in the high redshift limit ($z\gtrsim 200$) where Compton heating efficiently coupling the gas temperature to the photons, such that $\Tb\to 0$. For the coefficients $\TT$, $\TTT$ and $\TbT$ we have:
    \begin{equation}
        \TT,\TbT,\TTTT\propto \nHb\frac{\Tgb}{\Tgasb},\quad\quad \TTT,\TbTT\propto -\nHb\frac{\Tgb}{\Tgasb},
    \end{equation}
    which yields for $z\gtrsim 100$ the following relations between the coefficients:
    \begin{equation}
        \TT\simeq\TbT\simeq\TTTT\simeq-\TTT\simeq-\TbTT.
    \end{equation}
    Note that these coefficients do not tend to zero, since they are not suppressed by the factor $\Tgas-\Tg$. Instead they grow as:
    \begin{equation}
        \Tcm\propto \frac{\nH}{H}\propto (1+z)^{3/2},
    \end{equation}
    during matter domination via the dependence of the optical depth on the ratio $\nH/H$. The above behavior is indeed verified by the numerical solutions in Figure \ref{fig:Tijk}. 
    \item \textit{Ineffective Collisional Coupling}.---For $z\lesssim 50$, collisions become very inefficient and $\Ts$ approaches $\Tg$ again, which a small difference:
    \begin{equation}
        \Ts-\Tg\propto \nH\kappa_{10}^\mathrm{HH}(\Tgas).
    \end{equation}
    The brightness temperature then scales approximately as $\Tcm\propto \nH^2$, implying:
    \begin{equation}
        \Tb\simeq 2\Tbb\simeq 2\Tcmb,\quad\quad\quad \Tbbb\simeq 0.
    \end{equation}
    As $z\to 0$, the optical depth decreases rapidly and all coefficients tend to zero quickly. 
\end{itemize}

\subsubsection{Gas temperature}\label{sec:evodT}
The evolution equation for the gas temperature $\Tgas$ can be obtained from the first law of thermodynamics. Neglecting anything other than Compton heating, it reads \cite{Ali-Haimoud2014}:
\begin{equation}\label{eq:Tgasevo}
    \Tgasdot-\frac{2}{3}\frac{\nHdot}{\nH}\Tgas=\Gc\frac{\xe}{\xeb}(\Tg-\Tgas), 
\end{equation}
where we have defined the Compton heating rate $\Gc$ as:
\begin{equation}
    \Gc\equiv \frac{8\sigmaT a_r \Tg^4}{3(1+\xHe+\xe)\me}\xeb. 
\end{equation}
In the above equation, $\sigmaT$ is the Thomson cross section, $a_r$ is the radiation constant, $\me$ is the electron mass, $\nH\equiv \nHI+\np$ is the total density of Hydrogen (both in neutral and ionized form), $\xHe\equiv n_\mathrm{He}/\nH$ is the Helium fraction, $\xe\equiv n_e/\nH$ is the free electron fraction (and $\xeb$ its background value). Perturbing the above equation allows us to find a direct relationship between gas temperature fluctuations $\dT$ and baryon density fluctuations $\db$.

To obtain the evolution equation for $\dT$, one should in principle consistently include the coupling to fluctuations in $\Tg$, $\nH$, $\xe$ and $\xHe$. To simplify the analysis, we will make the same assumptions as in \cite{Ali-Haimoud2014}. As mentioned above, we consider small scales deep inside the horizon ($\kmin{} > 0.01$ Mpc$^{-1}$), so that photon temperature fluctuations are negligible and we set $\Tg=\Tgb$. Secondly, we assume the Helium fraction to be uniform, i.e. we take $\xHe=\xHeb$ and neglect any fluctuations. Finally, in the Compton heating rate $\Gc$, the free electron fraction only enters via the term $1+\xe+\xHe$. During the Dark Ages, $\xe\ll 1$, so that any electron fraction perturbation of $\Gc$ enters through $\xeb^2 \dxe{} \ll \dxe{} \ll 1$. Hence, we can neglect electron fraction perturbations in $\Gc$ such that it only depends on background values $\xeb,\xHeb$ and $\Tgb$.

The full non-linear evolution of the gas temperature fluctuation $\dT\tx$ can be obtained directly from equation \eqref{eq:Tgasevo} and reads \cite{Ali-Haimoud2014}:
\begin{equation}\label{eq:dTevo}
    \dTdot-\frac{2}{3}\dbdot\frac{1+\dT}{1+\db}+\frac{\Tgb}{\Tgasb}\Gc\dT=\Gc\big[(\Tgb/\Tgasb-1)\dxe+\dxe\dT\big]. 
\end{equation}
We obtain the background evolution of $\Tgasb$ as well as the recombination history $\xeb$ from \texttt{CAMB}.
Note that, as written in the form above, the right-hand-side couples the evolution of $\dT$ to the electron fraction fluctuation $\dxe$. In \cite{Munoz2015}, the coupling to $\dxe$ is neglected (i.e. the right-hand-side is taken to be zero), resulting in errors of order 10\% at linear order during the Dark Ages. In this work, we will include the effect of electron fraction perturbations up to third order. To do so, we require an additional equation describing the evolution of $\dxe$.

However, before deriving this additional equation, let us find out when the effect of $\dxe$ on gas temperature fluctuations is substantial by examining the r.h.s. of equation \eqref{eq:dTevo}. At high redshifts ($z\gtrsim 500$), Compton scattering efficiently couples the gas and CMB temperature, rendering the term proportional to $(\Tgb/\Tgasb-1)$ vanishingly small. In addition, the strong coupling combined with the fact that we consider scales on which $\Tg=\bar{T}_\gamma$ implies $\dT=0$, and hence the second term ($\propto \dxe\dT$) is negligibly as well. In the low redshift regime ($z\ll 200$), the gas cools adiabatically as $\Gc\ll H$ and fluctuations in the electron fraction also have no effect. However, at intermediate stages the effect is expected to be non-negligible.

\subsubsection{Free electron fraction}
Based on the discussion above, we only require an evolution equation for the free electron fraction that is accurate at late times ($z\lesssim 500$), and need not worry about the detailed recombination history at early times \cite{Lewis2007,Munoz2015}. We adapt the effective 3-level model for recombination \cite{Peebles1968}:
\begin{equation}\label{eq:evoxe}
    \xedot = -\Cp(\Tgas,\nH,\xe)\Big[\nH \xe^2\alphaB(\Tgas)-\betaB(\Tgas)(1-\xe)\;e^{-\Eonetwo/\kb\Tgas}\Big],
\end{equation}
where $\Eonetwo=10.2$ eV is the energy difference of the Ly-$\alpha$ transition, $\alphaB$ is the case-B effective recombination coefficient, $\betaB$ the corresponding photo-ionization rate. Finally, the Peebles factor $\Cp$ gives the ratio of the effective downward transition rate from the $n=2$ states to their effective lifetime:
\begin{equation}
    \Cp(\Tgas,\nH,\xe)=\frac{1+K\Ltwoone\nH(1-\xe)}{1+K\Ltwoone\nH(1-\xe)+K\betaB\nH(1-\xe)},
\end{equation}
where $K\equiv \lalpha^3/8\pi H$ and $\lalpha=121.5\;\mathrm{nm}$ is the Ly-$\alpha$ rest wavelength. 

Now, we simplify the evolution equation at late times. For $z\lesssim 500$, the second term in equation \eqref{eq:evoxe} describing the effect of photo-ionization, is completely negligible compared to the recombination term proportional to $\alphaB$. In addition, $\Cp\to 1$ for $z\lesssim 900$ \cite{Ali-Haimoud2014}, so that we obtain:
\begin{equation}\label{eq:xesimply}
    \xedot=-\alphaB\nH\xe^2,
\end{equation}
to excellent precision. For the recombination coefficient $\alphaB$, we use the fit \cite{Seager:1999km}:
\begin{equation}
    \alphaB(\Tgas)=F\frac{a_\alpha T_4^b}{1+cT_4^d},
\end{equation}
where $T_4\equiv\Tgas/10^4\;\mathrm{K}$, $a_\alpha=4.309\times 10^{-19}\;\mathrm{m}^3/\mathrm{s}$, $b=-0.6166$, $c=0.6703$ and $d=0.5300$. The Fudge factor $F=1.14$ is used to calibrate the effective 3-level result to a multi-level atom calculation \cite{Lewis2007}.

To obtain the evolution equation for $\dxe$, we perturb equation \eqref{eq:xesimply}. Notice that evolution of fluctuations in the electron fraction are coupled to those in the gas temperature and the baryon density via $\alphaB(\Tgas)$ and $\nH$, respectively. We expand $\alphaB$ in terms of $\dT$ up to cubic order as follows:
\begin{equation}
    \alphaB(\Tgas)=\alphaBbkg\Big[1+\sum_{n=1}^3 A_n\dT^n\Big],\;\;\;\;\;\;\;\; A_n\equiv \frac{1}{n!}\frac{\Tgasb^n}{\alphaBbkg}\frac{\partial^n\alphaBbkg}{\partial \Tgasb^n}, 
\end{equation}
where $\alphaBbkg\equiv \alphaB(\Tgasb)$. Upon defining $\Gr\equiv \alphaBbkg\nHb\xeb$ as the background recombination rate, we obtain the evolution of electron fraction perturbations up to terms of cubic order in baryon, gas temperature and electron fraction fluctuations:
\begin{align}
    \dxedot=-\Gr\Big[&\dxe+A_1\dT+\db\nonumber\\
    &+\dxe^2+A_2\dT^2+2A_1\dxe\dT+A_1\dT\db+2\dxe\db\nonumber\\
    &+A_3\dT^3+2A_2\dxe\dT^2+A_1\dxe^2\dT+A_2\dT^2\db+\dxe^2\db+2A_1\dxe\dT\db\Big]\label{eq:evodxe}. 
\end{align}

\subsection{Perturbative analysis of fluctuations}\label{sec:perturbations}
Recall that the goal of this section is to find the relation between $\dTcm$ and the underlying baryonic density and velocity field. At this point $\dTcm$ is still a function of fluctuations in the gas temperature as well. However, $\dT$ can be traded effectively for $\db$ by recognizing that gas temperature fluctuations trace fluctuations in the density field. The physical mechanism behind this tracing relationship is simple and may be explained as follows. Consider an overdense region in the gas ($\db>0$). Due to the higher density, the thermal motion of the particles in the gas in increased and hence the temperature will increase as well: $\dT$ traces $\db$. 

In order to find the tracing relationship between $\dT$ and $\db$, we start by writing the baryonic density contrast perturbatively, up to cubic order:
\begin{equation}
    \db=\sum_{n=1}^3\db^{(n)}\equiv \sum_{n=1}^3 \delta_n,
\end{equation}
where we assume $\db^{(n)}=\order\left[\left(\db^{(1)}\right)^n\right]$. In the second equality, we defined $\delta_n\equiv\delta_{b}^{(n)}$ for notational brevity. We assume that the time-dependence of the baryons is identical to that of CDM, i.e. $\db^{(n)}\propto a^{(n)}(t)$, which implies $\dbdot^{(n)}=nH\db^{(n)}$.\footnote{Note that we \textit{do not} assume the \textit{spatial} dependence of baryons and CDM to be identical. In fact, we will include the spatial dependence due to baryonic pressure, in contrast to \cite{Munoz2015}.} Similarly, we expand $\dT$ and $\dxe$ to cubic order as well:\footnote{In the notation below, the superscripts $\mathrm{T}$ and $x$ are labels, not powers.}
\begin{equation}\label{eq:expansiondTdxe}
    \dxe\tx=\sum_{n=1}^3 \dx_n\tx,\quad\quad\quad\dT\tx=\sum_{n=1}^3 \dTup_n\tx. 
\end{equation}

At each order, the evolution equations for the perturbations in the temperature and electron fraction can be obtained by inserting the above expansions into equation \eqref{eq:dTevo} and equation \eqref{eq:evodxe}, respectively. In App. \ref{app:evopert} we provide the resulting evolution equations and show how they can be used to solve for the coupling coefficients, relating $\dT$ and $\dxe$ to $\db$ up to third order:
\begin{align}
	\delta_\mathrm{T}\zx&=\CT_{1,1}(z)\;\delta_1+\CT_{2,1}(z)\;\delta_1^2+\CT_{2,2}(z)\;\delta_2+\CT_{3,1}(z)\;\delta_1^3+\CT_{3,2}(z)\;\delta_1\delta_2+\CT_{3,3}(z)\;\delta_3,\label{eq:dTexpansion}\\
	\delta_x\zx&=\Cx_{1,1}(z)\;\delta_1+\Cx_{2,1}(z)\;\delta_1^2+\Cx_{2,2}(z)\;\delta_2+\Cx_{3,1}(z)\;\delta_1^3+\Cx_{3,2}(z)\;\delta_1\delta_2+\Cx_{3,3}(z)\;\delta_3,\label{eq:dxexpansion}
\end{align}
where the coupling coefficients are functions of redshift only. In the coefficient $\mathcal{C}_{n,m}$, $n$ denotes the total order of the combination of density perturbations it multiplies, and $m=1,\dots,n$ labels the different coefficients at each order $n$. Note that $\mathcal{C}_{n,n}$ is the coefficient coupling to density perturbation $\delta_n$, whereas $\mathcal{C}_{n,m\neq n}$ couple to the product of lower order density fields with combined order $n$. At first order, we will often write $\mathcal{C}_1\equiv \mathcal{C}_{1,1}$ for notational simplicity.

\begin{figure}
    \centering
    \includegraphics[scale=0.9]{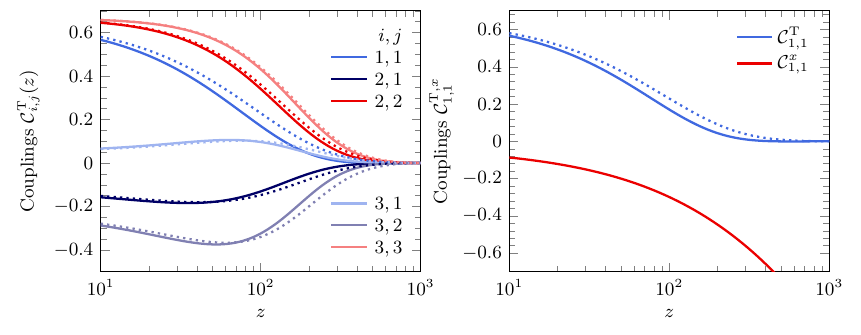}
    \caption{Left Panel: Coupling coefficients $\CT_{n,m}$ as a function of redshift. The dotted lines show the result in case electron fraction fluctuations are neglected (i.e. $\Cx_{n,m}\equiv 0$). Right Panel: First order coefficients as a function of redshift. Again, the dotted line shows $\CT_1$ in the absence of electron fraction fluctuations.} 
    \label{fig:CTcoeff}
\end{figure}

\subsubsection*{Evolution of Coupling Coefficients}
In Figure \ref{fig:CTcoeff}, we show the numerical solutions for the gas coupling coefficients $\CT_{n,m}$. The solid lines show the solutions including the effect of electron fraction perturbations. The dotted lines exclude their effect by setting $\Cx_{n,m}\equiv 0$ in all evolution equations. It is insightful to examine the numerical results at early (high-$z$) and late times (low-$z$), where the numerical analysis can be compared with analytic results:
\begin{itemize}
	\item \textit{High redshift limit}---At high redshifts, $z\gtrsim 500$, Compton heating is efficient ($\Gc/H\gg 1$) and keeps the matter and photons in equilibrium so that $\Tg=\Tgas$. Since we consider small scales on which we neglect photon temperature fluctuations, we therefore expect no gas temperature fluctuations. This is indeed verified by the solutions for $\CT_{n,m}$, which all tend to zero at high redshifts. Only when the gas temperature starts to decouple from the photon temperature at $z\lesssim 500$ due to adiabatic cooling, fluctuations in the former start growing. 
	\item \textit{Low redshift limit}---At low redshifts, Compton heating becomes completely inefficient, $\Gc/H\ll 1$, and adiabatic cooling completely determines the gas temperature:
	\begin{equation}
		\Tgasdot=\frac{2}{3}\frac{\nHdot}{\nH}\Tgas,
	\end{equation}
	which is just equation \eqref{eq:Tgasevo} in the absence of the Compton heating term (proportional to $\Gc$). From the above equation, we easily obtain:
	\begin{equation}
		\Tgas\propto\nH^{2/3},
	\end{equation}
	characterizing a gas that is cooling adiabatically. The above relationship between the gas temperature and Hydrogen density yields the following relationship between $\dT$ and $\db$:
	\begin{align}
		\dT &=\frac{2}{3}\db-\frac{1}{9}\db^2+\frac{4}{81}\db^3+\mathcal{O}(\db^4)\nonumber\\
		&=\frac{2}{3}(\delta_1+\delta_2+\delta_3)-\frac{1}{9}(\delta_1^2+2\delta_1\delta_2)+\frac{4}{81}\delta_1^3+\mathcal{O}(\delta_1^4),
	\end{align}
	where we have expanded $\db$ up the third order in the second line. Comparing to equation \eqref{eq:DT}, we find that the coefficients asymptote to constant values in the adiabatic or low redshift regime:
	\begin{equation}
		\CT_{1}=\CT_{2,2}=\CT_{3,3}\to 2/3,\quad\quad \CT_{2,1}\to -1/9,\quad\quad \CT_{3,1}\to 4/81,\quad\quad \CT_{3,2}\to -2/9, 
	\end{equation}
	which is indeed verified by the numerical solution. 
\end{itemize}

\noindent Below, we will first discuss the case in which electron fraction perturbations are neglected, and make contact with results obtained in literature. Then, we discuss the effect of electron fluctuations and provide a physical interpretation of their effect at first order. 

As mentioned above, excluding the effect of electron fraction fluctuations leads to the dotted curves for the gas temperature coefficients. Our results agree with those of \cite{Munoz2015}, in which the gas temperature coefficients where computed up to second order and in the absence of electron fluctuations. We find excellent agreement for the coefficients $\CT_{1}$, $\CT_{2,1}$ and $\CT_{2,2}$, which are called $C_1$, $C_2$ and $C_2'$ respectively in \cite{Munoz2015} and shown in their Figure 1. In addition, the first-order coupling coefficient agrees with the result obtained neglecting electron fraction fluctuations in \cite{Pillepich2006}, where it is called $g_1$ (see their Figure 4).   

Comparing the solid and dotted curves in Figure \ref{fig:CTcoeff}, we find that the effect of electron fraction perturbations is most pronounced at linear order, i.e. for $\CT_1$. At $z=30,50$ and $100$, the solution excluding electron fraction perturbations is larger by $8\%, 14\%$ and $34\%$, respectively. Our results confirm previous claims that neglecting $\dxe$ leads to errors of order $10\%$ for the linear evolution during the Dark Ages \cite{Munoz2015}. However, we also note that the effect of $\dxe$ rapidly becomes more significant at higher redshift during the Dark Ages, and should therefore be accounted for in a detailed analysis. At second and third order, the effect of electron fraction perturbations is smaller, but still starts to exceed the $10\%$ level at $z\gtrsim 100$, before decreasing again as $z \rightarrow 1100$.

In the low redshift limit, the solutions including and excluding $\dxe$ agree again, which can be understood as follows. Due to the expansion of the universe, the Hydrogen gas becomes diluted at low redshift, rendering the recombination rate small: $\Gr/H\ll 1$. The evolution equations for $\Cx_{n,m}$ then reduce to the form:
\begin{equation}\label{eq:limitCxevo}
	\frac{d\Cx_{n,m}}{da}\propto -\frac{\Cx_{n,m}}{a},
\end{equation}
implying that the electron fraction coefficients decay as $\Cx_{n,m}\propto a^{-1}$ at low redshift. In turn, the effect of electron fraction perturbations on the evolution of $\CT_{n,m}$ becomes small at low redshifts, and the dotted and solid curves become identical. Below, we will discuss the effect of $\dxe$ in more detail at first order.

At linear order, the effect of electron fraction fluctuations on the evolution of gas temperature fluctuations can be understood in an intuitive way from the physical principles at play \cite{Lewis2007}. Consider a region with a density that is higher than everage, so that $\delta_1>0$. In such a region, Hydrogen density is higher than average, so that the local recombination rate is slightly higher and hence more recombinations occur. As a consequence, the electron fraction will be lower than average, resulting in $\dx_1<0$ and reflected by $\Cx_{1}<0$. In turn, this reduces the coupling of the gas temperature to the photon temperature via Compton scattering, since there are less free electrons available to sustain the coupling. At background level, $\Tgas$ is below $\Tg$ during the Dark Ages due to adiabatic cooling. In an overdense region, the reduced coupling between $\Tg$ and $\Tgas$ makes the latter even smaller relative to the background difference. This results in $\dTup_1$ (or equivalently $\CT_1$) being smaller than it would have been if electron fraction fluctuations were neglected.

\subsection{Coupling 21-cm fluctuations to density and velocity fluctuations}
At this point, we are in the position to expand 21-cm fluctuations into fluctuations $\db$ and $\dv$, as we have uniquely related temperature fluctuations to density fluctuations. We will make this expansion explicit in real space first. Subsequently, we transform to momentum space, where we can relate $\dv$ to the velocity divergence $\theta_b$. 

\subsubsection*{Real Space}
\label{subsec:realspace}
\begin{figure}
    \centering
    \includegraphics[scale=1]{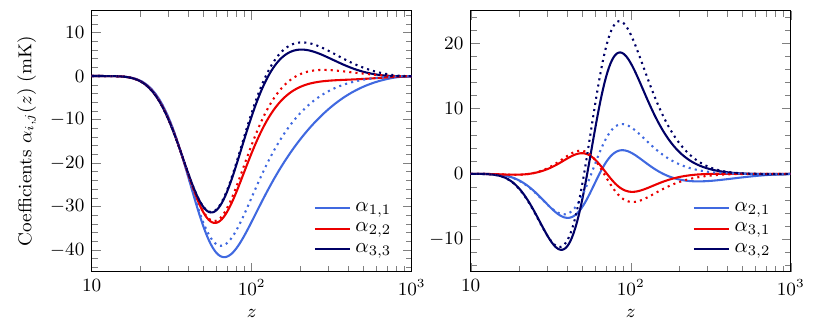}
    \caption{Coupling coefficients $\alpha_{n,n}$ (left panel) and $\alpha_{n,m\neq n }$ (right panel) as function of redshift. Again, the dotted lines exclude the effect of electron fluctuations.}
    \label{fig:alphacoeff}
\end{figure}

Now we have all the ingredients to compute the coupling between the 21-cm brightness temperature and density fluctuations up to third order. We will denote these couplings by $\alpha_{n,m}$, and they are defined in relation to $\dTcmtilde$ as:
\begin{align}
    \dTcmtilde\zx=\alpha_{1,1}(z)\delta_1+\alpha_{2,1}(z)\delta_1^2+\alpha_{2,2}(z)\delta_2+\alpha_{3,1}(z)\delta_1^3+\alpha_{3,2}(z)\delta_1\delta_2+\alpha_{3,3}(z)\delta_3. 
\end{align}
Using equation \eqref{eq:dTexpansion} and equation \eqref{eq:dxexpansion}, we can write the couplings $\alpha_{n,m}$ explicitly in terms of the $\T$ coefficients and the couplings $\CT_{n,m}$, yielding:
\begin{align}
    \alpha_{n,n}&\equiv \TT\CT_{n,n}+\Tb,\\
    \alpha_{2,1}&\equiv \TT\CT_{2,1}+\TTT[\CT_{1}]^2+\TbT\CT_1+\Tbb,\\
    \alpha_{3,1}&\equiv \TT\CT_{3,1}+2\TTT\CT_1\CT_{2,1}+\TbT\CT_{2,1}+\TTTT[\CT_1]^3+\TbTT[\CT_1]^2+\TbbT\CT_1+\Tbbb,\\
    \alpha_{3,2}&\equiv \TT\CT_{3,2}+2\TTT\CT_1\CT_{2,2}+\TbT(\CT_{2,2}+\CT_1)+2\Tbb.
\end{align}
We plot the couplings in Figure \ref{fig:alphacoeff} over the redshift range $10-1000$. Recall that in the adiabatic limit $\CT_{n,n}\to 2/3$ so that the couplings $\alpha_{n,n}$ become identical, which is indeed the case for $z\lesssim 50$. In addition, note that $\CT_{3,2}\simeq 2\CT_{2,1}$ and $\CT_{2,2}\simeq \CT_{1}$ in the adiabatic limit, such that $\alpha_{3,2}\simeq 2\alpha_{2,1}$, which is indeed the case for $z\lesssim 30$. Note that in the approach above, we have effectively traded the dependence on $\dT$ for dependence on $\db$. Ultimately, we are able to do so by using the first law of thermodynamics.

So far, we have worked with $\dTcmtilde$, i.e. we ignored the fluctuations $\dv$. To include the perturbations in $\dv$, we use the relation $\Tcm=\Tcmtilde/(1-\dv)$ and expand $\dv$ to third order, i.e. $\dv=\dv^{(1)}+\dv^{(2)}+\dv^{(3)}$.\footnote{Although the fluctuations $\dv$ appear similar to \emph{redshift space distortions} (RSD), they come from the velocity dependence of the optical depth and hence are of a qualitative different nature. In principle one should also include RSDs separately when fully modelling the 21-cm signal, introducing the linear growth rate $f = d\ln{D_+} / d\ln{a}$ as an additional parameter to marginalize over. We have chosen to ignore RSDs in our forecast as they do not introduce additional shape dependence and therefore overlap with the primordial shapes.} Assuming $|\dv|\ll 1$ and using a geometric series expansion, we obtain:
\begin{align}
	\dTcm\zx=&\;\Tcmb(z)\Big[\dv^{(1)}+\big[\dv^{(1)}\big]^2+\dv^{(2)}+\big[\dv^{(1)}\big]^3+2\dv^{(1)}\dv^{(2)}+\dv^{(3)}\Big]\zx\nonumber\\
	&+\alpha_{1,1}(z)\Big[1+\dv^{(1)}+\dv^{(2)}+\big[\dv^{(1)}\big]^2\Big]\db^{(1)}\zx\nonumber\\
	&+\alpha_{2,1}(z)\Big[1+\dv^{(1)}\Big]\big[\db^{(1)}\big]^2\zx+\alpha_{2,2}(z)\Big[1+\dv^{(1)}\Big]\db^{(2)}\zx\nonumber\\
	&+\alpha_{3,1}(z)\big[\db^{(1)}\zx\big]^3+\alpha_{3,2}(z)\big[\db^{(1)}\delta_b^{(2)}\big]\zx+\alpha_{3,3}(z)\db^{(3)}\zx.
\end{align}

\subsubsection*{Momentum Space}
In the end, we wish to obtain correlation functions of 21-cm fluctuations in momentum space. Therefore, we transform the above expression for $\dTcm\zx$ to momentum space. To do so we have to transform all different components separately. Products of perturbations in real space, e.g. the square of $\db^{(1)}$, become convolutions in momentum space. In addition, the transform of $\dv$ can be related directly to the transform of the velocity divergence $\theta_b\equiv \nabla\cdot \boldsymbol{v}_b$. The Fourier transform of $\dv$ is written in terms of that of $\theta_b$ as:
\begin{equation}
	\dv(z,\boldsymbol{k})=-\mu^2(\boldsymbol{k})\frac{\theta_b(z,\boldsymbol{k})}{\mathcal{H}}, 
\end{equation}
where $\mathcal{H}=aH$ and $\mu(\boldsymbol{k})\equiv \hat{\n}\cdot\boldsymbol{k}/k = k_\parallel / k$ is the cosine of the angle between the mode $\boldsymbol{k}$ and the line-of-sight $\boldsymbol{n}$. At first order, this expression can be simplified even further by invoking the continuity equation, to obtain \begin{equation}\label{eq:thdbfirstorder}
	\dv^{(1)}(z,\boldsymbol{k})=\mu^2(\boldsymbol{k})\db^{(1)}(z,\boldsymbol{k}). 
\end{equation}
Note that in transforming to momentum space, the angle between the line-of-sight and the Fourier mode is encoded in $\mu^2(\bk{})$. This leads to a distinct difference between coupling coefficients in momentum space and those in real space ($\Tcmb$ and $\alpha_{i,j}$); the former depend on the mode $\bk{}$ and time, whereas the latter depend on only on time and not on $\x$.

Now, we have all the relevant ingredients to compute the Fourier transform of the brightness fluctuation $\dTcm$. At each order, we couple the brightness perturbations to those in the baryon density and velocity divergence via the coefficients $c^{(i)}_j(z,\bk{},\q)$, of which some explicitly depend on the principal mode $\bk{}$ and internal modes $\q$ resulting from convolution integrals via the angle $\mu$ as explained above. The label $i$ denotes the total order of the quantity the coefficient couples to, and $j$ labels all different couplings at order $i$. Up to third order, the expansion of $\dTcm\zk$ is given by:
\begin{align}
	\dTcm\zk =&\;\dTcm^{(1)}\zk+\dTcm^{(2)}\zk+\dTcm^{(3)}\zk,
\end{align}
with $\dTcm^{(i)}\zk$ given in terms of the coefficients $c^{(i)}_j$ as:
\begin{align}\label{eq:dT21expansion}
	\dTcm^{(1)}\zk=&\;c_1^{(1)}\zk\db^{(1)}(\bk{})\nonumber\\
	\dTcm^{(2)}\zk=&\;c_1^{(2)}(z) \db^{(2)}(\bk{})-\frac{1}{\mathcal{H}}c_2^{(2)}\zk\theta_b^{(2)}(\bk{})+\qoneint c_3^{(2)}(z,\bk{},\qone)\db^{(1)}(\qone)\db^{(1)}(\bk{}-\qone)\nonumber\\
	\dTcm^{(3)}\zk=&\;c_1^{(3)}(z)\db^{(3)}(\bk{})-\frac{1}{\mathcal{H}}c_2^{(3)}\zk\theta_b^{(3)}(\bk{})\nonumber\\
	&+\int_{{\qone\qtwo}} c_3^{(3)}(z,\bk{},\qone,\qtwo)\db^{(1)}(\qone)\db^{(1)}(\qtwo)\db^{(1)}(\bk{}-\qonetwo)\nonumber\\
	&-\frac{1}{\mathcal{H}}\qoneint c_4^{(3)}(\bk{},\qone)\theta_b^{(2)}(\qone)\db^{(1)}(\bk{}-\qone)\nonumber\\
	&+\qoneint c_5^{(3)}(\bk{},\qone)\db^{(2)}(\qone)\db^{(1)}(\bk{}-\qone),
\end{align}
where we have suppressed the time dependence in $\db^{(n)}$ and $\theta_b^{(n)}$ for brevity. We explicitly excluded the factors of $1/\mathcal{H}$ from the definition of the couplings $c_j^{(i)}$, so that the latter all have the dimension of temperature. 

The couplings are a function of $\alpha_{i,j}$, $\Tcmb$ and $\mu(\bk{})$ and read:
\begin{align}
	c_1^{(1)}&\equiv \alpha_{1,1}(z)+\Tcmb(z)\mu^2(\bk{}),\nonumber\\
	c_1^{(2)}&\equiv \alpha_{2,2}(z),\nonumber\\
	c_2^{(2)}&\equiv \Tcmb\mu^2(\bk{}),\nonumber\\
	c_3^{(2)}&\equiv\Tcmb(z)\mu^2(\q)\mu^2(\bk{}-\q)+\alpha_{1,1}\mu^2(\q)+\alpha_{2,1}(z),\nonumber\\
	c_1^{(3)}&\equiv \alpha_{3,3}(z),\nonumber\\
	c_2^{(3)}&\equiv \Tcmb(z)\mu^2(\bk{}),\nonumber\\
	c_3^{(3)}&\equiv \Tcmb(z) \mu^2(\qone)\mu^2(\qtwo)\mu^2(\bk{}-\qonetwo)+\alpha_{1,1}(z)\mu^2(\qone)\mu^2(\qtwo)+\alpha_{2,1}(z)\mu^2(\qone)+\alpha_{3,1}(z),\nonumber\\
	c_4^{(3)}&\equiv 2\Tcmb(z)\mu^2(\q)\mu^2(\bk{}-\q)+\alpha_{1,1}(z)\mu^2(\q),\nonumber\\
	c_5^{(3)}&\equiv \alpha_{2,2}(z)\mu^2(\bk{}-\q)+\alpha_{3,2}(z). 
\end{align}
Note that since $\mu(\bk{})=\mu(-\bk{})$, all coefficient are symmetric under inverting all momentum arguments, e.g. $c_3^{(3)}(\bk{},\qone,\qtwo)=c_3^{(3)}(-\bk{},-\qone,-\qtwo)$. The first order coupling, $c_1^{(1)}$, actually comprises two couplings, to $\db^{(1)}$ and $\theta_b^{(1)}$, respectively. However, using equation \eqref{eq:thdbfirstorder} allows us to write down one overall coefficient coupling to $\db^{(1)}$. In summary, we have found a direct relationship between $\dTcm$ and $\db$ to third order in perturbations. \\

We are now able to express the statistics of 21-cm temperature fluctuations in terms of the underlying tracer field $\db$, which in turn traces primordial fluctuations, e.g. the power spectrum:
\begin{eqnarray}
P_{\dTcm}(\bk{1}) = \langle \dTcm(\bk{1}) \dTcm(-\bk{1}) \rangle' = \left(c_1^{(1)}(\bk{1})\right)^2\langle \db(\bk{1}) \db(-\bk{1}) \rangle' = \left(c_1^{(1)}(\bk{1})\mathcal{M}_b(\bk{1})\right)^2 P_\zeta(\bk{1}) \nonumber \\
\end{eqnarray}
where $\mathcal{M}_b$ is the linear transfer function of baryon fluctuations (obtainable through e.g. \camb{}) and the time (redshift) dependence of the prefactor is implicit.

\section{Non-Gaussianity of 21-cm brightness temperature anisotropies}
In this section we review the non-Gaussian contributions to the statistics of 21-cm brightness temperature fluctuations during the Dark Ages. First we will go over the details of primordial non-Gaussianity and present templates that can be used to extract imprints of inflation from 21-cm data. Subsequently, we discuss the generation of secondary non-Gaussianity in the 21-cm signal due to non-linear evolution of fluctuations, such as gravitational collapse.

\subsection{Primordial non-Gaussianity}
As was mentioned in the introduction, inflation naturally seeds structure formation through the generation of quantum fluctuations. Over time, higher density regions collapse under gravitational attraction, forming stars and eventually galaxies. The statistics of the initial conditions set by inflation contain a wealth of information about the physics driving the accelerated expansion, as well as any extra particles present during this period. Primordial (scalar) fluctuations are captured by the gauge invariant quantity $\zeta$ with mean zero. Generally, the $N$-point statistical correlation function of primordial fluctuations is given by:\footnote{The number of degrees of freedom of the $N$-point correlation function is reduced by homogeneity and isotropy to be $3(N-2)$.}
\begin{equation}
    \langle \zeta_{\kone} \zeta_{\ktwo} ... \zeta_{\boldsymbol{k}_N} \rangle = (2\pi)^3 \delta_D(\kone + \ktwo + ... +  \boldsymbol{k}_N)F^{(N)}_{\zeta}(\kone,\ktwo,...,\boldsymbol{k}_N) 
\end{equation}
where $F_\zeta^{(N)}$ encodes the size and \emph{shape} of the $N$-point correlation as a function of the momentum configuration. The lowest order statistic is known as the power spectrum:\footnote{From here on out we will use the primed correlator notation to remove the usual factor $(2\pi)^3 \delta_D(...)$.}
\begin{eqnarray}
    \langle \zeta_{\kone} \zeta_{\ktwo} \rangle' = P_\zeta(k_1)
\end{eqnarray}
Moving beyond the power spectrum, in this work we will be concerned with the first two non-Gaussian contributions $N=3$ (bispectrum) and $N=4$ (trispectrum):
\begin{eqnarray}
    \langle  \zeta_{\kone} \zeta_{\ktwo} \zeta_{\kthree} \rangle' &=& B_\zeta(k_1,k_2,k_3) \nonumber \\
     \langle  \zeta_{\kone} \zeta_{\ktwo} \zeta_{\kthree} \zeta_{\kfour} \rangle' &=& T_\zeta(k_1,k_2,k_3,s,t) 
\end{eqnarray}
where $s$ and $t$ are the Mandelstam-like variables as presented at the end of the introduction section. We will now discuss some common shapes of non-Gaussianity that arise in inflationary model building, first for the bispectrum, then the trispectrum.

\subsubsection{Primordial bispectra}
The simplest way to generate non-Gaussianity is by expanding fluctuations locally in terms of Gaussian fields as follows:
\begin{eqnarray}
    \zeta = \zeta_g + \frac{3}{5}\fnl{loc}\zeta_g^2
\end{eqnarray}
where the subscript $g$ denotes Gaussian fields, that have vanishing $(N>2)$-point functions. Such an expansion gives rise to the so called local shape of the primordial bispectrum:
\begin{equation}
    \label{eq:BLocal}
    B^{\text{loc}}(k_1,k_2,k_3) = \frac{6}{5}\fnl{loc} \left( P_\zeta(k_1)P_\zeta(k_2) + P_\zeta(k_2)P_\zeta(k_3) + P_\zeta(k_1)P_\zeta(k_3)\right)
\end{equation}
which peaks in the squeezed triangle configuration (e.g. $k_1 \ll k_2 \approx k_3$). Local non-Gaussianity naturally arises in multi-field models of inflation, where extra light fields are present that modulate the dynamics. When the inflaton is able to interact with itself, bispectra of the equilateral type are generated. Their shape is well described by the template:
\begin{eqnarray}
\label{eq:BEquil}
    B^{\text{equil}}(k_1,k_2,k_3) = \frac{18}{5} \fnl{equil} \Big[- (P_\zeta(k_1) P_\zeta(k_2) + \textrm{2 perms.})-2\Pz^{2/3}(k_1) \Pz^{2/3}(k_2) \Pz^{2/3}(k_3)\nonumber \\+  (P_\zeta(k_1) P_\zeta(k_2)^{1/3} P_\zeta(k_3)^{2/3} + \textrm{5 perms.})\Big] \nonumber \\
\end{eqnarray}
and peaks in the equilateral momentum configuration ($k_1 \approx k_2 \approx k_3$). All shapes that are due to self interactions can be systematically classified using the Effective Field Theory of Inflation \cite{Cheung2007}. Finally, a third shape that is often encountered is the orthogonal shape
\begin{eqnarray}
\label{eq:BOrtho}
   B^{\text{ortho}}(k_1,k_2,k_3) = \frac{18}{5} \fnl{ortho} \Big[- 3(P_\zeta(k_1) P_\zeta(k_2) + \textrm{2 perms.} ) -8\Pz^{2/3}(k_1) \Pz^{2/3}(k_2) \Pz^{2/3}(k_3) \nonumber \\ 3(P_\zeta(k_1) P_\zeta(k_2)^{1/3} P_\zeta(k_3)^{2/3} + \textrm{5 perms.})\Big] \nonumber \\
\end{eqnarray}
which is generated by more exotic models such as Galileon inflation. These three primordial shapes are the most commonly encountered bispectra in inflationary model building, hence in this work we will often refer to them together as the common primordial bispectra.\\
\\
One of the most exciting prospects of measuring non-Gaussianity ics the possibility of probing the particle spectrum during inflation. Heavy particles can naturally be present during this high energy phase, leaving a characteristic oscillatory shape in the squeezed limit of the primordial bispectrum, with frequency set by its mass \cite{Arkani-Hamed2015,Baumann2017}. More concretely, under exchange of a scalar particle with mass $m^2 > 3H/2$ this oscillation is given by \cite{Arkani-Hamed2015}:
\begin{eqnarray}
\lim_{\kthree \rightarrow 0} \langle \zeta_{\kone} \zeta_{\ktwo} \zeta_{\kthree} \rangle' \propto \frac{1}{k_1^3 k_2^3} \frac{\pi^2}{\cosh^2 \pi \mu} \left(\frac{k_3}{k_1}\right)^\frac{3}{2} \Big[ \left(\frac{k_3}{4k_1}\right)^{-i\mu} \frac{(1-i\sinh \pi\mu)(\frac{5}{2}-i\mu)(\frac{3}{2}-i\mu)\Gamma(i\mu)}{\Gamma(\frac{1}{2}+ i\mu)}\nonumber \\ + c.c \Big].\nonumber \\
\end{eqnarray}
where $\mu = \sqrt{m^2/H^2 - 9/4}$,\footnote{In order to stick to conventions in literature we use $\mu$ for both the line-of-sight angle $\mu(\bk{})$ and the mass parameter $\mu(m)$. This is unfortunate but their meaning should be sufficiently clear from the context.} clearly showing an oscillation in $\log(k_3/k_1)$ with amplitude, frequency and phase set by the mass of the scalar particle. The oscillations induced by these heavy fields, could be used to follow the evolution of the scale factor, possibly providing additional information about inflation. For this reason these fields are often referred to as \emph{primordial standard clocks} \cite{Chen:2012ja,Chen:2014cwa,Chen:2014joa,Chen:2015lza}. In \cite{Meerburg2016} a template is proposed for capturing the oscillatory behaviour of the bispectrum:
\begin{eqnarray}
\label{eq:BClock}
   B^{\text{clock}}(k_1,k_2,k_3) = \frac{3^{9/2}}{10}f_{\textrm{NL}}^{\textrm{clock}} \frac{A^2_{\zeta}}{(k_1 k_2 k_3)^2} \alpha_{123}^{-1/2} \sin\left(\mu \log\left(\frac{\alpha_{123}}{2} \right) + \delta \right)\Theta(\alpha_{123}-\alpha_0)\nonumber \\ + \textrm{2 perms.} 
\end{eqnarray}
where $\alpha_{123} = (k_1 + k_2)/k_3$ and the Heaviside step function $\Theta$ is included since the oscillation is only present for $\alpha_{123} > 2$ (or permutations). The authors consider a value of $\alpha_0 = 10$ in order to cut off near-equilateral configurations such that overlap with the equilateral shape is reduced. Such a cutoff also drastically reduces the amount of triangle configurations available to measure the signal, thereby lowering its signal-to-noise ratio (SNR). As we will see later in this work, the reduced overlap with the equilateral shape does not make up for the loss of SNR and a cutoff $\alpha_0 > 2$ is found to be optimal.\\
\\
If the exchange particle has a mass $0 \leq m^2 \leq 3H/2$ the oscillatory behaviour turns into a power-law \cite{Chen2009}, called the intermediate shape. Although there exists no analytical expression, the shape is captured well by the following template \cite{Chen2009,Meerburg2016}:
\begin{eqnarray}
\label{eq:BInt}
   B^\textrm{int}(k_1,k_2,k_3) = \frac{6}{5} A^2_\zeta f_\textrm{NL}^\textrm{int} 3^{\frac{7}{2}-3\nu} \frac{k_1^2 + k_2^2 + k_3^2}{(k_1 + k_2 + k_3)^{\frac{7}{2} - 3 \nu}} (k_1 k_2 k_3)^{-\frac{3}{2}-\nu} \tilde{\Theta}(k_1,k_2,k_3,\alpha_0)
\end{eqnarray}
where $\nu = \sqrt{9/4 - (m/H)^2} = -i\mu$ and $0 < v < 3/2$. This template interpolates between the behaviour of the local ($\nu > 3/2$) and equilateral ($\nu < 1/2$) shapes in the squeezed limit, hence coining the name intermediate. In order to reduce overlap with equilateral shapes, we introduce a cutoff similar to the clock template:
\begin{eqnarray}
   \tilde{\Theta}(k_1,k_2,k_3,\alpha_0) = \Theta(\alpha_{123} - \alpha_0) + \Theta(\alpha_{132} - \alpha_0) + \Theta(\alpha_{231} - \alpha_0)
\end{eqnarray}
In \cite{Meerburg2016} the cutoff was again set to $\alpha_0 = 10$ whereas in this work we will find $\alpha_0 = 2$ to maximize the SNR when marginalising over the other primordial shapes.\\
\\
Finally, a word about the normalization of the primordial bispectra. The common primordial shapes and intermediate shape are normalized such that $B(k,k,k)=\frac{18}{5}P_\zeta(k)^2$ in the equilateral configuration. The clock template is normalized such that its amplitude matches that of the intermediate shape for the transition mass $\nu = \mu = 0$ in the squeezed limit.

\subsubsection{Primordial trispectra}
The next order of correlation functions known as the trispectrum, correlate four fluctuations. As for the bispectrum, the simplest way of generating such a non-Gaussianity is by expanding the primordial field in terms of Gaussian fields to third order:
\begin{eqnarray}
    \zeta = \zeta_g + \frac{3}{5}\fnl{loc}\zeta_g^2 + \frac{9}{25}\gnl \zeta_g^3
\end{eqnarray}
in Fourier space, this expansion leads to two distinct local trispectra:
\begin{eqnarray}
    T_\zeta^{\taunl{}} &=& \tau^{\textrm{loc}}_{\textrm{NL}} \bigg[P_{\zeta}(k_1)P_{\zeta}(k_3)P_{\zeta}(s) + P_{\zeta}(k_1)P_{\zeta}(k_3)P_{\zeta}(t)\bigg] + \text{6 perms}.
\\
    T_\zeta^{\gnl{}} &=& \frac{54}{24} g_{\textrm{NL}}^{\textrm{loc}} P_{\zeta}(k_1)P_{\zeta}(k_2)P_{\zeta}(k_{3}) + \text{3 perms}.
\end{eqnarray}
where $s=|\boldsymbol{s}|=|\kone + \ktwo|$, $t = |\boldsymbol{t}| = |\kone + \kfour|$ and $u = |\boldsymbol{u}| = |\bk{1} + \bk{3}|$ are diagonal momenta.
For single-field models, the $\taunl{}$ is always generated in case of a nonzero local bispectrum, such that $\taunl{} \geq (\frac{6}{5}\fnl{loc})^2$, known as the Suyama-Yamaguchi bound \cite{Suyama2007}. Furthermore, this shape peaks in the collapsed limit, where one of the diagonals (e.g. $s$) becomes much smaller than the external momenta. The $\gnl{}$ amplitude is an independent variable and is characterized by a peak in the double squeezed limit, in which two external momenta become small. \\
\\
Quartic self-interactions of the inflaton field give rise to equilateral shapes, similar to the equilateral bispectrum. Such interactions can be studied systematically using the EFT of inflation \cite{Cheung2007}. This effective field theory approach allows one to study the interactions of the Goldstone of time translations $\pi = -\zeta/H$ in the decoupling limit (i.e. without gravity). Typically, quartic interactions for $\pi$ will imply the existence of cubic interactions that give rise to bispectra that have a much stronger signal than the corresponding trispectrum, unless some contrived mechanism is being considered \cite{Smith2015}. Hence, such interactions will likely be detected already by observing the bispectrum. In the presence of an additional light scalar $\sigma$, we instead study the EFT of multifield inflation \cite{Senatore2010}. In this case, it is possible to write down quartic interactions for $\sigma$ that generate a trispectrum for $\zeta$ without generating a larger bispectrum, making the trispectrum the leading order observable non-Gaussianity in such theories \cite{Senatore:2010jy}. At lowest order in derivatives one finds three interactions $\dot{\sigma}^4$, $\dot\sigma^2 (\partial \sigma)^2$ and $(\partial \sigma)^4$, which give rise to three distinct trispectra

\begin{eqnarray}
  T^{\dot{\sigma}^4}_\zeta &=& \frac{221184}{25} \geqone \frac{1}{k_1 k_2 k_3 k_4 k_t^5}, \\
    T^{\dot{\sigma}^2(\partial \sigma)^2}_\zeta &=& - \frac{27648}{325} \geqtwo \frac{k_t^2 + 3(k_3+k_4)k_t + 12k_3k_4}{k_1 k_2 (k_3 k_4)^3 k_t^5}(\kthree\cdot\kfour) + \text{5 perms}, \\
   T^{ (\partial \sigma)^4}_\zeta &=& \frac{165888}{2575} \geqthree \frac{2k_t^4 - 2k_t^2 \sum_i k_i^2 + k_t \sum_i k_i^3 + 12k_1k_2k_3k_4}{(k_1 k_2 k_3 k_4)^3 k_t^5}((\kone\cdot\ktwo)(\kthree\cdot\kfour) + \text{2 perms}), \nonumber \\
\end{eqnarray}
where $k_t = k_1 + k_2 + k_3 + k_4$. The amplitudes are normalized such that they match the amplitude of the local shape for $g_{\textrm{NL}} = 1$ in the tetrahedral configuration where $k_i = k$ and $\bk{i} \cdot \bk{j} = -k^2/3$: $\langle \zeta^4 \rangle' = \frac{216}{25} P_{\zeta}(k)^3$ \cite{Smith2015}. As we will see, these shapes have a significant overlap. Therefore, when searching for an equilateral trispectrum, it suffices to start of with just one shape.\\
\\
In \cite{Arkani-Hamed2015} it was noted that the oscillatory signature of additional heavy particles is also present in the collapsed limit of the trispectrum. For example, the exchange of a heavy scalar gives rise to a signal:
\begin{eqnarray}
   \lim_{s\rightarrow 0}\langle \zeta_{\kone} \zeta_{\ktwo} \zeta_{\kthree} 
   \zeta_{\kfour}\rangle' &=&  \frac{1}{4k_1^3 k_2^3 k_3^3 k_4^3} 
    \frac{1}{128\pi}(k_{12}k_{34})^{3/2} \times \nonumber \\ && \left[ \left(
    \frac{s^2}{4k_{12}k_{34}} \right)^{i \mu} (1 + i \sinh{\pi \mu}) (\frac{3}{2}+ i \mu)^2 (\frac{5}{2} + i \mu)^2 \Gamma(-i \mu)^2 \Gamma(\frac{1}{2} + i \mu)^2 + c.c \right] \nonumber \\
\end{eqnarray}
where $k_{12} = k_1 + k_2$ (note that $s \neq k_{12}$). Again the frequency, amplitude and phase of the oscillation, are set by the mass through $\mu$. In order to extract such an oscillation from data, we propose the template:
\begin{equation}
\label{eq:TClock}
   T^{\textrm{clock}}_\zeta = \frac{3^{\frac{3}{2}}}{16} g_{\textrm{NL}}^{\text{clock}} \frac{1}{k_1^3 k_2^3 k_3^3 k_4^3 } (k_{12}k_{34})^\frac{3}{2} \Theta(\alpha_{12} - \alpha_0) \Theta(\alpha_{34} - \alpha_{0})  \sin(\mu\ln(\alpha_{12,34}) + \delta) + \text{2 perms}.
\end{equation}
where $\alpha_{12,34} = \frac{s^2}{k_{12}k_{34}}$ and $\alpha_{12} = \frac{k_{12}}{s}$. For scalars with an intermediate mass, the oscillatory behaviour in the collapsed limit again turns into a power law as found in \cite{Assassi2012}:

\begin{eqnarray}
   \lim_{s\rightarrow 0}\langle \zeta_{\kone} \zeta_{\ktwo} \zeta_{\kthree} 
   \zeta_{\kfour}\rangle' = 4 g_{\textrm{NL}}^{\textrm{int}} P_\zeta(k_1)P_\zeta(k_2)P_\zeta(s)\left( \frac{s}{\sqrt{k_1 k_3}} \right)^{3-2\nu}
\end{eqnarray}
To constrain this type of non-Gaussianity we propose the use of a template:
\begin{eqnarray}
\label{eq:TInt}
   T^{\text{int}}_\zeta  = 4 \af 3^{\frac{3}{2}-\nu} g_{\textrm{NL}}^{\textrm{int}}  \frac{1}{(k_1 k_2 k_3 k_4)^{\frac{3}{2}}} \frac{1}{s^3} \alpha_{12,34}^{\frac{3}{2}-\nu}\Theta(\alpha_{12} - \alpha_0) \Theta(\alpha_{34} - \alpha_{0}) + \text{2 perms}.
\end{eqnarray}
As for the bispectrum, we included the Heaviside step functions in order to restrict to the collapsed limit. We will find $\alpha_0 = 2$ to be the cutoff value that optimises the signal-to-noise ratio for these templates. Furthermore, the intermediate template is normalized as to match the $\taunl{}$ shape in the tetrahedral configuration and the amplitude of the oscillatory shape in the collapsed limit is normalized to match the intermediate shape for the mass limit $\nu = \mu = 0$.

\subsection{Secondary non-Gaussianity}
As mentioned before, by the time we enter the Dark Ages non-linear evolution of the density field will generate non-Gaussianity even if we start out with a perfectly Gaussian field. Furthermore, the fact that 21-cm temperature fluctuations depend non-linearly on the underlying fields $\db,\dv$ and $\dT$ causes additional non-Gaussian imprints in the 21-cm signal. Since we are primarily interested in extracting (possible very weak) primordial non-Gaussianity from this signal, such secondary non-Gaussianity can be considered to be a confusion signal that should be accurately modeled and subtracted in order to adequately determine the primordial contribution. The magnitude and impact of these secondary non-Gaussianities were previously studied for the bispectrum in \cite{Munoz2015,Pillepich2006}. There, it was found that the secondary contribution is several orders of magnitude stronger than the primordial signal. Including the amplitude of the secondary bispectrum as nuisance parameters and marginalising over them reduces the signal-to-noise ratio of the primordial amplitude. In this work we improve upon some of the assumptions made in those previous works, by including the effects of electron fraction fluctuations and baryonic pressure. The former has been studied in section \ref{subsec:21cmfluctuations} and its effect is captured in the coefficients $\CT_{i,j}$. The latter will be studied next. After having set up the framework for higher order perturbation theory with baryonic pressure, we will be able to derive both the secondary bispectrum as well as for the first time the secondary trispectrum of 21-cm temperature fluctuations during the Dark Ages.

\subsubsection{Perturbation theory including baryonic pressure}
By means of equation \eqref{eq:dT21expansion}, we have direct perturbative relation between 21-cm fluctuations and baryon density contrast and velocity divergence up to third order. However, we have so far been agnostic about the functional form of $\db^{(n)}$ and $\theta_b^{(n)}$. In earlier work \cite{Pillepich2006,Munoz2015}, two assumptions are made about the functional form. First, it assumed that the time-dependence of the baryonic fluctuations is identical to that of CDM fluctuations, so that the $n$-th order perturbations can be written as \cite{Bernardeau2001}:
\begin{equation}
    \delta_b^{(n)}\zk=a^n(z)\delta_{n,b}(\bk{}),\quad\quad\quad \theta_b^{(n)}\zk=\mathcal{H}(z)a^n(z) \theta_{n,b}(\bk{}).
\end{equation}
Note that the functional form is product separable in temporal and spatial (momentum) dependence. The latter is contained in $\delta_n$ and $\theta_n$, which in standard perturbation theory read:
\begin{eqnarray}
    \delta_{n,b}(\bk{})=\int_{\qone\cdots\q_n}(2\pi)^3\diracd{}(\bk{}-\q_{1\cdots n}) F_{n,b}(\qone,\dots,\q_n)\delta_{1,b}(\qone)\cdots\delta_{1,b}(\q_n) \\
    \theta_{n,b}(\bk{})=\int_{\qone\cdots\q_n}(2\pi)^3\diracd{}(\bk{}-\q_{1\cdots n}) G_{n,b}(\qone,\dots,\q_n)\delta_{1,b}(\qone)\cdots\delta_{1,b}(\q_n),
\end{eqnarray}
where $\q_{1\cdots n}\equiv \qone+\cdots+\q_n$ and the \emph{baryonic} momentum kernels $F_{n,b}$ and $G_{n,b}$ are symmetric under interchanging the momenta. The second assumption made in previous works, is that the baryonic kernels are equivalent to that of CDM so that they can be replaced by the latter: $F_{n,b}\rightarrow F_{n,c}$ and $G_{n,b}\rightarrow G_{n,c}$, which can trivially be obtained from a simple recursion relation \cite{Bernardeau2001}.\\

However, in reality baryons are very different from CDM in their physical nature. Where CDM is pressureless, baryons constitute pressure. In particular, on scales comparable to the so-called baryonic Jeans scale $\lambda_{\mathrm{J}}$, the pressure becomes considerable and suppresses the growth of fluctuations due to gravity. On scales smaller than the Jeans scale, pressure competes with gravity to generate acoustic sound waves and the notion of growing density fluctuations becomes ill-defined.\\

To take the first step in the direction of accounting for pressure effects, we use the formalism for baryonic fluctuations constructed in \cite{Shoji2009}.\footnote{We will only highlight the required essentials of the formalism, for details, we refer to \cite{Shoji2009}.} Within this formalism, the Jeans scale is assumed to be constant, i.e. $\lambda_{\mathrm{J}}\neq \lambda_{\mathrm{J}}(z)$. Although strictly speaking this assumption becomes incorrect once the gas temperature decouples from the photon temperature, it has the convenient consequence that the time-dependence of baryonic fluctuations is still equivalent to that of CDM, $\db^{(n)}\propto a^n$, and the results of section \ref{sec:perturbations} are still applicable. Then, the baryonic fields $\delta_b$ and $\theta_b$ can be expanded as:
\begin{equation}
	\delta_b\zk =\sum_{n=1}^\infty a^n(z)g_n(\bk{})\delta_{n,c}(\bk{}),\quad\quad\quad \theta_b\zk=\sum_{n=1}^\infty \H(z) a^n(z)h_n(\bk{})\theta_{n,c}(\bk{}),
\end{equation}
where we have defined the \emph{filtering} functions:\footnote{In principle, these filtering functions depend on time. Therefore, they are often defined as: $g_n\zk=\db^{(n)}\zk/\delta_c^{(n)}\zk$, including time dependence. However, as shown in \cite{Shoji2009}, when decaying modes are omitted and the Jeans scale is taken to be constant, the filtering functions become time-independent as well, justifying this definition.}
\begin{equation}
    g_n(\bk{})=\frac{\delta_{n,b}(\bk{})}{\delta_{n,c}(\bk{})},\quad\quad\quad h_n(\bk{})=\frac{\theta_{n,b}(\bk{})}{\theta_{n,c}(\bk{})}. 
\end{equation}
Note that while we solely require $g_1$ at first order (as the linear density contrast and velocity divergence are trivially related as $\theta_b=-\mathcal{H}\delta_b$), at higher order we require different filtering functions for $\delta_b$ (i.e. $g_n$) and $\theta_b$ ($h_n$). 

We expect the filtering functions to be strongly decreasing for scales around and beyond the Jeans scale ($k/\kJ\gg 1$), as pressure suppresses the evolution of baryonic density fluctuations relative to those in CDM on these scales. In addition, far away from the Jeans scale ($k/\kJ\ll 1$), we assume the behaviour of baryons and CDM to be equivalent, i.e. $g_n,h_n\to 1$. At first order, the filtering function is given by:
\begin{equation}
	g_1(k)=\frac{1}{1+k^2/\kJ^2}, 
\end{equation}
which indeed contains the limiting behavior described above. At second order, the function $g_2$ can be written in terms of $g_1$ as:
\begin{equation}
	g_2(\bk{})=\sigma(k)\bigg[1+\frac{7}{3}\frac{\delta'_{2,c}(\bk{})}{\delta_{2,c}(\bk{})}\bigg],\quad\quad\quad\sigma(k)\equiv \frac{1}{10/3+(k/\kJ)^2},
\end{equation}
where $\delta'_{2,c}$ is given by (the prime does not denote a time derivative):
\begin{equation}
	\delta'_{2,c}(\bk{})\equiv\int_{\qone\qtwo}(2\pi)^3\diracd(\bk{}-\qonetwo)\;\mathcal{F}_2^{(s)}(\qone,\qtwo)\;\delta_{1,c}(\qone)\delta_{1,c}(\qtwo).  
\end{equation}
The modified kernel $\mathcal{F}_2^{(s)}(\qone,\qtwo)$ (where the superscript (s) denote the symmetrization of the kernel) reads:
\begin{equation}
	\mathcal{F}_2^{(s)}(\qone,\qtwo)\equiv \left[F_2^{(s)}(\qone,\qtwo)+\frac{3}{14}\frac{k^2}{\kJ^2}\right]g_1(\qone)g_1(\qtwo), 
\end{equation}
which reduces to the CDM kernel $F_2^{(s)}(\qone,\qtwo)$ in the limit $k/\kJ\ll 1$. The function $h_2(\bk{})$ can be written as:
\begin{align}
	h_2(\bk{})=&\;\frac{1}{\theta_{2,c}(\bk{})}\bigg[\int_{\qone\qtwo} (2\pi)^3\diracd(\bk{}-\qonetwo)\Big[1+\vartheta(\qone,\qtwo)\Big]\delta_{1,b}(\qone)\delta_{1,b}(\qtwo)\bigg]
	-2\frac{\delta_{2,b}(\bk{})}{\theta_{2,c}(\bk{})}, 
\end{align}
where we have defined $\vartheta(\qone,\qtwo)\equiv (\qone\cdot\qtwo)(q_1^2+q_2^2)/2q_1^2q_2^2$. In the limit $k/\kJ\to 0$, we obtain $h_2\to 1$ as expected.\\

The higher order perturbations $\delta^{(3)}_b,\theta^{(3)}_b$ that show up in the secondary trispectrum will also be modified as compared to the CDM expressions.\footnote{See Appendix B of \cite{Shoji2009}, Eqs. B23 and B25.} However, as we will see later, when comparing forecasts of the bispectrum with and without baryonic pressure effects, the differences are small. Hence, for computational simplicity we will not include baryonic pressure in our forecasts for the trispectrum.
With the above perturbation expansions, we are now able to write down the secondary contributions to the non-Gaussianity of the 21-cm signal.

\subsubsection{Secondary bispectrum}
The lowest order non-Gaussian statistic of 21-cm brightness temperature fluctuations, i.e. the 21-cm bispectrum $B_{\delta T}$, now consists of a primary, primordial contribution:
\begin{eqnarray}
\label{eq:21cmBispectrum}
\langle \dTcm^{(1)}(\bk{1}) \dTcm^{(1)}(\bk{2}) \dTcm^{(1)}(\bk{3}) \rangle' = \left(\prod_{i=1}^3 c_1^{(1)}(\bk{i}) \mathcal{M}_b(\bk{i}) \right)\times B_\zeta(\bk{1},\bk{2},\bk{3}) \nonumber \\
\end{eqnarray}
where $\mathcal{M}_b(\bk{})$ is the linear transfer function for baryonic fluctuations. Moreover, there is also a higher order, secondary contribution:
\begin{align}\label{eq:secbispectrum}
	\langle \dTcm &(\bk{1})\dTcm(\ktwo)\dTcm(\kthree)\rangle_{\mathrm{sec.}}' = \langle \dTcm^{(1)} (\bk{1})\dTcm^{(1)}(\ktwo)\dTcm^{(2)}(\kthree)\rangle' +2\;\mathrm{perms.}\nonumber\\
	&=c_1^{(1)}(\bk{1})c_1^{(1)}(\ktwo)c_1^{(2)}(\bk{3})\times 2\tilde{\mathcal{F}}_2^{(s)}(\bk{1},\ktwo)P_c(\bk{1})P_c(\bk{2})\nonumber\\
	&+c_1^{(1)}(\bk{1})c_1^{(1)}(\ktwo)c_2^{(2)}(\kthree)\times 2\tilde{\mathcal{G}}_2^{(s)}(\bk{1},\ktwo)P_c(\bk{1})P_c(\bk{2})\nonumber\\
	&+c_1^{(1)}(\bk{1})c_1^{(1)}(\ktwo)\left[c_3^{(2)}(\kthree,-\bk{1})+c_3^{(2)}(\kthree,-\ktwo)\right]g_1^2(k_1)g_1^2(k_2)P_c(\bk{1})P_c(\bk{2})\nonumber\\
	&+2\;\mathrm{perms.}
\end{align}
We have defined the modified kernels $\tilde{\mathcal{F}}_2^{(s)}$ and $\tilde{\mathcal{G}}_2^{(s)}$ as:
\begin{align}
    \tilde{\mathcal{F}}_2^{(s)}(\kone,\ktwo)\equiv&\; g_1(k_1)g_1(k_2)\sigma_2(|\bk{12}|)\left[F_2^{(s)}(\kone,\ktwo)+\frac{7}{3}\mathcal{F}_2^{(s)}(\kone,\ktwo)\right],\nonumber\\
    \tilde{\mathcal{G}}_2^{(s)}(\kone,\ktwo)\equiv &\; 2g_1(k_1)g_1(k_2)	\sigma_2(|\boldsymbol{k}_{12}|)\left[F_2^{(s)}(\kone,\ktwo)+\frac{7}{3}\mathcal{F}_2^{(s)}(\kone,\ktwo)\right]\nonumber\\
	&-g^2_1(k_1)g^2_1(k_2)[1+\vartheta(\kone,\ktwo)],
\end{align}
In the limit $\lambda_J \rightarrow \infty$ and $P_c \rightarrow P_b$ this restores to the previously derived secondary bispectrum of 21-cm brightness temperature fluctuations \cite{Pillepich2006,Munoz2015}.

\subsubsection{Secondary trispectrum}
The secondary trispectrum is composed of 11 different contributions, which can be summarized schematically as:
\begin{align}
	\langle\delta T_{21}(\kone)\delta T_{21}(\ktwo)\delta T_{21}(\kthree)\delta T_{21}(\kfour)\rangle'_{\mathrm{sec.}}&=T_{\delta_1\delta_1\delta_2\delta_2}+T_{\delta_1\delta_1\theta_2\theta_2}+T_{\delta_1\delta_1\delta_2\theta_2}\nonumber\\&+T_{\delta_1\delta_1\delta_2[\delta_1]^2}+T_{\delta_1\delta_1\theta_2[\delta_1]^2}+T_{\delta_1\delta_1[\delta_1]^2[\delta_1]^2}\nonumber\\
&+T_{\delta_1\delta_1\delta_1[\delta_1]^3}+T_{\delta_1\delta_1\delta_1[\delta_1\delta_2]}+T_{\delta_1\delta_1\delta_1[\delta_1\theta_2]}\nonumber\\
	&+T_{\delta_1\delta_1\delta_1\delta_3}+T_{\delta_1\delta_1\delta_1\theta_3},
\end{align}
where for example
\begin{eqnarray}
T_{\delta_1\delta_1\delta_2\theta_2} = f(\bk{1},\bk{2},\bk{3},\bk{4}) \langle \delta^{(1)}_b(\bk{1}) \delta^{(1)}_b(\bk{2})  \delta^{(2)}_b(\bk{3}) \theta^{(2)}_b(\bk{4}) \rangle +11\;\mathrm{perms.} 
\end{eqnarray}
Here the function $f$ is composed of the coefficients $c_i^{(j)}$ introduced above. The explicit expressions for all 11 contributions can be found in Appendix \ref{app:Trisp}.

\section{The information content of the Dark Ages}
\label{sec:totalinformation}
In this section we will determine the amount of information in 21-cm brightness temperature fluctuations during the Dark Ages available to constrain primordial non-Gaussianity.\footnote{The code used to perform the numerical calculations is openly available \href{https://github.com/tsfloss/21-cm-Dark-Ages-Fisher-Forecast}{on GitHub via this link}.}  First we will determine the total amount of information available when neglecting any form of signal confusion (either instrumental, systematic or secondary), giving an ultimate upper bound for the sensitivities that can be achieved. Subsequently, using our improved model of the 21-cm signal, we determine the amount of information when including signal confusion due to secondary non-Gaussianities. Finally, we will determine the sensitivity of different experimental scenarios.

\subsection{Total primordial information content}
In order to determine the total information content, we will assume a cosmic variance limited experiment of the Dark Ages between $30 \leq z \leq 100$ and take the largest scale to be $\kmin = 0.01 \text{ Mpc}^{-1}$, since on larger scales one has to include fluctuations in the CMB temperature and non-relativistic perturbation theory breaks down.\footnote{The survey volume sets the largest scale to be only slightly bigger at $k = 0.005 \text{ Mpc}^{-1}$. Hence, from an observational perspective one gains very little from going beyond our choice of $\kmin{}$.} When including the secondary bispectrum and trispectrum, following \cite{Chen2016a} we consider 14 redshift bins (i.e. $\Delta z = 5$), assuming that they are sufficiently uncorrelated such that the covariance matrix can be taken to be diagonal in redshift space. More details about the Fisher analysis and how to evaluate the Fisher matrix for both the bispectrum and trispectrum, we refer to Appendix \ref{app:Fisher}.\\

The results for the common primordial bispectra are summarized in Table \ref{tab:TotalInformationCommonBispectra}. The sensitivity to the amplitude of the primordial bispectrum is given on the diagonal. The increased sensitivity to the local shape as compared to the other two common shapes, can be understood as due to a slightly enhanced scaling with $\kmax$ \cite{Kalaja2020}. Furthermore, on the off-diagonal we summarize the overlap coefficient between the common primordial bispectra.\\

\begin{table}[H]
\centering
\begin{tabular}{|c|c|c|c|c|}
\hline
                &  $\fnl{loc}$  & $\fnl{equil}$ & $\fnl{ortho}$ \\ \hline
 $ \fnl{loc}$ &  $4.22 \times 10^{-6}$ & $0.29$ & $-0.33$  \\ \hline
 $ \fnl{equil}$ &  & $2.88\times 10^{-5}$ & $0.19$  \\ \hline
 $ \fnl{ortho}$ &  &  & $2.10 \times 10^{-5}$  \\ \hline
\end{tabular}
\caption{\emph{Diagonal}: sensitivity to the amplitude of the common primordial shapes. \emph{Off-diagonal}: cosine \eqref{eq:cosine} between common primordial shapes. We assume a cosmic variance limited experiment between $30 \leq z \leq 100$ and $ 0.01 \leq k \leq 300 \textrm{ Mpc}^{-1}$.}
\label{tab:TotalInformationCommonBispectra}
\end{table}

\begin{figure}[H]
    \centering
    \includegraphics[scale=0.5]{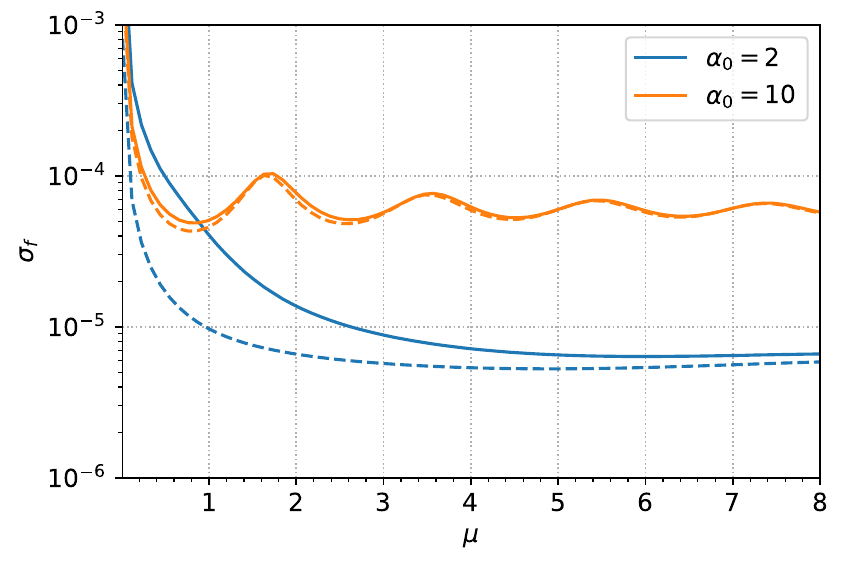}
    \includegraphics[scale=0.5]{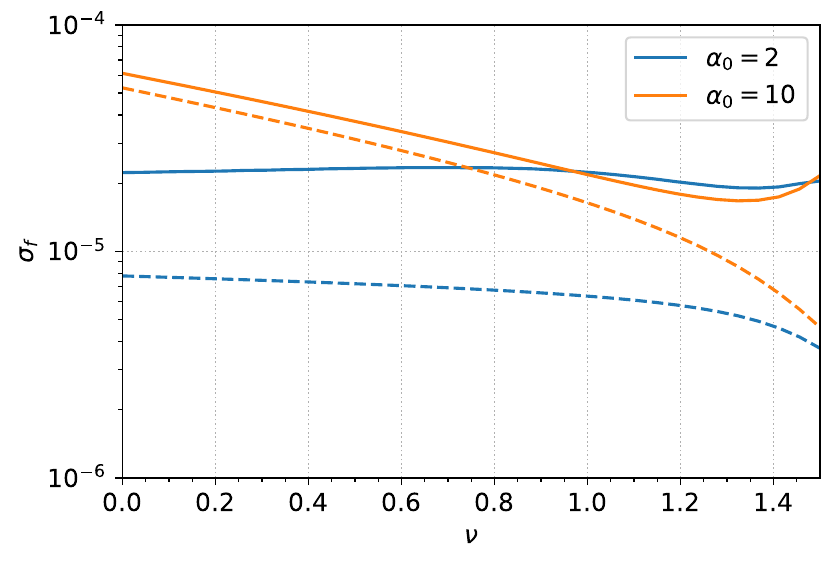}
    \caption{Sensitivity to the amplitude of the massive exchange templates in a cosmic variance limited experiment between $30 \leq z \leq 100$ and $ 0.01 \leq k \leq 300 \textrm{ Mpc}^{-1}$, before (dashed) and after (solid) marginalising over the common primordial bispectra, for different cutoff values $\alpha_0$. \emph{Left}: Clock template, equation \eqref{eq:BClock}. \emph{Right}: Intermediate template, equation \eqref{eq:BInt}}
    \label{fig:ExchangeBispectraError}
\end{figure}

\begin{figure}[H]
    \centering
    \includegraphics[scale=0.5]{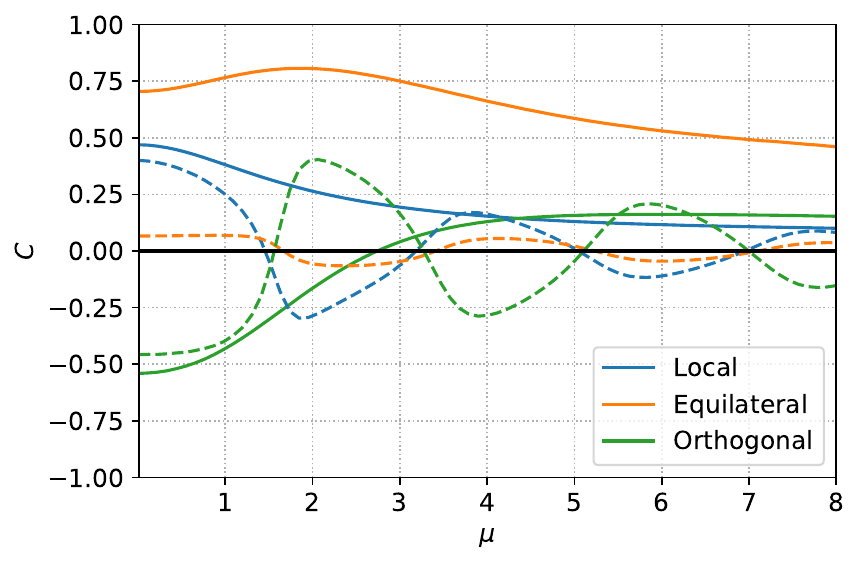}
    \includegraphics[scale=0.5]{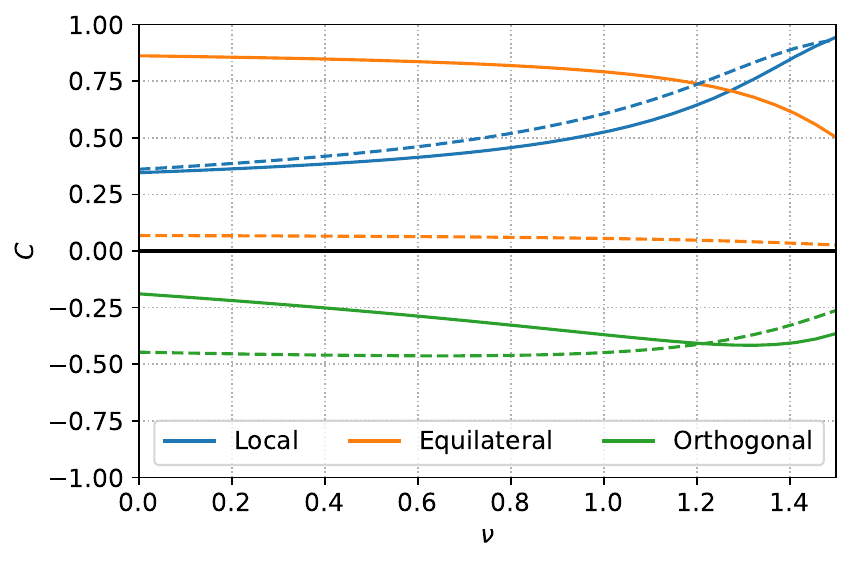}
    \caption{Cosine \eqref{eq:cosine} between the massive exchange templates and common primordial shapes for different mass in a cosmic variance limited experiment between $30 \leq z \leq 100$ and $ 0.01 \leq k \leq 300 \textrm{ Mpc}^{-1}$. Solid lines correspond to the template with $\alpha_0 = 2$. Dashed lines correspond to the template with $\alpha_0 =  10$. \emph{Left}: Clock template equation \eqref{eq:BClock}. \emph{Right}: Intermediate template equation \eqref{eq:BInt}}
    \label{fig:ExchangeBispectraOverlap}
\end{figure}

We now consider the massive scalar exchange template for $m \geq 3H/2$ (i.e. equation \eqref{eq:BClock}). As mentioned, the authors of \cite{Meerburg2016} suggested the use of a cutoff $\alpha_0 = 10$ in order to remove overlap with the equilateral shape equation \eqref{eq:BEquil}. Such a cutoff also drastically reduces the amount of triangles available to observe, hence reducing the signal-to-noise ratio. Using the Fisher analysis we are able to determine the amount of overlap between the clock template and the common primordial shapes that we aim to look for in future experiments. In Figure \ref{fig:ExchangeBispectraOverlap}a we present this overlap as a function of $\mu$ for both the cutoff $\alpha_0 = 10$ and $\alpha_0 = 2$. We find that although the overlap with the equilateral shape is almost completely removed for the former case, the overlap with the other shapes is increased for some mass values compared to the latter. In \ref{fig:ExchangeBispectraError}a we show the sensitivity to the amplitude of the clock template, which clearly shows that a larger cutoff is favorable only for masses $\mu \lesssim 1$. A similar observation can be made for the intermediate template (Figure \ref{fig:ExchangeBispectraOverlap}b and \ref{fig:ExchangeBispectraError}b) where we again find $\alpha_0 = 2$ to optimise the sensitivity except for some small improvement with $\alpha_0 = 10$ when $\nu \gtrsim 1$. Hence, in what follows we will take $\alpha_0 = 2$.\\

Proceeding to the trispectrum, in Table \ref{tab:TotalInformationTrispectra} (on the diagonal) we present the ultimate sensitivity of 21-cm temperature fluctuations from the Dark Ages to the amplitude of the local and equilateral trispectra. We find the $\taunl{}$ shape to be much better constrained than the other shapes. This can again be understood as due to an enhanced scaling \cite{Kalaja2020}, which we will comment on later in this subsection. The overlap between these primordial trispectra are given by the off diagonals, where we see that the enhanced scaling of the $\taunl{}$ shape makes it completely orthogonal to the other shapes, making it a very clean signal to look for.\\

\begin{table}[H]
\centering
\begin{tabular}{|c|c|c|c|c|c|c|}
\hline
 &  $\taunl$ & $\gnl$ & $\geqone$ & $\geqtwo$ & $\geqthree$  \\ \hline
 $\taunl$ & $1.79\times 10^{-7}$ & $0.00$ & $0.00$ & $0.00$ & $0.00$ \\ \hline
 $\gnl$ &  & $0.02$ & $0.13$ & $0.11$ & $0.06$ \\ \hline
 $\geqone$ &  &  & $0.229$  & $0.95$ & $0.65$ \\ \hline
 $\geqtwo$ &  &  & & $0.254$ & $0.82$\\ \hline
 $\geqthree $ &  &  &  & & $0.0609$ \\ \hline
\end{tabular}
\caption{\emph{Diagonal}: sensitivity $\sigma$ to the amplitude of primordial trispectra. \emph{Off-diagonal}: cosine \eqref{eq:cosine} between common primordial trispectra. We assume a cosmic variance limited experiment between $30 \leq z \leq 100$ and $0.01 \leq k \leq 300 \Mpc^{-1}$.}
\label{tab:TotalInformationTrispectra}
\end{table}

\begin{figure}[H]
    \centering
    \includegraphics[scale=0.5]{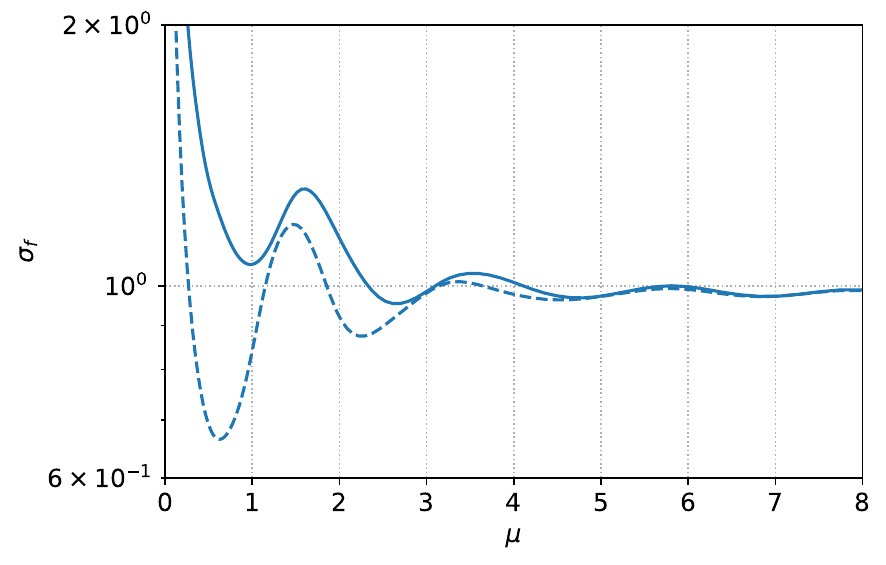}
    \includegraphics[scale=0.5]{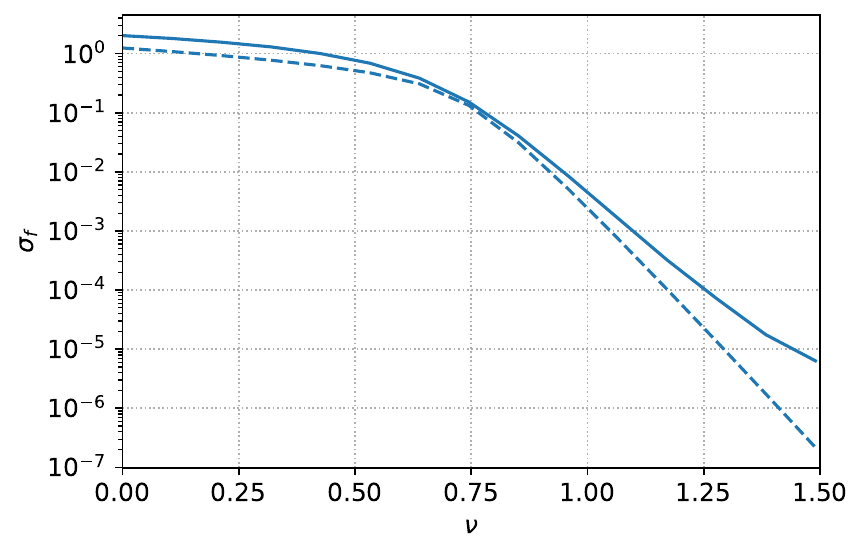}
    \caption{Sensitivity to the amplitude of the exchange trispectrum templates in a cosmic variance limited experiment between $30 \leq z \leq 100$ and $ 0.01 \leq k \leq 300 \textrm{ Mpc}^{-1}$, before (dashed) and after (solid) marginalising over the common primordial trispectra. \emph{Left}: Clock template equation \eqref{eq:TClock}. \emph{Right}: Intermediate template equation \eqref{eq:TInt}, here the black line indicates the mass value $\nu = 3/4$ at which the enhanced scaling sets in \cite{Kalaja2020}.} 
    \label{fig:ExchangeTrispectraError}
\end{figure}

\begin{figure}[H]
    \centering
    \includegraphics[scale=0.5]{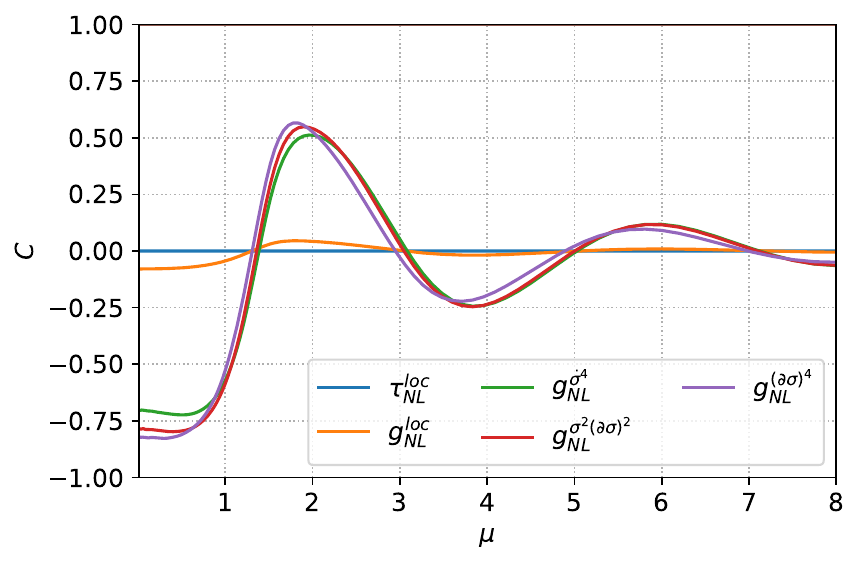}
    \includegraphics[scale=0.5]{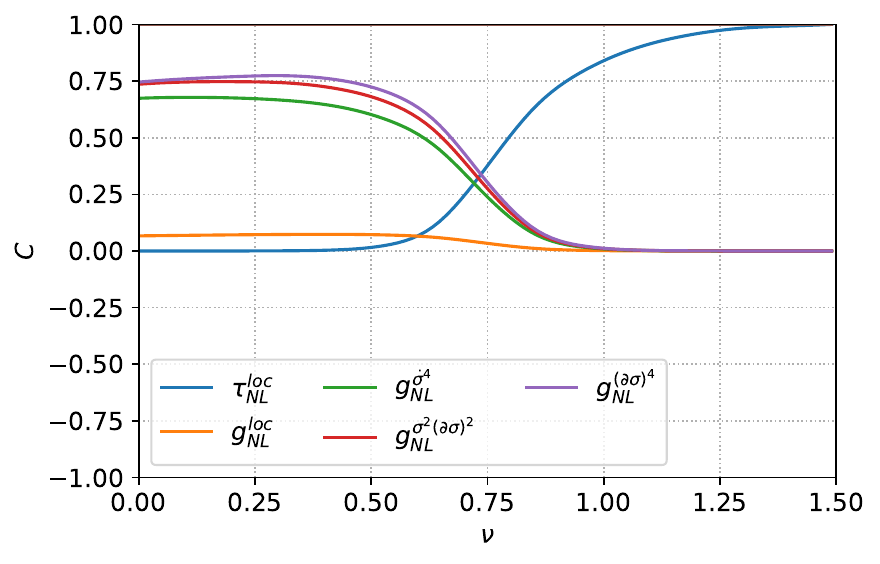}
    \caption{Cosine \eqref{eq:cosine} between the exchange templates and the other primordial shapes for different mass in a cosmic variance limited experiment between $30 \leq z \leq 100$ and $ 0.01 \leq k \leq 300 \textrm{ Mpc}^{-1}$. \emph{Left}: Clock template equation \eqref{eq:TClock}. \emph{Right}: Intermediate template equation \eqref{eq:TInt}}
    \label{fig:ExchangeTrispectraOverlap}
\end{figure}

The ultimate sensitivities for the massive exchange trispectra are given in Figure \ref{fig:ExchangeTrispectraError}. Although it's clear that the clock trispectrum does not outperform the clock bispectrum, we see that the intermediate template enjoys an enhanced scaling for masses $\nu > 3/4$. Such a scaling can improve the sensitivity to the trispectrum past the sensitivity to the bispectrum. However, for $\nu$ close to $3/2$, the shape acquires significant overlap with the $\taunl{}$ shape, which drastically reduces the sensitivity after marginalization making the sensitivity on par to that of the bispectrum. The overlap of the exchange trispectra with the local and equilateral trispectra is summarized in Figure \ref{fig:ExchangeTrispectraOverlap}, which shows how the intermediate template interpolates between the local and equilateral shapes depending on the mass $\nu$.\\

The results presented in this subsection give an indication of how well 21-cm brightness temperature fluctuations from the Dark Ages are able to constrain primordial cosmology if we were able to harvest all of its information down to the Jeans scale $\kmax = 300 \Mpc^{-1}$, demonstrating the ability of 21-cm cosmology to achieve the exquisite sensitivity necessary to advance our understanding of the primordial universe, possibly even paving a way towards the Cosmological Collider as noticed previously by \cite{Meerburg2016}.

\subsection{Including secondary non-Guassianity}
Although 21-cm brightness temperature fluctuations are still a relatively pristine tracer of the underlying density field sourced by inflation, non-linear evolution such as due to gravitational collapse or velocity effects, will make even perfectly Gaussian fluctuations non-Gaussian over time, and between the surface of last scattering at $z \approx 1100$ and the Dark Ages $z \approx 200$ there is a considerable amount of time for such secondary non-Gaussianities to be generated. When trying to extract primordial non-Gaussianity from 21-cm observations, such secondary effects enter as confusion signals, introducing extra nuisance parameters that should be modeled and accounted for when analyzing data. Secondary non-Gaussianity of the 21-cm signal has already been studied in the context of the bispectrum before \cite{Pillepich2006,Munoz2015}. In the previous sections we have improved upon some of the assumptions that went into those works in order to more accurately model the secondary bispectrum, as well as to extend this to the secondary trispectrum. In this subsection we will study the impact of secondary non-Gaussianity on the constraining power of the 21-cm signal from the Dark Ages. \\

The amount of additional nuisance parameters depends on how accurately we are able to model the secondary non-Gaussianity. Most optimally, we might be able to model the internal parameters of secondary non-Gaussianity ($\bar{T}_{21}$ and $\alpha_{i,j}$ in \ref{subsec:realspace}) to such precision that the only free parameter of the secondary $N$-point function is its amplitude. In that case the secondary $N$-point function can be treated as a single shape. More realistically, we obtain a best fit for the internal parameters such that their residuals become the free parameters of the secondary correlation function \cite{Munoz2015}:
\begin{eqnarray}
\label{eq:BestFitSecondaryCorrelationFunction}
        F_{\delta T}(\bk{1},...,\bk{N}) = F^{\textrm{sec,0}}_{\delta T}(\bk{1},...,\bk{N}) + \sum_{i} \Delta A_i \frac{\partial F_{\delta T}^{\textrm{sec}}}{\partial A_i}
\end{eqnarray}
such that we end up with an individual shape $\frac{\partial F_{\delta T}^{\textrm{sec}}}{\partial A_i}$ for every internal parameter. Most pessimistically, one could consider every independent shape present in the secondary $N$-point function. Due to the large amount of shapes (already 21 for the bispectrum), calculating the full Fisher matrix quickly becomes intractable for higher $N$-point functions. In the following we will adopt the best fit approach of \cite{Munoz2015}, giving rise to a four parameter secondary bispectrum and seven parameter secondary trispectrum. We have improved upon previous work by including free electron fraction perturbations to model the internal parameters of secondary non-Gaussianity, and we include baryonic pressure (in the bispectrum).\\

We quantify the impact of including secondary non-Gaussianity on the signal-to-noise ratio using the signal-to-noise degradation factor (SND) defined in equation \eqref{eq:SND}. It's interesting to note that although the signal-to-noise degradation at each redshift might be relatively large, when co-adding the redshifts the degradation is significantly reduced. This can be understood due to the difference in how the secondary shapes scale with $z$ as compared to the primordial shapes: going from equation \eqref{eq:FisherBispectrum} to equation \eqref{eq:FisherPrimordialBispectrum}, we see that the amount of signal from the primordial shape does not depend on $z$ whereas the secondary shapes (including off-diagonal contributions to the Fisher matrix) do. Hence, when adding information from redshift slices, one can also start distinguishing the primordial and secondary shapes by their redshift dependence.\\

\begin{figure}[H]
    \centering
    \includegraphics[scale=0.345]{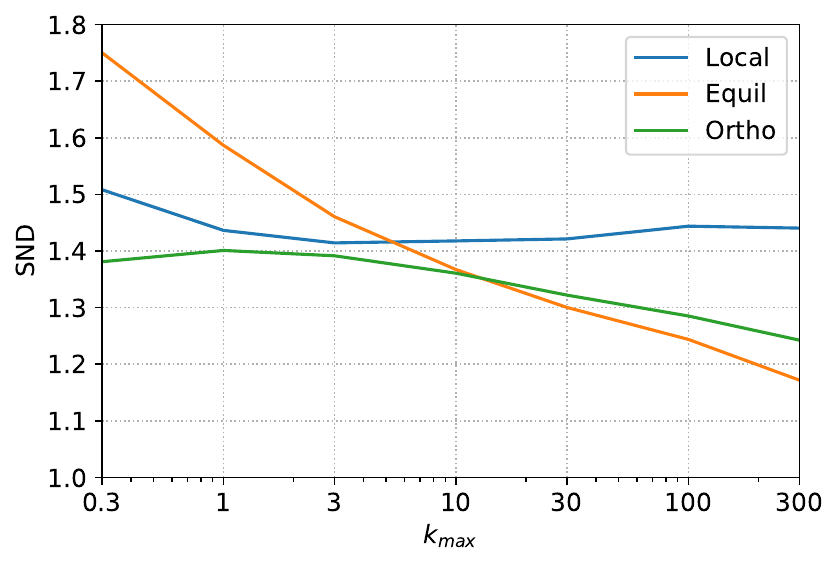}
    \includegraphics[scale=0.345]{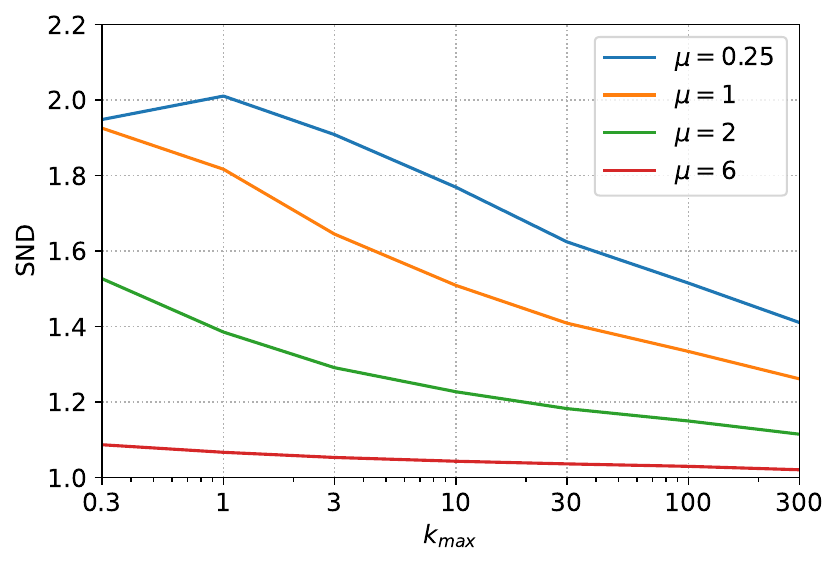}
    \includegraphics[scale=0.345]{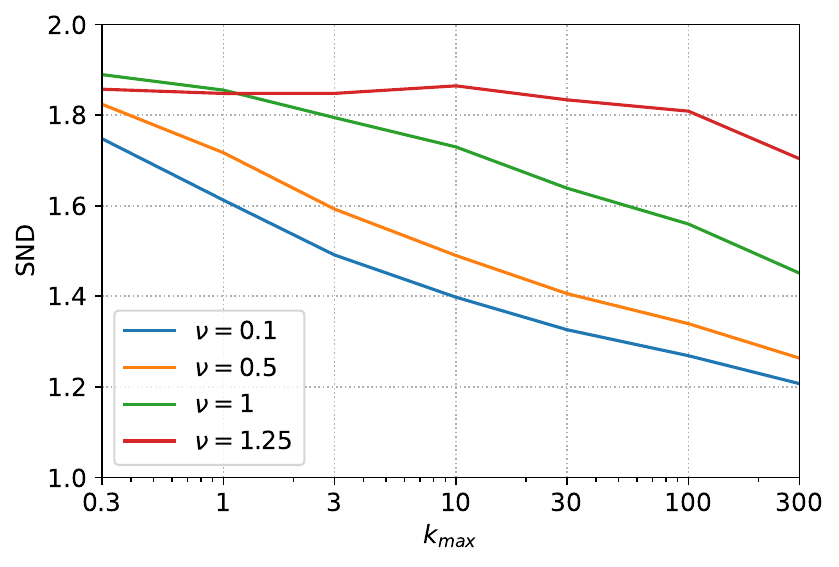}
    \caption{The signal-to-noise degradation factor after marginalising over the 4 parameter secondary bispectrum as a function of $k_{\textrm{max}}$ for different primordial bispectra. Note the different scales of the vertical axes.} 
    \label{fig:SNDBispectra}
\end{figure}

In Figure \ref{fig:SNDBispectra} we plot the amount of degradation as a function of $\kmax$ for the primordial bispectra. We find that including baryonic pressure slightly worsens the signal-to-noise degradation as compared to the simplified approach of \cite{Munoz2015}, but not by much. We expect the same to be true for the trispectrum, so in order to keep the trispectrum calculations more tractable and efficient, we neglect baryonic pressure effects in the upcoming trispectrum analysis. Nevertheless, in an actual data analysis one should model the bi- and trispectrum as accurately as possible, in order to reliably extract the primordial signal. Furthermore, we find the four parameters of the secondary bispectrum to contribute evenly to the signal-to-noise degradation (meaning their cosine \eqref{eq:cosine} with $\fnl{}$ is similar).\\

\begin{figure}[H]
\centering
    \includegraphics[scale=0.5]{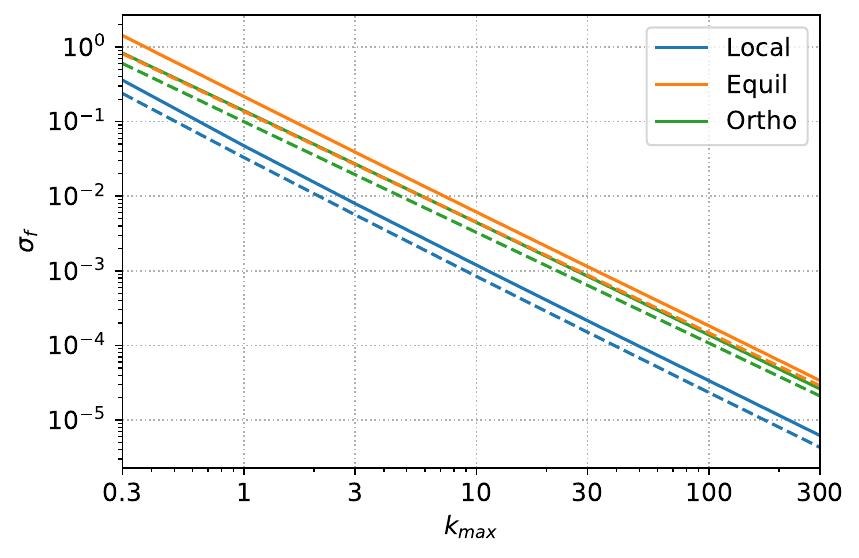}
    \caption{The sensitivity to the amplitude of the common primordial bispectra as a function of $k_{\text{max}}$ before (dashed) and after (solid) marginalising over the 4 parameter secondary bispectrum including baryonic pressure effects. The solid green line coincides with the dashed orange line.} 
    \label{fig:CommonBispectraErrorkmax}
\end{figure}

\begin{figure}[H]
    \centering
    \includegraphics[scale=0.525]{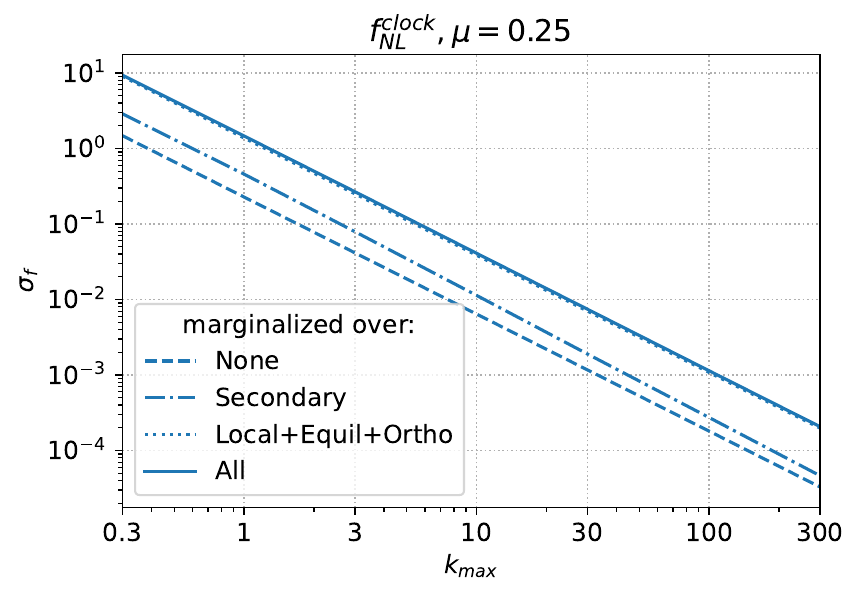}
    \includegraphics[scale=0.525]{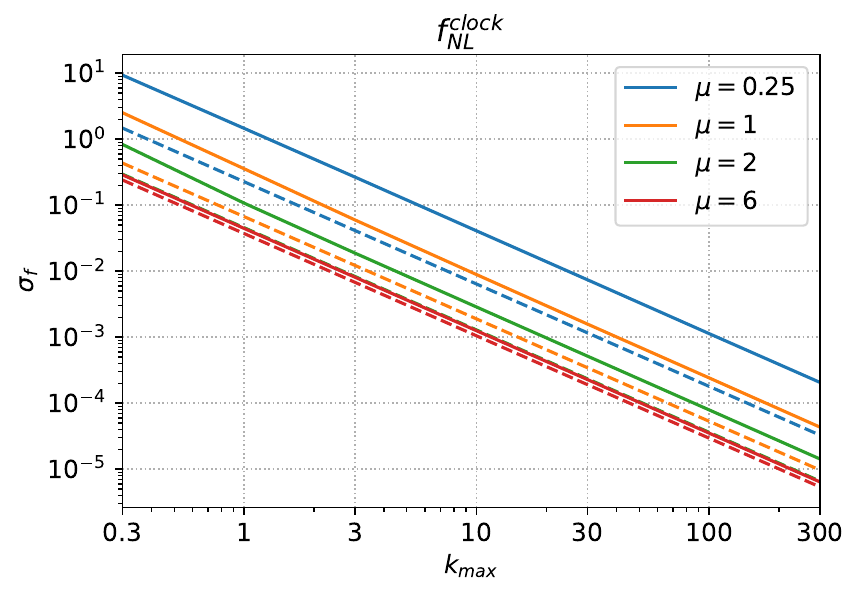}
    \caption{\emph{Left}: sensitivity to the amplitude of the bispectrum clock template with $\mu = 0.1$ as a function of $\kmax$ without marginalising (dashed), after marginalising over the 4 parameter secondary bispectrum (dashed-dotted), after marginalising over the common primordial bispectra (dotted) and after marginalising over both secondary and common primordial bispectra (solid). \emph{Right}: sensitivity to the amplitude of the clock template for different $\mu$ as function of $\kmax$ after marginalising over both secondary and common primordial bispectra.}
    \label{fig:BispectraClockErrorkmax}
\end{figure}

\begin{figure}[H]
\centering
    \includegraphics[scale=0.525]{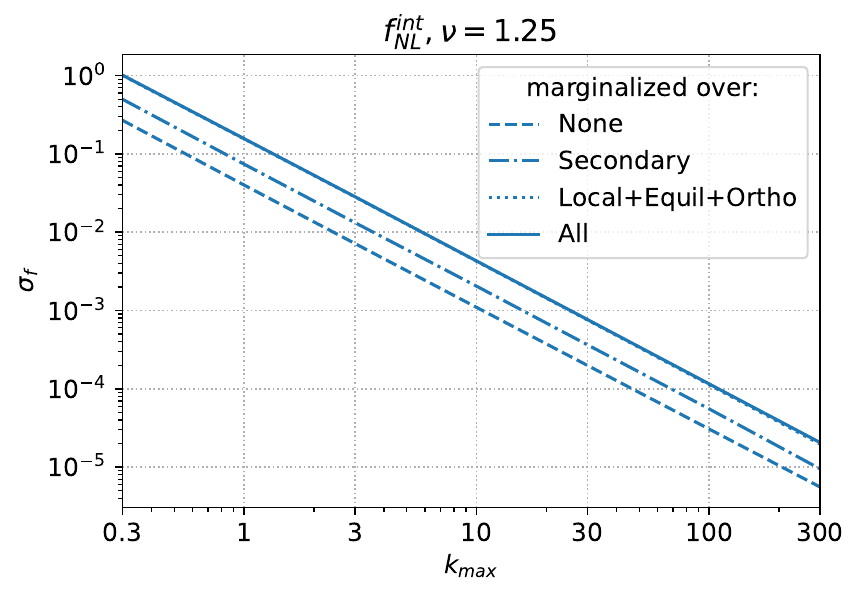}
    \includegraphics[scale=0.525]{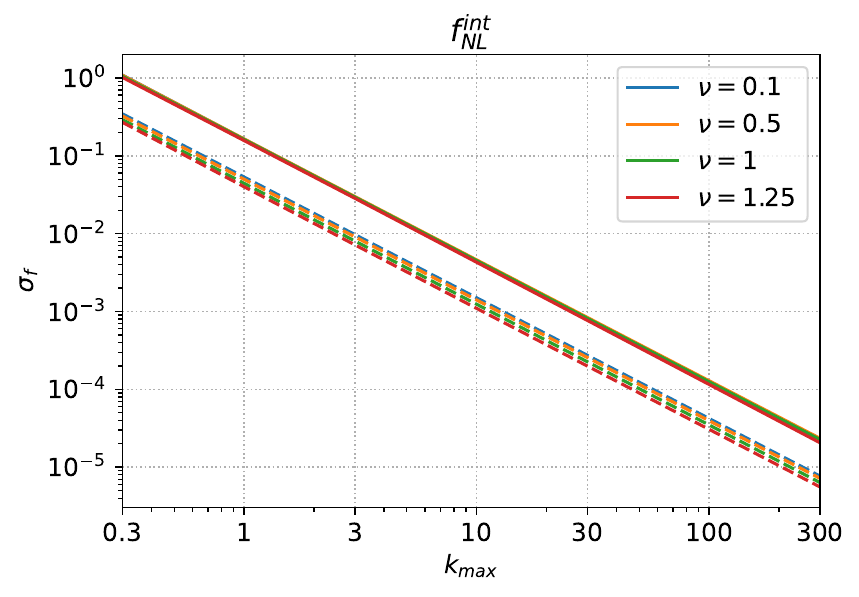} 
    \caption{\emph{Left}: sensitivity to the amplitude of the bispectrum intermediate template with $\nu = 1.4$ as a function of $\kmax$ without marginalising (dashed), after marginalising over the 4 parameter secondary bispectrum (dashed-dotted), after marginalising over the common primordial bispectra (dotted) and after marginalising over both secondary and common primordial bispectra (solid). \emph{Middle}: sensitivity to the amplitude of the int template for different $\nu$ as function of $\kmax$ after marginalising over secondary and common primordial bispectra. \emph{Right}: middle plot zoomed in on $100 \leq \kmax \leq 300$ to distinguish between the various masses.}
    \label{fig:BispectraIntErrorkmax}
\end{figure}

We summarize the sensitivity to the amplitude of the common primordial bispectra as a function of $\kmax{}$ in Figure \ref{fig:CommonBispectraErrorkmax}, clearly showing the small loss of signal-to-noise due to the secondary bispectrum. Similar results are presented for the clock and intermediate template in Figure \ref{fig:BispectraClockErrorkmax} and Figure \ref{fig:BispectraIntErrorkmax} respectively. On the left side of figures we see that once overlap with the common primordial shapes is taken into account, the additional effect of the secondary bispectrum (difference between dotted and solid line in the figures) becomes negligible. We thus conclude that for the exchange templates, the overlap with other primordial shapes is more severe than any overlap with secondary shapes. \\ 

\begin{figure}[H]
    \centering
    \includegraphics[scale=0.345]{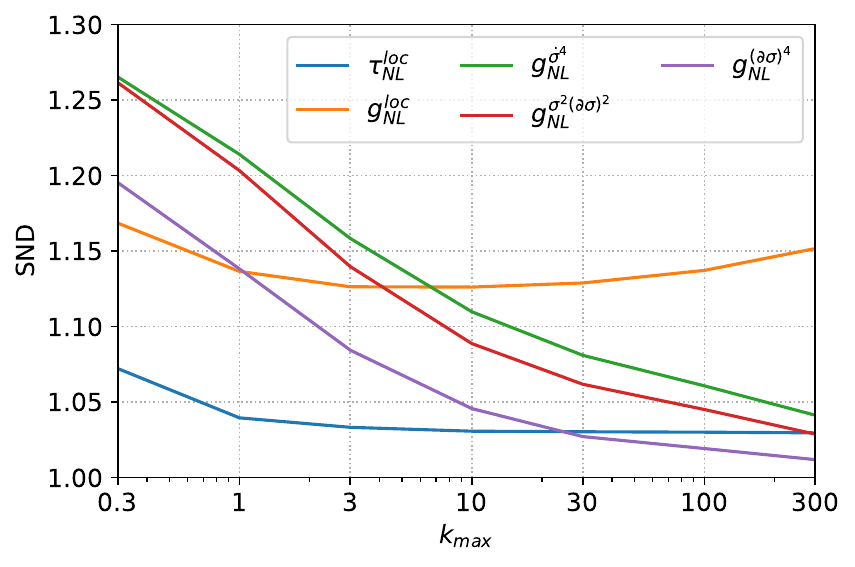}
    \includegraphics[scale=0.345]{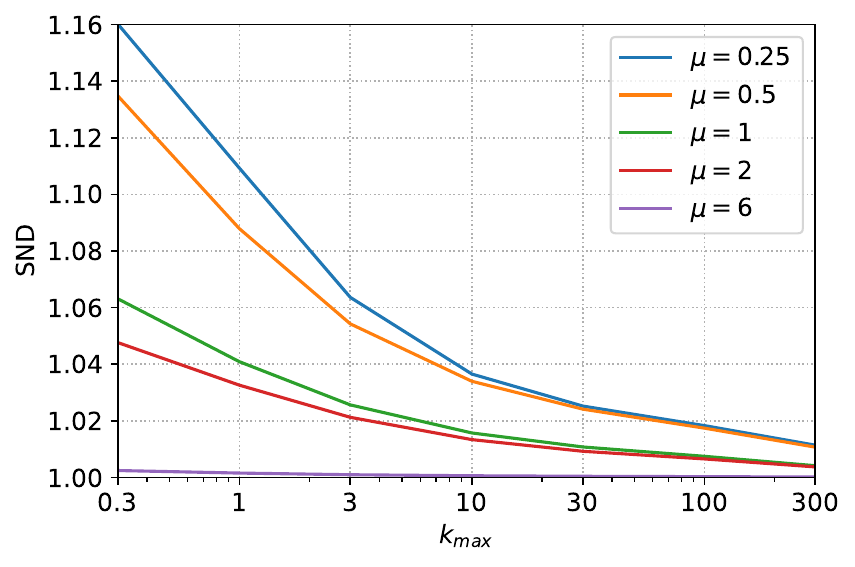}
    \includegraphics[scale=0.345]{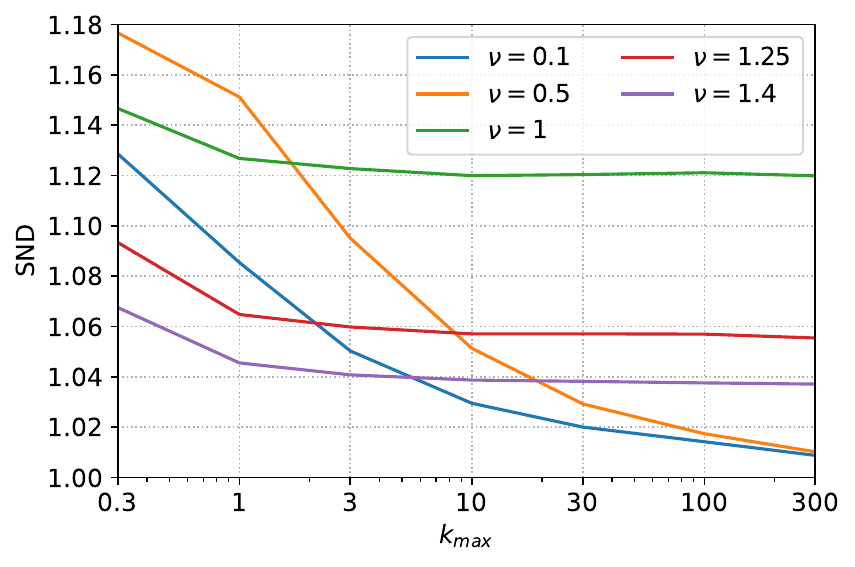}
    \caption{The signal-to-noise degradation factor after marginalising over the 7 parameter secondary trispectrum as a function of $k_{\textrm{max}}$ for different primordial trispectra. Note the different scales of the vertical axes.} 
    \label{fig:SNDTrispectra}
\end{figure}

\begin{figure}
\centering
    \includegraphics[scale=0.5]{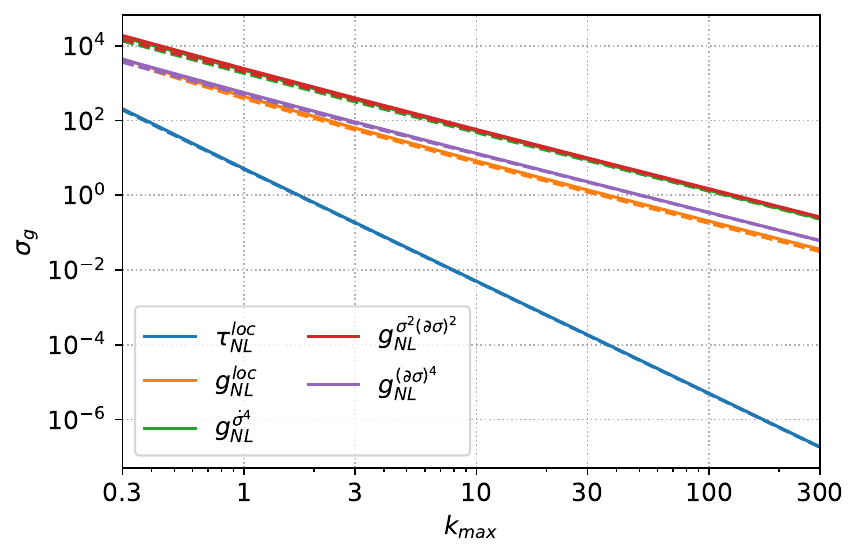}
    \caption{The sensitivity to the amplitude of the common primordial trispectra as a function of 
    $\kmax{}$ before (dashed) and after (solid) marginalising over the 7 parameter secondary trispectrum. Green and red lines coincide. The enhanced scaling of the $\taunl{}$ shape is clearly visible.} 
    \label{fig:CommonTrispectraErrorkmax}
\end{figure}

\begin{figure}[H]
\centering
    \includegraphics[scale=0.5]{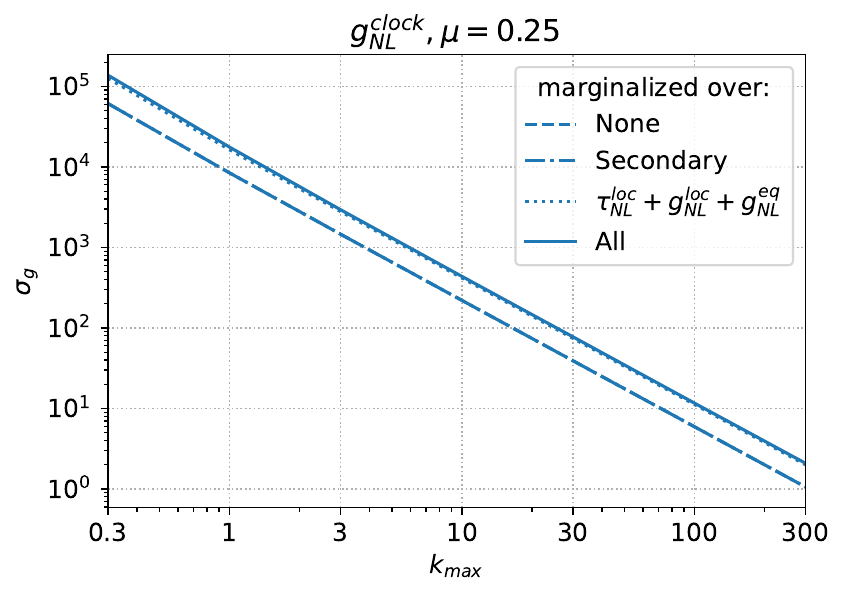}
    \includegraphics[scale=0.5]{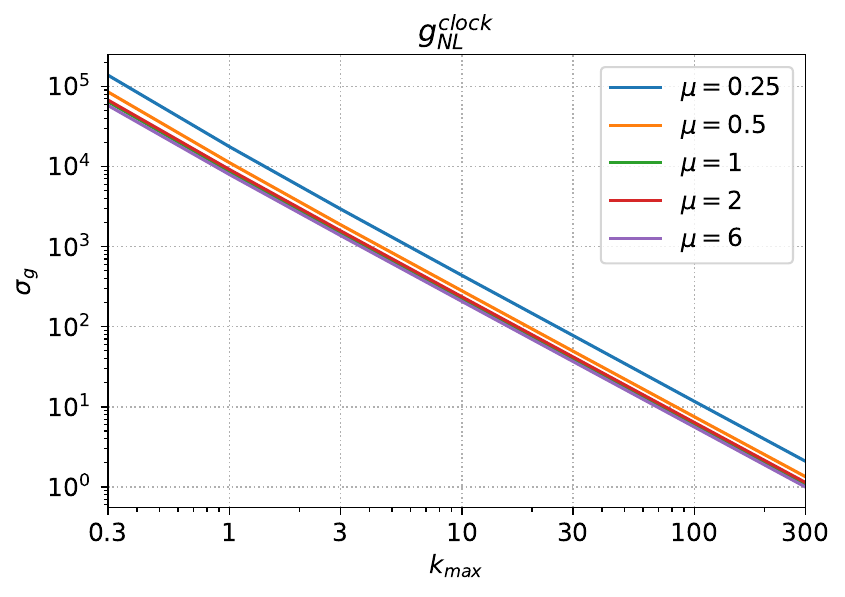}
    \caption{\emph{Left}: sensitivity to the amplitude of the trispectrum clock template with $\mu = 0.25$ as a function of $\kmax$ without marginalising (dashed), after marginalising over the 7 parameter secondary trispectrum (dashed-dotted), after marginalising over the common primordial trispectra (dotted) and after marginalising over both secondary and common primordial trispectra (solid). \emph{Right}: sensitivity to the amplitude of the clock template for different $\mu$ as function of $\kmax$ after marginalising over both secondary and common primordial trispectra.}
    \label{fig:TrispectraClockErrorkmax}
\end{figure}

\begin{figure}[H]
    \centering
    \includegraphics[scale=0.5]{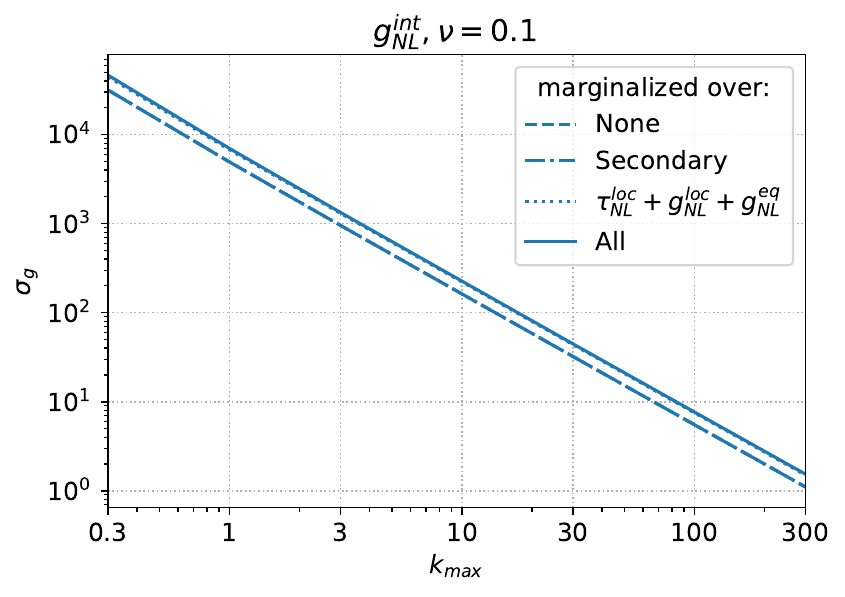}
    \includegraphics[scale=0.5]{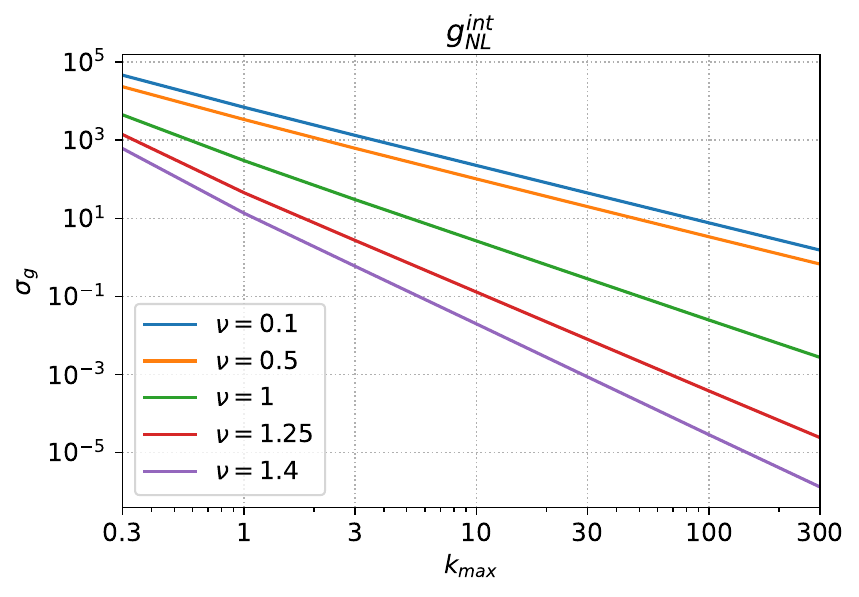}
    \caption{\emph{Left}: sensitivity to the amplitude of the trispectrum intermediate template with $\nu = 0.1$ as a function of $\kmax$ without marginalising (dashed), after marginalising over the 7 parameter secondary trispectrum (dashed-dotted), after marginalising over the common primordial trispectra (dotted) and after marginalising over both secondary and common primordial trispectra (solid). \emph{Right}: sensitivity to the amplitude of the intermediate template for different $\nu$ as function of $\kmax$ after marginalising over both secondary and common primordial trispectra.}
    \label{fig:TrispectraIntErrorkmax}
\end{figure}

We now investigate the impact of the secondary trispectrum on the previously obtained sensitivities to primordial trispectra. In Figure \ref{fig:SNDTrispectra} we summarize the signal-to-noise degradation after marginalising over the seven residual parameters of the secondary trispectrum. Due to the more complex shapes of the trispectra in $k$-space as compared to the bispectra, we see little overlap with the secondary trispectra, especially for higher values of $\kmax{}$. The small overlap is reflected in the small loss of sensitivity to the amplitude of the common shapes in Figure \ref{fig:CommonTrispectraErrorkmax}. As for the bispectrum, we find similar overlap for all parameters. Results for the clock and intermediate template are summarized in Figure \ref{fig:TrispectraClockErrorkmax} and Figure \ref{fig:TrispectraIntErrorkmax} respectively. Similar to the case of the bispectrum, in the left figures we see that when including the overlap of the common primordial trispectra, the additional effect of the secondary trispectra is negligible. What is different from the case of the bispectrum is the enhanced scaling for the intermediate template at mass values $\nu > 0.75$.

\subsection{Forecasting experiments}
\label{sec:BaselineForecasts}
\begin{figure}[H]
    \centering
    \includegraphics[scale=0.52]{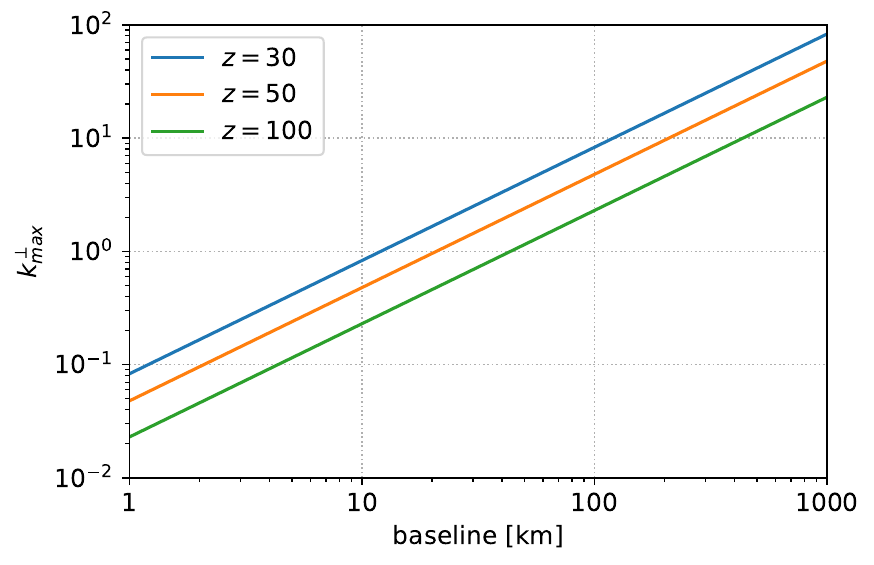}
    \includegraphics[scale=0.52]{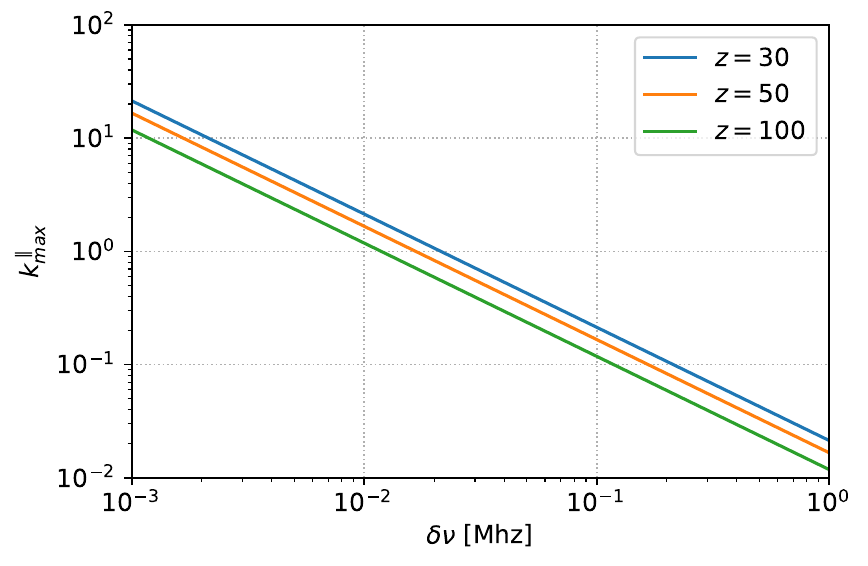}
    \caption{The maximal wavenumber $\kmax{}$ that can be probed in a certain direction for several redshifts. \emph{Left:} perpendicular $k_{\rm max}^\perp$ as a function of baseline in kilometers. \emph{Right:} parallel (line-of-sight) $k_{\rm max}^\parallel$ as a function of window size $\delta \nu$.}
    \label{fig:kmax}
\end{figure}

Until now we have studied the signal-to-noise with which we can measure primordial non-Gaussianity as a function of the maximum accessible momentum $\kmax{}$. In reality however, there is a difference between the radial (or redshift) component of the momenta along the line-of-sight $k_\parallel$ and the perpendicular (or angular) component $k_\perp$. The resolution along these components is set by different properties of the experimental setup. The line-of-sight resolution is determined by our ability to distinguish redshift slices, which in practise corresponds the size of the frequency bins set by the window size $\delta \nu$ \cite{Chen2016a}:

\begin{eqnarray}
\label{eqn:klosmax}
k_{\text{max}}^\parallel \approx \sqrt{\frac{17}{3}} \frac{1}{20 \delta \nu \sqrt{1+z}}
\end{eqnarray}
with $\delta \nu$ in units of MHz. Furthermore, the angular resolution R is set by the baseline $b$ (the distance between two receivers in an interferometer) of the experiment through:
\begin{eqnarray}
R = \frac{\lambda_{\textrm{obs}}}{b}
\end{eqnarray}
which yields the maximum perpendicular mode
\begin{eqnarray}
k_{\text{max}}^\perp \approx   \frac{2 \pi \nu_0 b}{d(z)(1+z)c}
\end{eqnarray}
with $b$ in units of km and where $\nu_0 \approx 1420 \times 10^6$ Hz is the frequency of the 21-cm signal, $c$ the speed of light in km/s and $d(z)$ the comoving distance to redshift $z$ in units of Mpc. In practise it is much easier to improve on the window $\delta \nu$ than it is to increase the baseline. In Figure \ref{fig:kmax} we show $\kmax{}$ as a function of baseline and window size at several redshifts.

\begin{figure}[H]
\centering
    \includegraphics[scale=0.345]{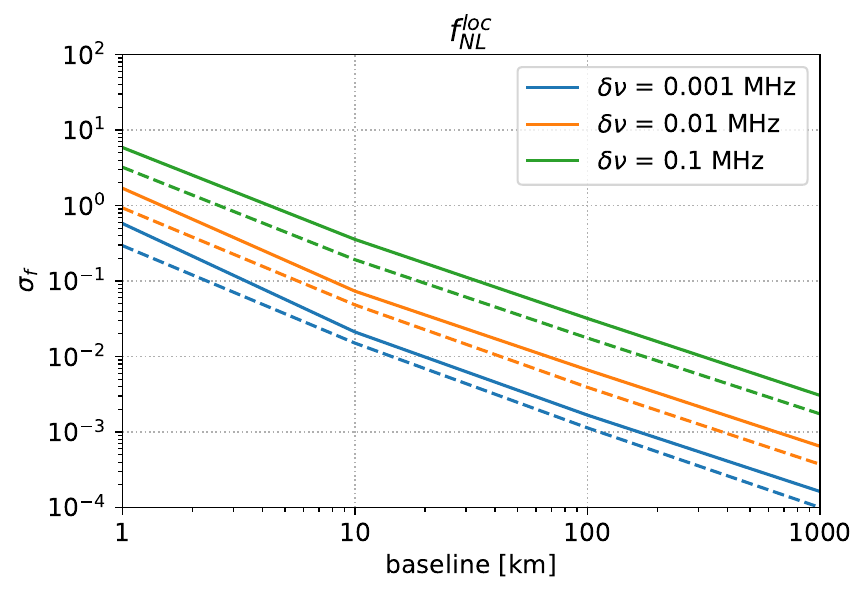}
    \includegraphics[scale=0.345]{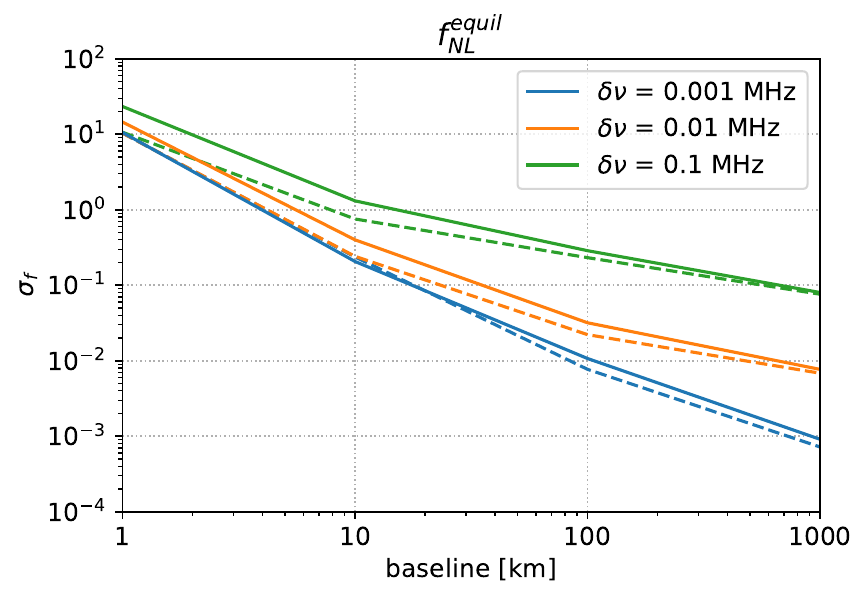}
    \includegraphics[scale=0.345]{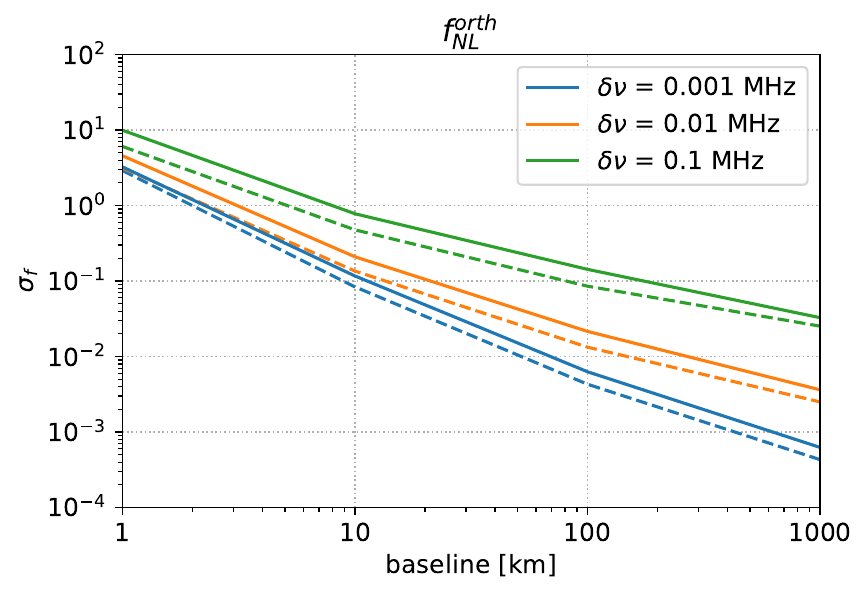}
    \caption{Forecast sensitivity to the amplitude of the common primordial bispectra for different baseline and window size, before (dashed) and after (solid) marginalising over the 4 parameter secondary bispectrum.} 
    \label{fig:BispectrumCommonBaseline}
\end{figure}

Figure \ref{fig:BispectrumCommonBaseline} shows the forecast sensitivity to the amplitude of the three common primordial bispectra. We find that already for a reasonably sized array with a baseline of several kilometers 21-cm observations from the Dark Ages can in improve current as well as future CMB constraints by orders of magnitude. Notably, $\sigma_{f^{\rm loc}_{\rm NL}} \sim O(10^{-2})$ can in principle be achieved with a baseline smaller than 10 kilometers, allowing for a test of the famous Maldacena consistency condition.

\begin{figure}[H]
\centering
    \includegraphics[scale=0.345]{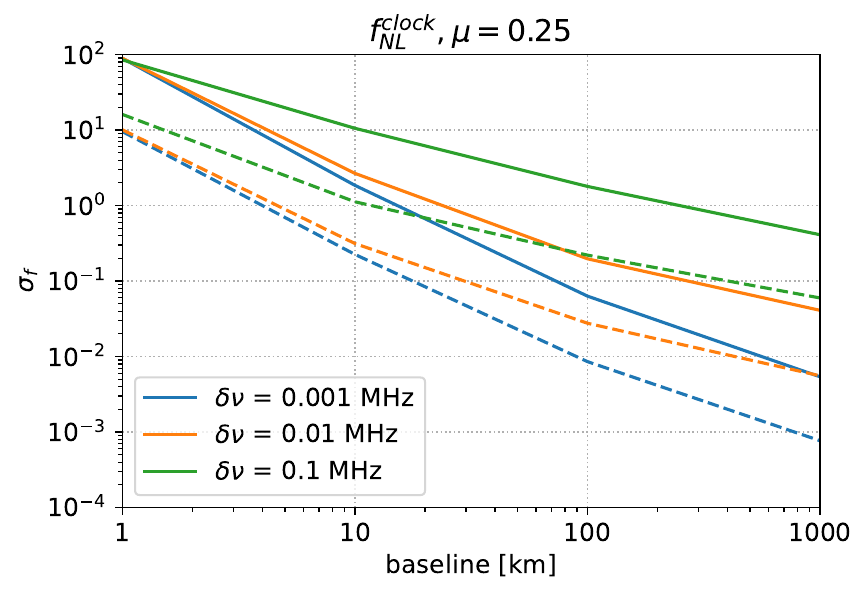}
    \includegraphics[scale=0.345]{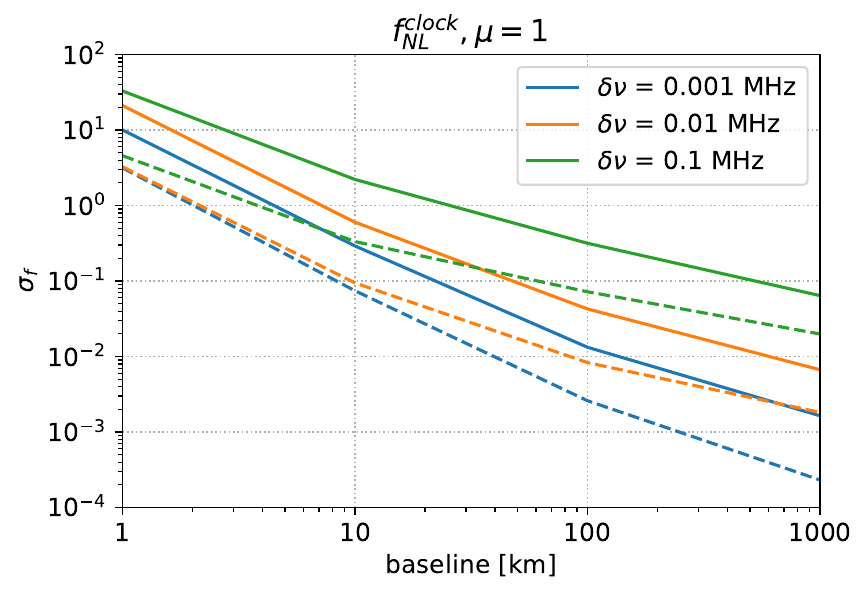}
    \includegraphics[scale=0.345]{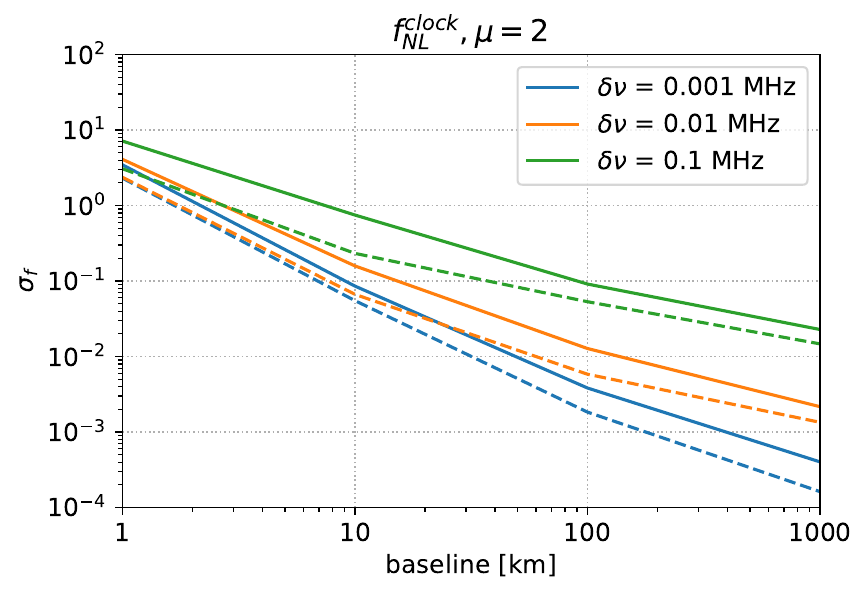}
    \includegraphics[scale=0.345]{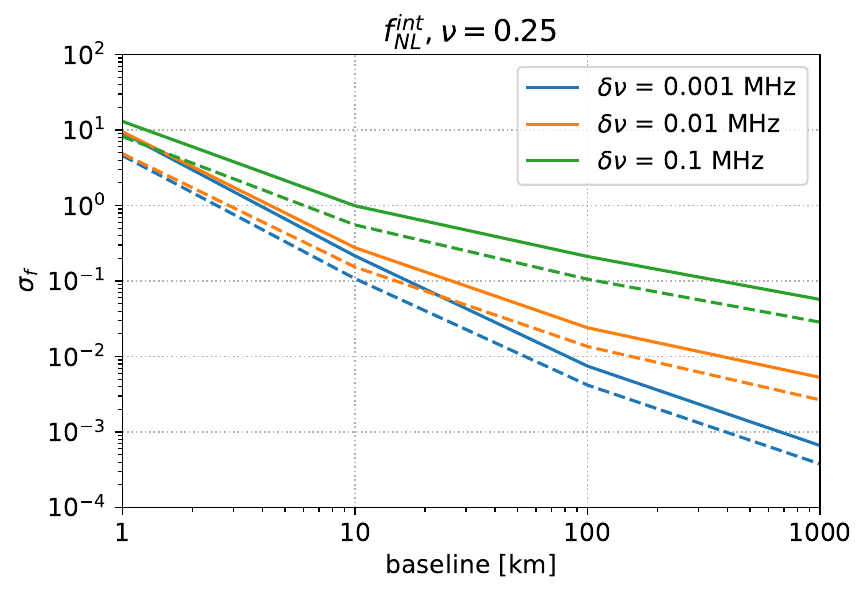}
    \includegraphics[scale=0.345]{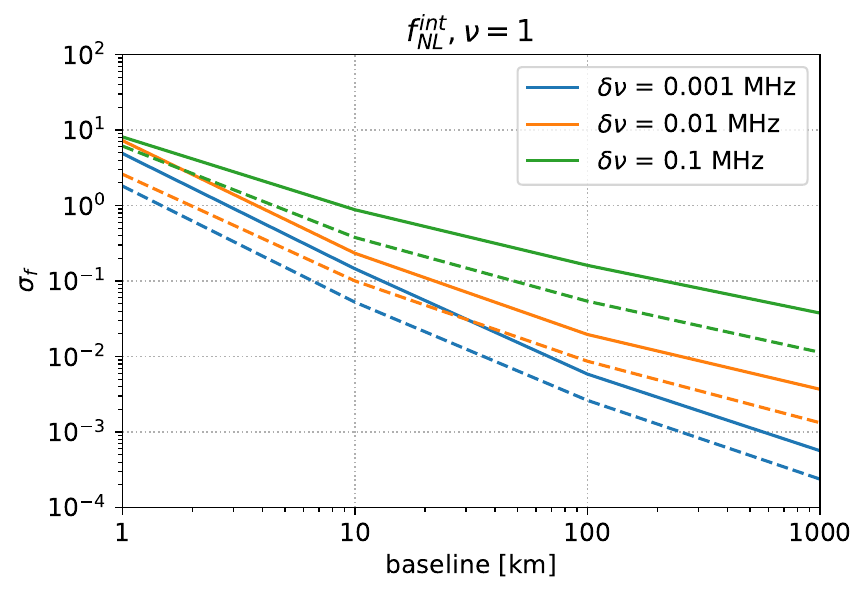}
    \includegraphics[scale=0.345]{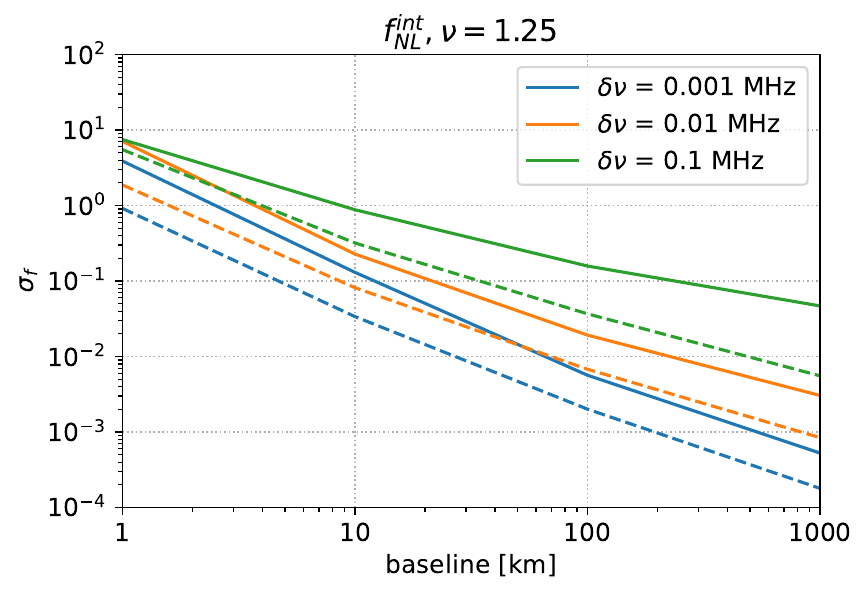}
    \caption{Forecast sensitivity to the amplitude of some massive exchange bispectra for different baseline and window, before (dashed) and after (solid) marginalising over the 4 parameter secondary bispectrum and common primordial bispectra. Top row is the Clock template, bottom row is the intermediate template.} 
    \label{fig:BispectrumExchangeBaseline}
\end{figure}

Figure \ref{fig:BispectrumExchangeBaseline} summarizes the ability to detect primordial features due to the exchange of massive particles during inflation for different baseline and window size. Depending on the nature of the exchange coupling the size of the amplitude of non-Gaussian signal can be anywhere between $\fnl{} \ll 10^{-2}$ and $\fnl{} > O(1)$ (see section E. of \cite{Meerburg2016} for a clear discussion on the different possibilities). We see that a baseline of $100 - 1000$ km could already provide a clear picture of the particle content of the early universe, opening up an entirely new window into the physics of inflation.

\begin{figure}[H]
\centering
    \includegraphics[scale=0.345]{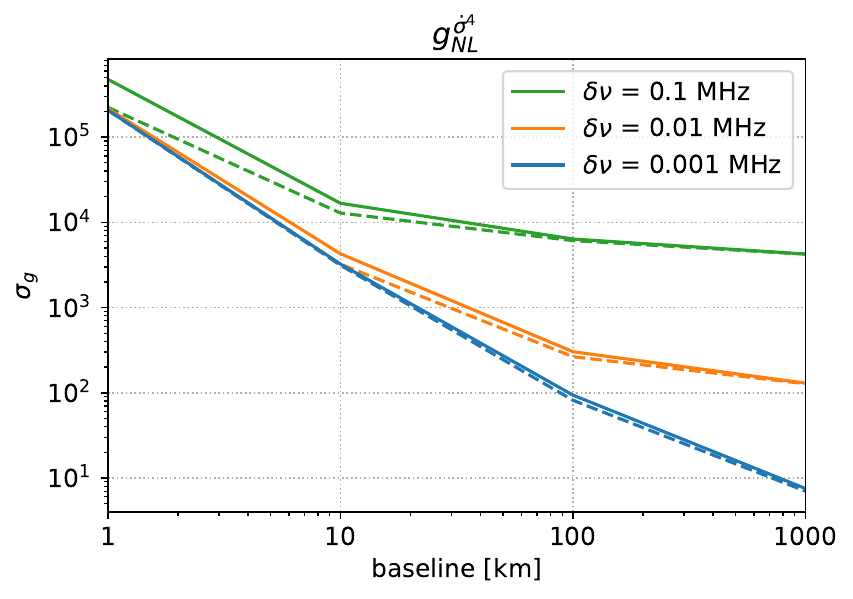}
    \includegraphics[scale=0.345]{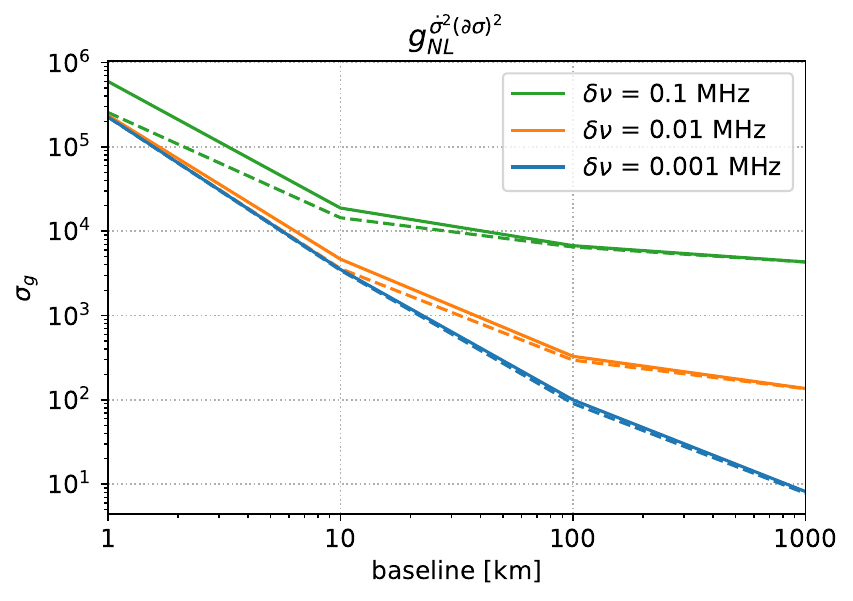}
    \includegraphics[scale=0.345]{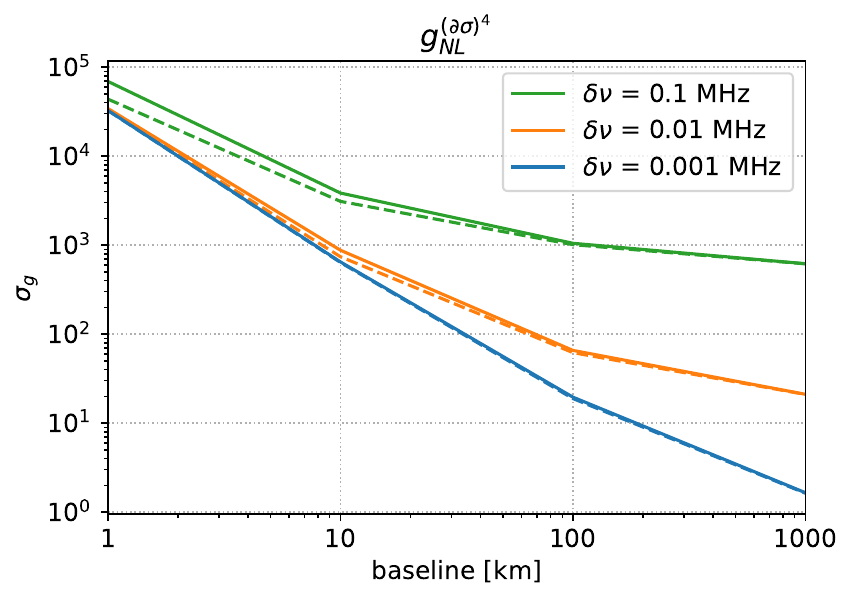}
    \includegraphics[scale=0.345]{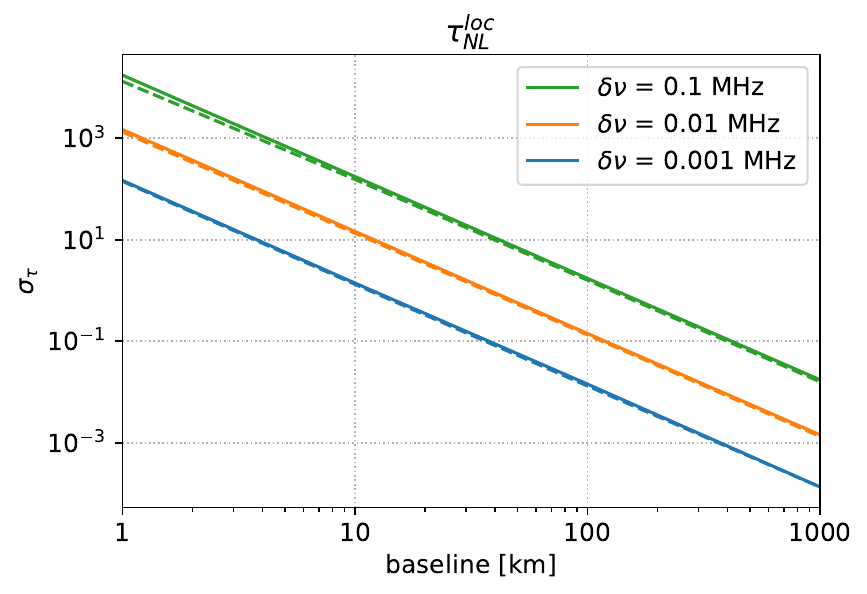}
    \includegraphics[scale=0.345]{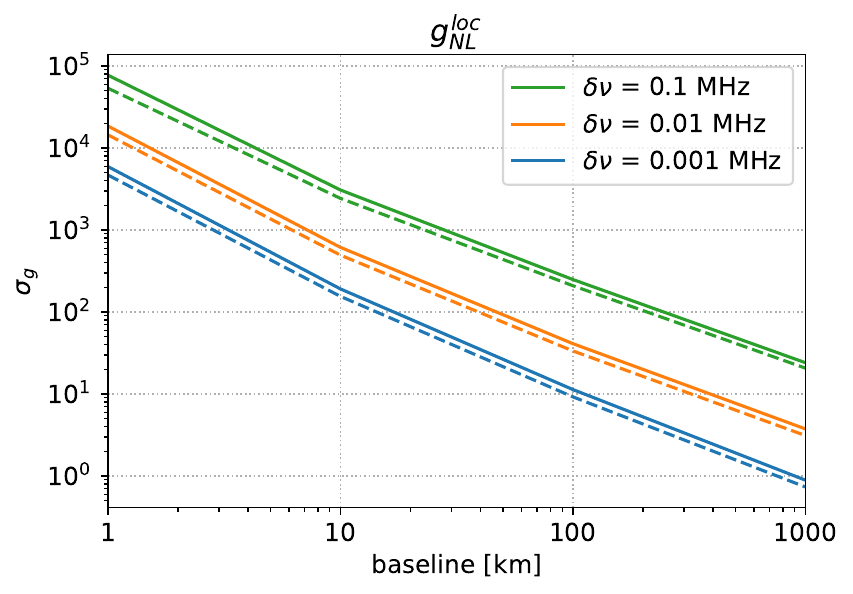}
    \caption{Forecast sensitivity to the amplitude of the primordial trispectra for different baseline and window size, before (dashed) and after (solid) marginalising over the 7 parameter secondary trispectrum.} 
    \label{fig:TrispectrumCommonBaseline}
\end{figure}

In Figure \ref{fig:TrispectrumCommonBaseline} we present for the first time forecasts of the sensitivity of different experimental setups to the amplitude of primordial trispectra, taking into account the secondary trispectrum of 21-cm brightness temperature fluctuations. We find that for most trispectra it proves hard to do better than $\sigma_g \sim O(1)$, a clear exception being the $\taunl{}$ local shape, that exhibits enhanced scaling with $\kmax{}$ resulting in sensitivities on par with that of the bispectrum.

\begin{figure}[H]
\centering
    \includegraphics[scale=0.345]{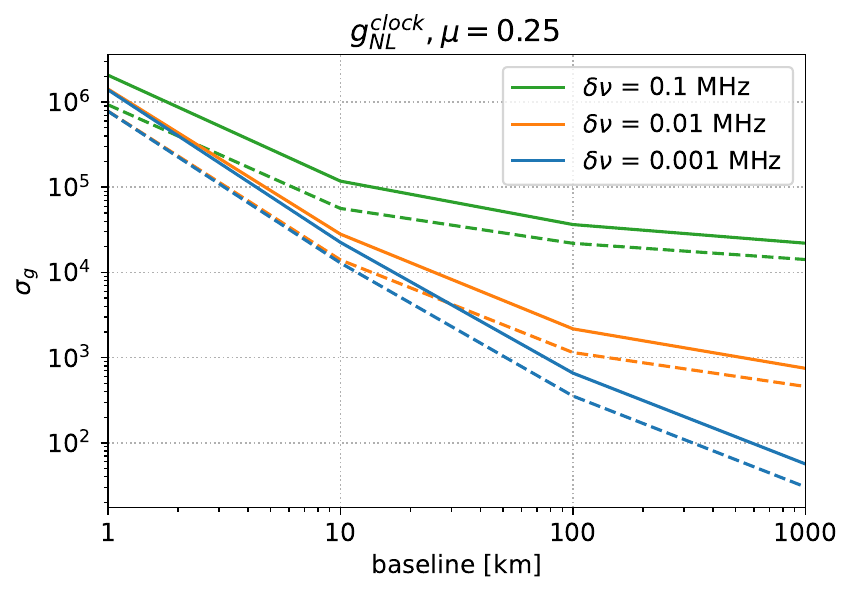}
    \includegraphics[scale=0.345]{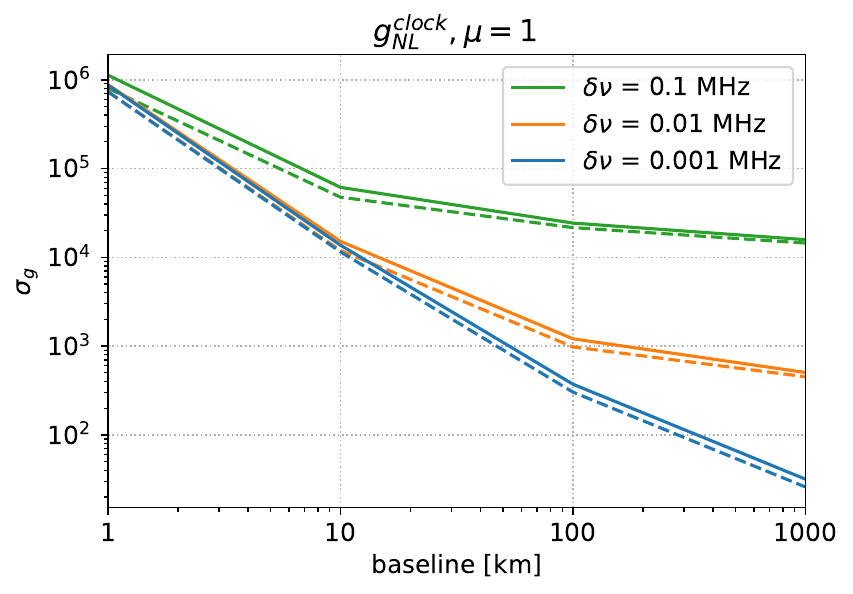}
    \includegraphics[scale=0.345]{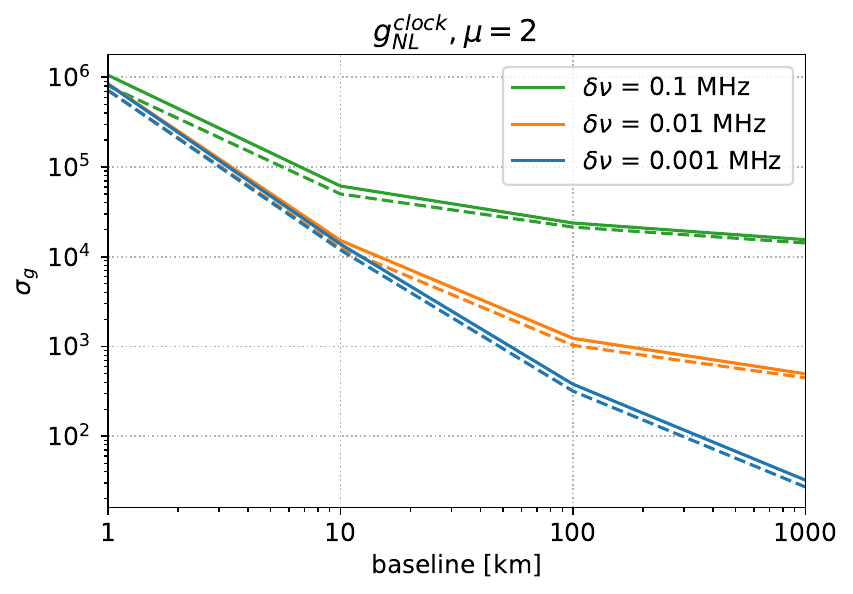}
    \includegraphics[scale=0.345]{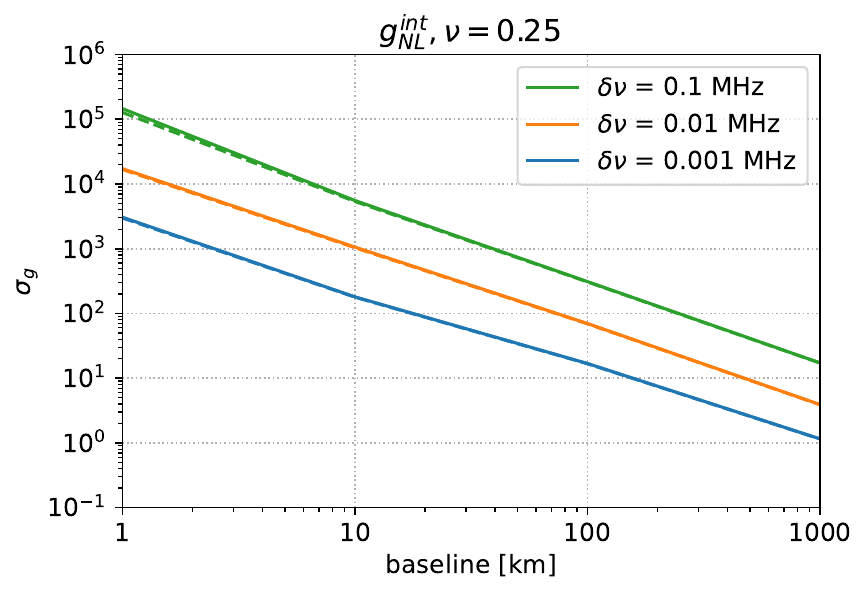}
    \includegraphics[scale=0.345]{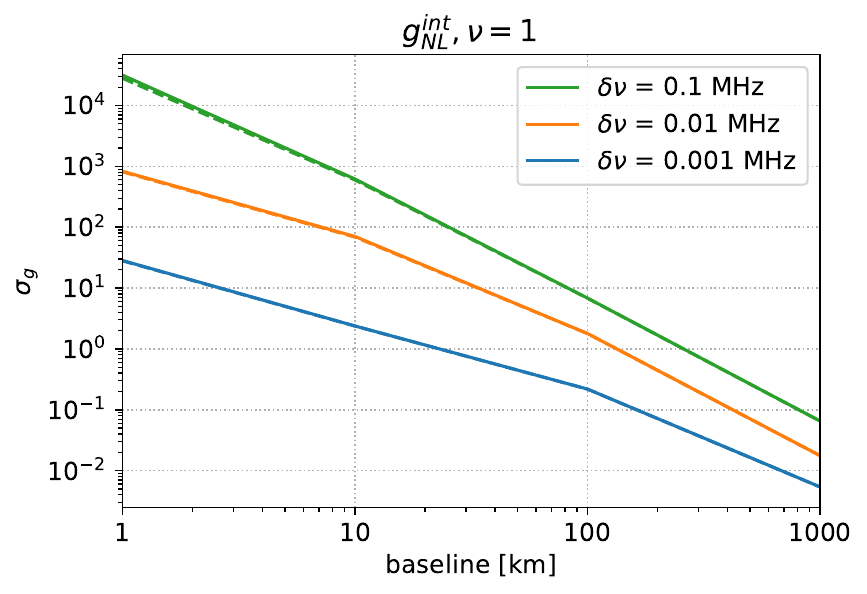}
    \includegraphics[scale=0.345]{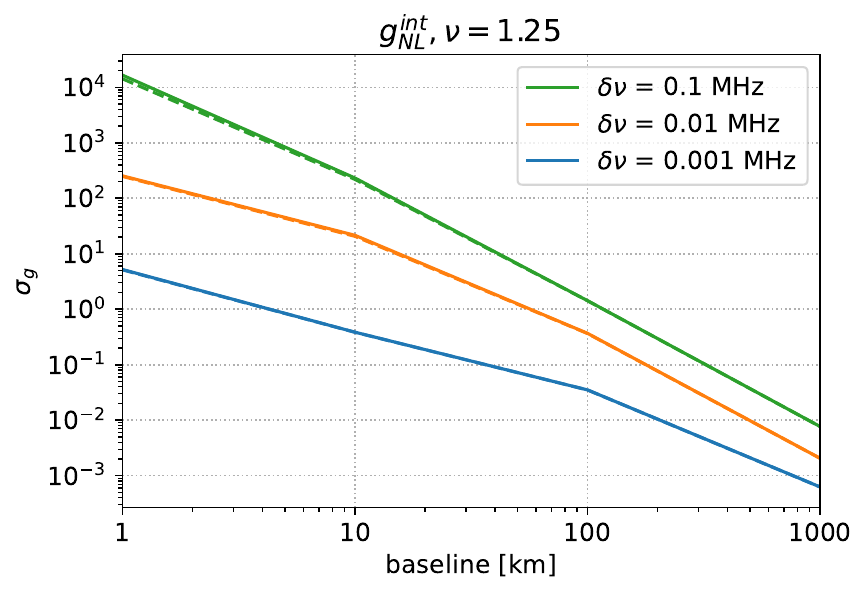}
    \caption{Forecast sensitivity to the amplitude of some massive exchange trispectra for different baseline and window size, before (dashed) and after (solid) marginalising over the 7 parameter secondary trispectrum and local and equilateral trispectra.}
    \label{fig:TrispectrumExchangeBaseline}
\end{figure}

Moving on to the trispectrum of massive exchange, we show the results in Figure \ref{fig:TrispectrumExchangeBaseline}. We conclude that the oscillatory clock template can realistically only be constrained up to a sensitivity of $\sigma_g \sim O(10)$, still many orders of magnitude better than CMB observations (likely $\sim O(10^4)$ in the future). However, the intermediate trispectrum of massive exchange can reach sensitivities that are much more similar to the bispectrum sensitivities, allowing one to probe the same masses through both the bispectrum and trispectrum. Although we have not explicitly included spinning particles in this work, having two independent probes of the same mass could be used to break degeneracy between the mass and spin of the particle, that both contribute to the amplitude of the signal \cite{Arkani-Hamed2015}.

\section{Conclusions and Outlook}
Measurements of primordial non-Gaussianity play a key role in unraveling the mysteries of the early universe. Until now, the most stringent constraints on the amount of pnG come from observations of the cosmic microwave background radiation, which are still consistent with zero. Future CMB experiments (e.g. SO and CMBS4) could improve constraints by an order of magnitude, possibly zeroing in on non-zero pnG. However, in order to accurately distinguish different non-Gaussian signatures such as the imprint of massive particles present during inflation, sensitivities are required that can simply not be achieved with CMB measurements, due to the limited amount of modes available. To advance our understanding of the primordial universe, we ought to employ additional probes. Arguably the most promising avenue to this end is the use of 21-cm brightness temperature fluctuations. During a part of the cosmological Dark Ages (between $30 \leq z \leq 200$), these tiny fluctuation can in principle be observed in absorption to the CMB radiation. The result is a large observable 3D volume of linear modes that trace the primordial initial conditions, which could constrain primordial non-Gaussianity with unprecedented precision, opening up an exciting new window into the physics of the early universe. \\

In order to extract such primordial information from the tracer field, it's required to accurately model the physics of the hydrogen gas. Furthermore, the non-linear relation between the initial conditions and the 21-cm brightness temperature fluctuations as well as gravitational collapse of the matter field induce secondary non-Gaussianity, swamping to primordial signal that we aim to observe. In this paper, we (re)address some of these aspects and made several advances. Specifically, we have:

\begin{itemize}
    \item improved analytical modelling of the hydrogen gas by including the effect of free electron fraction perturbations, which turns out to be sizable (Figure \ref{fig:CTcoeff}) and hence should be included in any detailed analysis of 21-cm observations of the Dark Ages
    \item derived the perturbation expansion of the 21-cm tracer field up to third order in the underlying matter field, while including baryonic pressure effects on scales close to the Jeans scale
    \item thereby obtained an improved expression for the secondary bispectrum of 21-cm brightness temperature fluctuations during the Dark Ages
    \item derived, for the first time, the secondary trispectrum of the 21-cm tracer field
\end{itemize}
With this improved modelling of the tracer field and the newly obtained secondary bi- and trispectrum, we asses the information content of 21-cm fluctuations to constrain primordial physics, under the most optimal conditions. More specifically, we perform a Fisher analysis to determine the error bars on the amplitude of various primordial bi- and trispectra in a potential 21-cm Dark Ages experiment, marginalizing over the parameters of secondary non-Gaussianity. From this analysis we draw the following conclusions:

\begin{itemize}
    \item although secondary non-Gaussianity swamps the primordial signal, the primordial signal can be extracted without a big loss of signal-to-noise both for the bispectrum and the trispectrum
    \item tomography of 21-cm fluctuations can in principle improve constraints on primordial non-Gaussianity by several orders of magnitude, already for relatively low $\kmax{}$, as noted before in refs. \cite{Pillepich2006,Munoz2015,Meerburg2016}
    \item the enhanced scaling of some trispectra with $\kmax{}$ \cite{Kalaja2020} indeed manifests itself in 21-cm observations.
    \item our Fisher forecasts demonstrate the importance of marginalizing over primordial shapes as compared to secondary shapes when considering the massive particle exchange templates
\end{itemize}
Our results clearly show the potential of the 21-cm signal from the Dark Ages to improve constraints on primordial non-Gaussianity and thereby our understanding of the infant universe. Many interesting directions of research remain in order to better understand the intricacies of the 21-cm Dark Ages signal. We would like to reiterate some of the assumptions made in this work that can be improved upon as well as some other possible directions of future research:

\begin{itemize}
    \item we have included baryonic pressure effects to the secondary bispectrum along the lines of \cite{Shoji2009}. One of the assumptions made there is the constancy of the baryonic Jeans scale $k_J$. In reality this scale depends on time (redshift) and a more accurate modelling of secondary non-Gaussianity of the 21-cm signal should include this time-dependence
    \item Our expression for the 21-cm trispectrum can be used to include non-Gaussian covariance in an analysis of the 21-cm power spectrum such as \cite{Chen2016a} (see \cite{Bertolini:2015fya} for a similar analysis in the context of LSS). In principle non-Gaussian covariance will also impact the forecasts presented in this work. We leave this to future work.
    \item By modelling the effects of the first stars' Lyman-$\alpha$ radiation on the 21-cm brightness temperature, one could extend our analysis from the Dark Ages into the epoch of Cosmic Dawn. Earth-based experiments measuring the 21-cm signal from this epoch are already being prepared and could provide constraints on primordial non-Gaussianity in the nearer future \cite{SKA,HERA}.
\end{itemize}
Finally it's imperative to once more mention the practical and technological challenges involved in order to measure such a signal. Although the 21-cm signal from the epoch of reionization can be observed from earth, once we pass into the Dark Ages ($z\geq 30$), the signal has redshifted into wavelengths that are largely blocked by earth's ionosphere, requiring us to design and build a lunar or space based observatory. Nevertheless, serious efforts are being made for establishing a lunar based observatory with future NASA and ESA moon missions \cite{Silk2020,DAPPER,KoopmansPeering}. Furthermore, a host of foreground sources obscure the pristine 21-cm signal, of which synchrotron emission in our own galaxy the most dominant and the amplitude of these foregrounds increasing towards lower frequencies (higher redshifts). In theory, foregrounds can be avoided when looking at fluctuations as the foreground mode appears initially as a line of sight mode on large scales. Interferometers then introduce additional mode coupling that leads to the famous wedge \cite{Parsons:2012qh,Liu:2014bba,Liu:2014yxa,Ghosh:2017woo}. The future of 21-cm as a competitive cosmological probe will rely critically on whether we are capable of extracting the signal away from the wedge (although it might be possible to reconstruct some of the modes in the wedge, see e.g. \cite{Modi2019}). However these challenges are part of the ongoing efforts in this field, both at low redshifts \cite{CHIME,TIANLAI,HIRAX,PUMA}, reinoization redshifts \cite{HERA,LOFAR,LWARe,MWA,SKA} and high redshift probes \cite{NenuFAR, LWADawn, SKA}.\footnote{Besides challenges related to signal extraction for the very large interferometers there are also computational challenges which will have to be addresses due to the shear number of baselines that can be correlated, see e.g. \cite{PUMA} and references therein.} Finally, we have ignored any form of noise as due to the experimental setup (e.g. thermal noise, systematics etc.) as well as any technical challenges such as integration time. We leave a more realistic forecast including these considerations to future work. The main goal of this work is to show the potential of the 21-cm signal at high redshift and to establish the importance of this probe as a way to explore the nature of the early universe in the future. Reaching sensitivities that allow us to answer questions related to new particles and new fields will be limited by the number of available modes \cite{Kalaja2020} and the epoch of  the dark ages in principle contains the largest number of these pristine modes.

\acknowledgments

The authors thank Matteo Biagetti, Alba Kalaja, Eiichiro Komatsu and Guilherme Pimentel for insightful discussions. We thank Julian Mu\~noz and Moritz Münchmeyer for discussions and providing feedback on the manuscript.  We thank the Center for Information Technology of the University of Groningen for their support and for providing access to the Peregrine high performance computing cluster. T.F is supported by the Fundamentals of the Universe research program within the University of Groningen. P.D.M acknowledges support from the Netherlands organization for scientific research (NWO) VIDI grant (dossier 639.042.730). L.V.E.K acknowledges the financial support from
the European Research Council (ERC) under 
the European Union’s Horizon 2020 research 
and innovation programme (Grant agreement 
No. 884760, "CoDEX"). 

\appendix
\section{Evolution of perturbations}\label{app:evopert}
In this appendix, we will provide the explicit evolution equations for perturbations in the free electron fraction and gas temperature, used to numerically solve for the coupling coefficients $\mathcal{C}_{i,j}^{x,\mathrm{T}}$ (defined in equation \eqref{eq:dTexpansion} and equation \eqref{eq:dxexpansion}).

\subsubsection*{Free electron fraction}
To obtain the evolution equations for $\dx_n$, we substitute the expansions for $\dxe$ and $\dT$ (equation \eqref{eq:expansiondTdxe}) into the evolution equation for $\dxe$ (equation \eqref{eq:evodxe}). Order by order, we then obtain the following evolution equations: 
\begin{align}
    \dxdot_1=-\Gr(&\dx_1+A_1\dTup_1+\delta_1),\\
    \dxdot_2=-\Gr\Big[&\dx_2+A_2\dTup_2+\delta_2+(\dx_1)^2+2\dx_1\delta_1+A_1\dTup_1(\delta_1+\dx_1)+A_2(\dTup_1)^2\Big],\\
    \dxdot_3=-\Gr\Big[&\dx_3+A_3\dTup_3+\delta_3+2\dx_1\dx_2+2(\dx_2\delta_1+\dx_1\delta_2)+(\dx_1)^2\delta_1\nonumber\\
    &+2A_1(\dx_1\dTup_2+\dx_2\dTup_1)+A_1\dTup_1\dx_1(\dx_1+2\delta_1)+A_1(\dTup_1\delta_2+\dTup_2\delta_1)\nonumber\\
    &+A_2(\dTup_1)^2(\delta_1+2\dx_1)+2A_2\dTup_1\dTup_2+A_3(\dTup_1)^3\Big].
\end{align}

\subsubsection*{Gas temperature}
Similarly, we obtain the evolution equations for $\dTup_n$ by substituting equation \eqref{eq:expansiondTdxe} into equation \eqref{eq:dTevo} to get:
\begin{align}
    \dTupdot_1&=\Theta_1,\\
    \dTupdot_2&=\Theta_2+\frac{2}{3}H\delta_1(\dTup_1-\delta_1)-\Gc\dx_1\dTup_1,\\
    \dTupdot_3&=\Theta_3+\frac{2}{3}H\delta_1\big[(\delta_1)^2+\dTup_2-3\delta_2\big]+\frac{2}{3}H\big[2\delta_2-(\delta_1)^2\big]\dTup_1-\Gc[\dx_2\dTup_1+\dx_1\dTup_2],
\end{align}
where we have used that $\dot{\delta}_n=nH\delta_n$ and defined $\Theta_n$ as:
\begin{equation}
    \Theta_n\equiv \Gc\Big[(\Tgb/\Tgasb-1)\dx_n-(\Tgb/\Tgasb)\dTup_n\Big]+\frac{2}{3}Hn\delta_n. 
\end{equation}

\subsubsection*{Coupling coefficients}
Note that the evolution equations for $\dxTup_n$ are coupled \emph{linear} differential equations, we can turn them into coupled differential equations for the coupling coefficients. Compactly, the evolution equations for the couplings can be written as:
\begin{equation}
    \frac{d}{da}\CxT_{n,m}=\DxT_{n,m}+\SxT_{n,m},
\end{equation}
in terms $a\equiv (1+z)^{-1}$ as evolution variable. In the compact notation above, $\DxT_{n,m}$ denote all terms contributing to the evolution at the order $n$, whereas $\SxT_{n,m}$ encode the (combined) contribution of lower order perturbations. Explicitly, $\DxT_{n,m}$ are defined as:
\begin{align}\label{eq:DT}
	\DT_{n,m}&\equiv -\bigg[n+\frac{\Gc}{H}\frac{\Tgb}{\Tgasb}\bigg]\frac{\CT_{n,m}}{a}+\frac{\Gc}{H}\bigg[\frac{\Tgb}{\Tgasb}-1\bigg]\frac{\Cx_{n,m}}{a}+\frac{2n}{3a}\;\delta_{nm}, \\
	\Dx_{n,m}&\equiv -\bigg[n+\frac{\Gr}{H}\bigg]\frac{\Cx_{n,m}}{a}-\frac{1}{a}\frac{\Gr}{H}\big[A_1\CT_{n,m}+\delta_{nm}\big]. 
\end{align}
By construction, the evolution of the coefficients $\CxT_{n,n}$ is only sourced by $\CxT_{n,n}$, so that $\SxT_{n,n}=0$. The cases $n\neq m$ are non-zero and read:
\begin{align}
    \ST_{2,1}\equiv&-\frac{1}{a}\frac{\Gc}{H}\CT_{1,1}\Cx_{1,1}+\frac{2}{3a}\Big[\CT_{1,1}-1\Big],\\
    \ST_{3,1}\equiv&-\frac{1}{a}\frac{\Gc}{H}\Big[\CT_{1,1}\Cx_{2,1}+\CT_{2,1}\Cx_{1,1}\Big]+\frac{2}{3a}\Big[1-\CT_{1,1}+\CT_{2,1}\Big],\\
    \ST_{3,2}\equiv&-\frac{1}{a}\frac{\Gc}{H}\Big[\CT_{1,1}\Cx_{2,2}+\CT_{2,2}\Cx_{1,1}\Big]+\frac{2}{3a}\Big[-3+2\CT_{1,1}+\CT_{2,2}\Big],\\
    \Sx_{2,1}\equiv& -\frac{1}{a}\frac{\Gr}{H}\Big[\Cx_{1,1}(\Cx_{1,1}+2)+A_1\CT_{1,1}(1+2\Cx_{1,1})+A_2(\CT_{1,1})^2\Big],\\
    \Sx_{3,1}\equiv&-\frac{1}{a}\frac{\Gr}{H}\Big[2\Cx_{2,1}(1+\Cx_{1,1})+(\Cx_{1,1})^2+A_1\CT_{2,1}(1+2\Cx_{1,1})+A_1\CT_{1,1}([\Cx_{1,1}]^2+2\Cx_{2,1}+2\Cx_{1,1})\Big]\nonumber\\
    &-\frac{1}{a}\frac{\Gr}{H}\Big[A_2(\CT_{1,1})^2(1+2\Cx_{1,1})+2A_2\CT_{1,1}\CT_{2,1}+A_3(\CT_{1,1})^3\Big],\\
    \Sx_{3,2}\equiv&-\frac{1}{a}\frac{\Gr}{H}\Big[2\Cx_{2,2}(1+\Cx_{1,1})+2\Cx_{1,1}+A_1\CT_{1,1}(1+2\Cx_{2,2})+A_1\CT_{2,2}(1+2\Cx_{1,1})+2A_2\CT_{1,1}\CT_{2,2}\Big]. 
\end{align}

\subsubsection*{Initial conditions}
To solve the above system of equations, we  set the initial conditions following the reasoning of \cite{Ali-Haimoud2014}. At high redshifts ($z_i=1000$) we have $\Tgas=\Tg$ to high accuracy. Since we neglect fluctuations in the photon temperature, we conclude that at $z_i$ gas temperature fluctuations must vanish as well, $\dT(z_i,\boldsymbol{x})=0$, or equivalently $\CT_{n,m}(z_i)=0$. Regarding electron fraction fluctuations $\dxe$ (or equivalently $\Cx_{n,m}$), one should in principle start computing the evolution at earlier times to get the appropriate initial conditions at the initial redshift $z_i$. However, we are interested in the effect of $\dxe$ at late times (during the Dark Ages) on the 21-cm signal, which only enters via the coupling to gas temperature fluctuations. At late times, this coupling is insensitive to the initial value $\dxe(z_i)$, since perturbations grow after recombination and the initial value is quickly forgotten \cite{Ali-Haimoud2014}. Therefore, we may set $\dxe(z_i)=0$, corresponding to $\Cx_{n,m}(z_i)=0$.

\section{Fisher information matrix}
\label{app:Fisher}
We summarize the relevant equations used to perform the forecasts in this paper. Following \cite{Chen2016a} consider a cosmic variance limited survey over 14 redshift bins between $30\leq z \leq 100$, assuming these are sufficiently uncorrelated. For the 21-cm bispectrum the (cosmic variance limited) Fisher information matrix for a redshift bin $z_i$ is defined by \cite{Meerburg2016,Scoccimarro2003,Baldauf2016}:
\begin{eqnarray}
\label{eq:FisherBispectrum}
\label{eq:Fisher21cmBispectrum}
            F_{\alpha\beta}(z_i) = \int_{\bk{1}>\bk{2}>\bk{3}} \frac{\partial B_{\delta T}(\bk{1},\bk{2},\bk{3},z_i)}{\partial p_\alpha} \frac{(2\pi)^6 \delta_{D}^2(\bk{1} + \bk{2} + \bk{3})}{P_{\delta T}(\bk{1},z_i)P_{\delta T}(\bk{2},z_i)P_{\delta T}(\bk{3},z_i)} \frac{\partial B_{\delta T}(\bk{1},\bk{2},\bk{3},z_i)}{\partial p_\beta} \nonumber \\
\end{eqnarray}
where $p_\alpha,p_\beta$ are the bispectrum parameters of interest. The total Fisher information is then simply given by the sum over redshift bins. When considering primordial bispectra one sees that all $z$ and line-of-sight momentum dependence in the numerator cancel against the denominator and we find the simplified Fisher information matrix:
\begin{eqnarray}
        F_{\alpha\beta} = V_{\textrm{tot}} \int_{\bk{1}>\bk{2}>\bk{3}} \frac{\partial B_{\zeta}(\bk{1},\bk{2},\bk{3})}{\partial p_\alpha}\frac{(2\pi)^3 \diracd{}(\bk{1}+\bk{2} + \bk{3})}{P_\zeta(k_1)P_\zeta(k_2)P_\zeta(k_3)} \frac{\partial B_{\zeta} (\bk{1},\bk{2},\bk{3})}{\partial p_\beta}
\end{eqnarray}
where we used $\diracd{}(\boldsymbol{0}) = V/(2\pi)^3$ where $V_{\textrm{tot}}$ is the total comoving survey volume. Now no binning of redshift is necessary. Furthermore, for (isotropic) primordial bispectra, the fisher matrix can be further simplified by integrating out the delta function, to read \cite{Fergusson2012}

\begin{eqnarray}
\label{eq:FisherPrimordialBispectrum}
        F_{\alpha\beta} = \frac{V_{\textrm{tot}}}{8\pi^4} \int_{k_\textrm{min}}^{k_{\textrm{max}}} dk_1 \int_{k_1/2}^{k_1} dk_2 \int_{k^*_\textrm{min}}^{k_2} dk_3 \frac{\partial B_\zeta(k_1,k_2,k_3)}{\partial p_\alpha}\frac{k_1 k_2 k_3}{P_\zeta(k_1)P_\zeta(k_2)P_\zeta(k_3)} \frac{\partial B_\zeta(k_1,k_2,k_3)}{\partial p_\beta} \nonumber \\
\end{eqnarray}
where $k_{\text{min}}^* = \text{Max}(\kmin{},k_1 - k_2)$. Using this expression we determine the total amount of information 
 in \ref{sec:totalinformation}.
Contrary to the primordial bispectrum, the 21-cm secondary bispectrum (see e.g. \eqref{eq:21cmBispectrum}) depends explicitly on the line-of-sight momentum through the angles $\mu(\bk{}) = k_\parallel/k$. Furthermore, the redshift dependence in the numerator and denominator of the Fisher matrix equation no longer cancel. Starting from equation \eqref{eq:FisherBispectrum} and integrating out the delta function we find

\begin{eqnarray}
F_{\alpha\beta}(z_i) = &&\frac{1}{3!}\frac{V_i}{(2\pi)^6}\int_{\kmin{}}^{\kmax{}} d^2\bk{1 \perp} \af d^2\bk{2 \perp} \af dk_{1\parallel} \af dk_{2\parallel} \nonumber \\ &&\frac{\partial B_{\delta T}(\bk{1},\bk{2},\bk{3},z_i)}{\partial p_\alpha}\frac{1}{P_{\delta T}(\bk{1},z_i)P_{\delta T}(\bk{2},z_i)P_{\delta T}(\bk{3},z_i)} \frac{\partial B_{\delta T}(\bk{1},\bk{2},\bk{3},z_i)}{\partial p_\beta} \nonumber \\
\end{eqnarray}
where now $\bk{3} = -\bk{1} - \bk{2}$ and the factor $1/3!$ in front is there to count every triangle configuration only once.
Going to cylindrical coordinates, we can get rid of one more (angular) degree of freedom by writing $k_{3\perp}^2 = k_{1\perp}^2 + k_{2\perp}^2 + 2 k_{1\perp} k_{2\perp} \cos(\phi_{12})$ and invoking isotropy in the perpendicular direction:

\begin{eqnarray}
F_{\alpha\beta}(z_i) = && \frac{V_i}{3!(2\pi)^5} \int_{0}^{2\pi}d\phi_{12} \int_{\kmin{}}^{\kmax{}} dk_{1\perp} \af dk_{2\perp} \af dk_{1\parallel} \af dk_{2\parallel} \af k_{1\perp} k_{2\perp} \nonumber \\ 
&&\frac{\partial B_{\delta T}(\bk{1},\bk{2},\bk{3},z_i)}{\partial p_\alpha}\frac{1}{P_{\delta T}(\bk{1},z_i)P_{\delta T}(\bk{2},z_i)P_{\delta T}(\bk{3},z_i)} \frac{\partial B_{\delta T}(\bk{1},\bk{2},\bk{3},z_i)}{\partial p_\beta} \nonumber \\
\end{eqnarray}
We use this expression to evaluate the Fisher matrix for the 21-cm secondary bispectrum. The integration ranges can be straightforwardly modified to distinguish between $k_{\text{max}}^\parallel$ and $k_{\text{max}}^\perp$ as we do in section \ref{sec:BaselineForecasts}.

Analogous to the bispectrum, the Fisher matrix for the trispectrum is defined as:
\begin{eqnarray}
\label{eq:FisherTrispectrum}
            F_{\alpha\beta}(z_i) = &&\int_{\bk{1}>\bk{2}>\bk{3}>\bk{4}} \frac{\partial T_{\delta T}(\bk{1},\bk{2},\bk{3},\bk{4},z_i)}{\partial p_\alpha} \times \nonumber \\ 
            && \frac{(2\pi)^6 \delta_{D}^2(\bk{1} + \bk{2} + \bk{3} + \bk{4})}{P_{\delta T}(\bk{1},z_i)P_{\delta T}(\bk{2},z_i)P_{\delta T}(\bk{3},z_i),P_{\delta T}(\bk{4},z_i)} \frac{\partial T_{\delta T}(\bk{1},\bk{2},\bk{3},\bk{4},z_i)}{\partial p_\beta} \nonumber \\
\end{eqnarray}
When considering only primordial trispectra, direction- and z-dependence again drop out. The quadrilateral can then be parametrized by the length of four sides and two diagonals and the Fisher matrix can be shown to take the following form \cite{Fergusson2012}:

\begin{eqnarray}
\label{eq:FisherPrimordialTrispectrum}
F_{\alpha\beta} = \frac{V_\textrm{tot}}{4!(2\pi)^3 2\pi^4} \int_{\mathcal{V}_T} dk_1 \af dk_2 \af dk_3 \af dk_4 \af ds \af dt \af \frac{\partial T_{\zeta}(k_1,k_2,k_3,k_4,s,t)}{\partial p_\alpha} \times \nonumber \\ \frac{k_1 k_2 k_3 k_4 s t}{\sqrt{g_1} P_\zeta(k_1)P_\zeta(k_2)P_\zeta(k_3)P_\zeta(k_4)}\frac{\partial T_{\zeta}(k_1,k_2,k_3,k_4,s,t)}{\partial p_\beta}
\end{eqnarray}
where $\mathcal{V}_T$ is the tetrahedral region spanned by the quadrilateral, which can be enforced by triangle conditions on every side of the tetrahedron \cite{Chen2009tri}. Furthermore, the function $g_1$ is given by \cite{Fergusson2012}:
\begin{eqnarray}
g_1 = s^2 t^2 \left(\sum_i k_i^2 -s^2 -t^2\right) - s^2\kappa_{23}\kappa_{14} + t^2 \kappa_{12} \kappa_{32} - (k_1^2 k_3^2 - k_2^2 k_4^2)(\kappa_{12} + \kappa_{34})
\end{eqnarray}.
Using equation \eqref{eq:FisherPrimordialTrispectrum} we forecast the amount of information available to constrain the trispectrum from the Dark Ages in section \ref{sec:totalinformation}.
For the 21-cm secondary trispectrum, we again distinguish between perpendicular and line-of-sight momenta. One way to parametrize these quadrilaterals is using three of its sides:
\begin{eqnarray}
F_{\alpha\beta}(z_i) =  \frac{1}{4!} \frac{V_i}{(2\pi)^9} && \int d^2\bk{1 \perp} d^2\bk{2 \perp} d^2\bk{3 \perp} dk_{1\parallel} dk_{2\parallel} dk_{3\parallel} \af \nonumber \\ 
&& \frac{\partial T_{\delta T}(\bk{1},\bk{2},\bk{3},\bk{4},z_i)}{\partial p_\alpha} 
             \left(\prod_{j=1}^4 \frac{1}{P_{\delta T}(\bk{j},z_i)}\right) \frac{\partial T_{\delta T}(\bk{1},\bk{2},\bk{3},\bk{4},z_i)}{\partial p_\beta} \nonumber \\
\end{eqnarray}
where it is now understood that $\bk{4} = -\bk{1}-\bk{2}-\bk{3}$. Furthermore, in cylindrical coordinates we can write 
\begin{eqnarray}
k_{4\perp}^2 = k_{1\perp}^2 + k_{2\perp}^2 + k_{3\perp}^2 + 2k_{1\perp} k_{2\perp} \cos{(\phi_1-\phi_2)} + 2k_{1\perp} k_{3\perp} \cos{(\phi_1-\phi_3)} \nonumber \\ + 2k_{2\perp} k_{3\perp} \cos{(\phi_2-\phi_3)}
\end{eqnarray}
and we can get rid of one angular degree of freedom by isotropy, leading to the following expression for the Fisher matrix:
\begin{eqnarray}
\label{eq:FisherTrispectrumSecondary}
F_{\alpha\beta}(z_i) =  \frac{V_i}{4!(2\pi)^8} && \int_{0}^{2\pi} d\phi_{2} \af  d\phi_{3} \int_{\kmin{}}^{\kmax{}} dk_{1\perp} \af  dk_{2\perp} \af dk_{3\perp} \af dk_{1\parallel} \af dk_{2\parallel} \af dk_{3\parallel} \af k_{1\perp} k_{2\perp} k_{3\perp} \nonumber \\ 
&& \frac{\partial T_{\delta T}(\bk{1},\bk{2},\bk{3},\bk{4},z_i)}{\partial p_\alpha}
            \left(\prod_{j=1}^4 \frac{1}{P_{\delta T}(\bk{j},z_i)}\right) \frac{\partial T_{\delta T}(\bk{1},\bk{2},\bk{3},\bk{4},z_i)}{\partial p_\beta} \nonumber \\
\end{eqnarray}
Using this expression we evaluate the Fisher information matrix including the 21-cm secondary trispectra. Note that the integration ranges can be modified to distinguish between $k_{\text{max}}^\parallel$ and $k_{\text{max}}^\perp$ as we do in section \ref{sec:BaselineForecasts}.

Some of the trispectra considered in this work, such as the $\taunl{}$ and intermediate shape, have a diverging behaviour in the collapsed limit, where one of the diagonals becomes very small. In order for the VEGAS integration algorithm \cite{Lepage2020} to be able to properly evaluate the Fisher matrix for these trispectra, it is more favorable to parametrize the quadrilateral in terms of one side and two diagonals instead, leading to the alternative expression:
\begin{eqnarray}
\label{eq:FisherTrispectrumSecondaryAlternative}
F_{\alpha\beta}(z_i) =  \frac{V_i}{4!(2\pi)^8} && \int_{0}^{2\pi} d\phi_{s} \af  d\phi_{t} \int_{\kmin{}}^{\kmax{}} dk_{1\perp} \af  ds_\perp \af dt_\perp \af dk_{1\parallel} \af ds_{\parallel} \af dt_{\parallel} \af k_{1\perp} s_\perp t_\perp \nonumber \\ 
&& \frac{\partial T_{\delta T}(\bk{1},\bk{2},\bk{3},\bk{4},z_i)}{\partial p_\alpha}
            \left(\prod_{j=1}^4 \frac{1}{P_{\delta T}(\bk{j},z_i)}\right) \frac{\partial T_{\delta T}(\bk{1},\bk{2},\bk{3},\bk{4},z_i)}{\partial p_\beta} \nonumber \\
\end{eqnarray}
where the sides of the quadrilateral now read:
\begin{eqnarray}
k_{2\perp}^2 &=& |\boldsymbol{s}_\perp - \bk{1 \perp}|^2 = s_{\perp}^2 + k_{1\perp}^2 - 2s_{\perp}k_{1\perp}\cos{(\phi_1 - \phi_s)},\nonumber \\
k_{2\parallel} &=& s_\parallel - k_{1\parallel} \nonumber \\
k_{3\perp}^2 &=& |\bk{1 \perp} - \boldsymbol{s}_\perp -  \boldsymbol{t}_\perp|^2 = k_{1\perp}^2 + s_{\perp}^2 + t_{\perp}^2 - 2 k_{1\perp} s_{\perp} \cos{(\phi_1 - \phi_s)} - 2 k_{1\perp} s_{\perp} \cos{(\phi_1 - \phi_s)} \nonumber \\ && - 2 s_\perp t_\perp \cos{(\phi_s - \phi_t)}\nonumber \\
k_{3\parallel} &=& k_{1\parallel} - s_\parallel - t_\parallel \nonumber \\
k_{4\perp}^2 &=& |\boldsymbol{t} - \bk{1}|^2 = t_{\perp}^2 + k_{1\perp}^2 - 2t_{\perp}k_{1\perp}\cos{(\phi_1 - \phi_t)},\nonumber \\
k_{2\parallel} &=& s_\parallel - k_{1\parallel}
\end{eqnarray}

Once the Fisher information matrix has been evaluated, the error with which we can determine the parameter $p_\alpha$ is then given by:
\begin{eqnarray}
   \sigma_{p_\alpha} = F_{\alpha\alpha}^{-1/2}
\end{eqnarray}
while the signal-to-noise is given by its inverse:
\begin{eqnarray}
   \left(\frac{S}{N}\right)_{p_\alpha} = \sqrt{F_{\alpha\alpha}}
\end{eqnarray}
When trying to constrain multiple parameters the error for each parameters is instead determined by \emph{marginalising} over the parameter space, which in practise means inverting the Fisher information matrix to obtain the marginalised error:
\begin{eqnarray}
        \sigma_{p_\alpha}^{\textrm{marg}} = \sqrt{(F^{-1})_{\alpha \alpha}}.
\end{eqnarray}
The amount of signal that is lost due to the marginalisation can then be quantified by the signal-to-noise degradation (SND) factor:\footnote{Note that we use a slightly different definition of the SND factor as compared to \cite{Meerburg2016}, where the SND factor is defined as $\textrm{SND} = \sqrt{(F^{-1})_{\alpha\alpha}F_{\alpha\alpha}} -1$.} 
\begin{eqnarray}
\label{eq:SND}
    \textrm{SND}_{\p_\alpha} = \frac{\sigma^{\textrm{marg}}_{p_\alpha}}{\sigma_{p_\alpha}} =    \sqrt{(F^{-1})_{\alpha\alpha}F_{\alpha\alpha}}.
\end{eqnarray}
which depends on how much parameters are correlated. This correlation can be quantified by the Fisher information matrix elements for the parameters considered:
\begin{eqnarray}
\label{eq:cosine}
        C_{\alpha \beta} = \frac{F_{\alpha\beta}}{\sqrt{F_{\alpha\alpha}F_{\beta\beta}}}.
\end{eqnarray}
which will take a value between $-1$ and $1$ and is often refered to as the \emph{overlap} or \emph{cosine} between two parameters. As an example, for two parameters the SND factor and correlation matrix are then simply related by:
\begin{eqnarray}
\textrm{SND}_{\p_\alpha} = \sqrt{\frac{1}{1-(C_{\alpha \beta})^2}}.        
\end{eqnarray}
Hence, an overlap of about $C_{\alpha \beta} = 0.995$ is needed to reduce the sensitivity by a factor of $10$.

\section{21-cm secondary trispectrum}
\label{app:Trisp}
We present the explicit expressions of the 11 contributions to the secondary trispectrum of 21-cm brightness temperature fluctuations. The contributions from two second order fluctuations are:
\begin{align}
	T_{\delta_1\delta_1\delta_2\delta_2}&=4c_1^{(1)}(\kone)c_1^{(1)}(\ktwo)c_1^{(2)}(\kthree)c_1^{(2)}(\kfour)\Big[F_2^{(s)}(\kone,\bk{24})F_2^{(s)}(-\ktwo,\bk{24})P_1P_2P_{24}+\kone\leftrightarrow\ktwo\Big]\nonumber\\
	&+5\;\mathrm{p.}\nonumber\\
	T_{\delta_1\delta_1\theta_2\theta_2}&=4c_1^{(1)}(\kone)c_1^{(1)}(\ktwo)c_2^{(2)}(\kthree)c_2^{(2)}(\kfour)\Big[G_2^{(s)}(\kone,\bk{24})G_2^{(s)}(-\ktwo,\bk{24})P_1P_2P_{24}+\kone\leftrightarrow\ktwo\Big]\nonumber\\&+5\;\mathrm{p.}\nonumber\\
	T_{\delta_1\delta_1\delta_2\theta_2}&=4c_1^{(1)}(\kone)c_1^{(1)}(\ktwo)c_1^{(2)}(\kthree)c_2^{(2)}(\kfour)\Big[F_2^{(s)}(\kone,\bk{24})G_2^{(s)}(-\ktwo,\bk{24})P_1P_2P_{24}+\kone\leftrightarrow\ktwo\Big]\nonumber\\
	&+11\;\mathrm{p.}\nonumber\\
	T_{\delta_1\delta_1\delta_2[\delta_1]^2}&=2c_1^{(1)}(\kone)c_1^{(1)}(\ktwo)c_1^{(2)}(\kthree)\Big\{\Big[c_3^{(2)}(\kthree,\bk{23})+c_3^{(2)}(\kthree,-\ktwo)\Big]F_2^{(s)}(\kone,\bk{23})P_1P_2P_{23}\nonumber\\
	&+\kone\leftrightarrow\ktwo\Big\}+11\;\mathrm{p.}\nonumber\\
	T_{\delta_1\delta_1\theta_2[\delta_1]^2}&=2c_1^{(1)}(\kone)c_1^{(1)}(\ktwo)c_2^{(2)}(\kthree)\Big\{\Big[c_3^{(2)}(\kthree,\bk{23})+c_3^{(2)}(\kthree,-\ktwo)\Big]G_2^{(s)}(\kone,\bk{23})P_1P_2P_{23}\nonumber\\
	&+\kone\leftrightarrow\ktwo\Big\}+11\;\mathrm{p.}\nonumber\\
	T_{\delta_1\delta_1[\delta_1]^2[\delta_1]^2}&=c_1^{(1)}(\kone)c_1^{(1)}(\ktwo)\Big\{\Big[ c_3^{(2)}(\kthree,\bk{13})+c_3^{(2)}(\kthree,-\kone)\Big]\Big[c_3^{(2)}(\kfour,\bk{24})+c_3^{(2)}(\kfour,-\ktwo)\Big]\nonumber\\
	&\times P_1P_2P_{13}+\kthree\leftrightarrow\kfour\Big\}+5\;\mathrm{p.}\nonumber\\
\end{align}
The contributions from one third order fluctuation are:
\begin{align}
	T_{\delta_1\delta_1\delta_1\delta_3}&=6c_1^{(1)}(\kone)c_1^{(1)}(\ktwo)c_1^{(1)}(\kthree)c_1^{(3)}(\kfour)\;F_3^{(s)}(\kone,\ktwo,\kthree)P_1P_2P_3+3\;\mathrm{p.}\nonumber\\
	T_{\delta_1\delta_1\delta_1\theta_3}&=6c_1^{(1)}(\kone)c_1^{(1)}(\ktwo)c_1^{(1)}(\kthree)c_2^{(3)}(\kfour)\;G_3^{(s)}(\kone,\ktwo,\kthree)P_1P_2P_3+3\;\mathrm{p.}\nonumber\\
	T_{\delta_1\delta_1\delta_1[\delta_1]^3}&=c_1^{(1)}(\kone)c_1^{(1)}(\ktwo)c_1^{(1)}(\kthree)\Big\{\Big[c_3^{(3)}(\kfour,-\ktwo,-\kthree)+\ktwo\leftrightarrow\kthree\Big]+2\;\mathrm{c.p.}\Big\}P_1P_2P_3+3\;\mathrm{p.}\nonumber\\
	T_{\delta_1\delta_1\delta_1[\delta_1\delta_2]}&=2c_1^{(1)}(\kone)c_1^{(1)}(\ktwo)c_1^{(1)}(\kthree)\Big[c_5^{(3)}(\kfour,\bk{14})F_2^{(s)}(\ktwo,\kthree)+2\;\mathrm{c.p.}\Big]P_1P_2P_3+3\;\mathrm{p.}\nonumber\\
	T_{\delta_1\delta_1\delta_1[\delta_1\theta_2]}&=2c_1^{(1)}(\kone)c_1^{(1)}(\ktwo)c_1^{(1)}(\kthree)\Big[c_4^{(3)}(\kfour,\bk{14})G_2^{(s)}(\ktwo,\kthree)+2\;\mathrm{c.p.}\Big]P_1P_2P_3+3\;\mathrm{p.}\nonumber\\
\end{align}
where we used the condensed notation $P_i = P_b(\bk{i})$ for the baryonic power spectra, c.p. denotes cyclic permutation. \clearpage

\bibliographystyle{JHEP}
\bibliography{bibliography.bib}

\providecommand{\href}[2]{#2}\begingroup\raggedright\begin{thebibliography}{10}

\bibitem{Planck2018Inflation}
Y.~Akrami, F.~Arroja, M.~Ashdown, J.~Aumont, C.~Baccigalupi, M.~Ballardini
  et~al., \emph{Planck2018 results},
  \href{http://dx.doi.org/10.1051/0004-6361/201833887}{\emph{Astronomy \&
  Astrophysics} {\bf 641} (Sep, 2020) A10}.

\bibitem{Planck2018PNG}
{\scshape Planck} collaboration, Y.~Akrami et~al., \emph{{Planck 2018 results.
  IX. Constraints on primordial non-Gaussianity}},
  \href{http://dx.doi.org/10.1051/0004-6361/201935891}{\emph{Astron.
  Astrophys.} {\bf 641} (2020) A9},
  [\href{http://arxiv.org/abs/1905.05697}{{\tt 1905.05697}}].

\bibitem{Meerburg2019}
P.~D. Meerburg et~al., \emph{{Primordial Non-Gaussianity}},
  \href{http://arxiv.org/abs/1903.04409}{{\tt 1903.04409}}.

\bibitem{Maldacena2002}
J.~M. Maldacena, \emph{{Non-Gaussian features of primordial fluctuations in
  single field inflationary models}},
  \href{http://dx.doi.org/10.1088/1126-6708/2003/05/013}{\emph{JHEP} {\bf 05}
  (2003) 013}, [\href{http://arxiv.org/abs/astro-ph/0210603}{{\tt
  astro-ph/0210603}}].

\bibitem{Creminelli:2003iq}
P.~Creminelli, \emph{{On non-Gaussianities in single-field inflation}},
  \href{http://dx.doi.org/10.1088/1475-7516/2003/10/003}{\emph{JCAP} {\bf 10}
  (2003) 003}, [\href{http://arxiv.org/abs/astro-ph/0306122}{{\tt
  astro-ph/0306122}}].

\bibitem{Creminelli2004}
P.~Creminelli and M.~Zaldarriaga, \emph{{Single field consistency relation for
  the 3-point function}},
  \href{http://dx.doi.org/10.1088/1475-7516/2004/10/006}{\emph{JCAP} {\bf 10}
  (2004) 006}, [\href{http://arxiv.org/abs/astro-ph/0407059}{{\tt
  astro-ph/0407059}}].

\bibitem{Arkani-Hamed2015}
N.~Arkani-Hamed and J.~Maldacena, \emph{{Cosmological Collider Physics}},
  \href{http://arxiv.org/abs/1503.08043}{{\tt 1503.08043}}.

\bibitem{Lee:2016vti}
H.~Lee, D.~Baumann and G.~L. Pimentel, \emph{{Non-Gaussianity as a Particle
  Detector}}, \href{http://dx.doi.org/10.1007/JHEP12(2016)040}{\emph{JHEP} {\bf
  12} (2016) 040}, [\href{http://arxiv.org/abs/1607.03735}{{\tt 1607.03735}}].

\bibitem{Baumann2017}
D.~Baumann, G.~Goon, H.~Lee and G.~L. Pimentel, \emph{{Partially Massless
  Fields During Inflation}},
  \href{http://dx.doi.org/10.1007/JHEP04(2018)140}{\emph{JHEP} {\bf 04} (2018)
  140}, [\href{http://arxiv.org/abs/1712.06624}{{\tt 1712.06624}}].

\bibitem{Cooray2006}
A.~Cooray, \emph{{21-cm Background Anisotropies Can Discern Primordial
  Non-Gaussianity}},
  \href{http://dx.doi.org/10.1103/PhysRevLett.97.261301}{\emph{Phys. Rev.
  Lett.} {\bf 97} (2006) 261301},
  [\href{http://arxiv.org/abs/astro-ph/0610257}{{\tt astro-ph/0610257}}].

\bibitem{Pillepich2006}
A.~Pillepich, C.~Porciani and S.~Matarrese, \emph{{The bispectrum of redshifted
  21-cm fluctuations from the dark ages}},
  \href{http://dx.doi.org/10.1086/517963}{\emph{Astrophys. J.} {\bf 662} (2007)
  1--14}, [\href{http://arxiv.org/abs/astro-ph/0611126}{{\tt
  astro-ph/0611126}}].

\bibitem{Meerburg2016}
P.~D. Meerburg, M.~M\"unchmeyer, J.~B. Mu\~noz and X.~Chen, \emph{{Prospects
  for Cosmological Collider Physics}},
  \href{http://dx.doi.org/10.1088/1475-7516/2017/03/050}{\emph{JCAP} {\bf 03}
  (2017) 050}, [\href{http://arxiv.org/abs/1610.06559}{{\tt 1610.06559}}].

\bibitem{Silk2020}
J.~Silk, \emph{{The limits of cosmology: role of the Moon}},
  \href{http://dx.doi.org/10.1098/rsta.2019.0561}{\emph{Phil. Trans. A. Math.
  Phys. Eng. Sci.} {\bf 379} (2021) 20190561},
  [\href{http://arxiv.org/abs/2011.04671}{{\tt 2011.04671}}].

\bibitem{Munoz2015}
J.~B. Mu\~noz, Y.~Ali-Ha\"\i{}moud and M.~Kamionkowski, \emph{{Primordial
  non-gaussianity from the bispectrum of 21-cm fluctuations in the dark ages}},
  \href{http://dx.doi.org/10.1103/PhysRevD.92.083508}{\emph{Phys. Rev. D} {\bf
  92} (2015) 083508}, [\href{http://arxiv.org/abs/1506.04152}{{\tt
  1506.04152}}].

\bibitem{Kalaja2020}
A.~Kalaja, P.~D. Meerburg, G.~L. Pimentel and W.~R. Coulton, \emph{{Fundamental
  limits on constraining primordial non-Gaussianity}},
  \href{http://dx.doi.org/10.1088/1475-7516/2021/04/050}{\emph{JCAP} {\bf 04}
  (2021) 050}, [\href{http://arxiv.org/abs/2011.09461}{{\tt 2011.09461}}].

\bibitem{Planck2018Parameters}
{\scshape Planck} collaboration, N.~Aghanim et~al., \emph{{Planck 2018 results.
  VI. Cosmological parameters}},
  \href{http://dx.doi.org/10.1051/0004-6361/201833910}{\emph{Astron.
  Astrophys.} {\bf 641} (2020) A6},
  [\href{http://arxiv.org/abs/1807.06209}{{\tt 1807.06209}}].

\bibitem{Furlanetto:2006jb}
S.~Furlanetto, S.~P. Oh and F.~Briggs, \emph{{Cosmology at Low Frequencies: The
  21 cm Transition and the High-Redshift Universe}},
  \href{http://dx.doi.org/10.1016/j.physrep.2006.08.002}{\emph{Phys. Rept.}
  {\bf 433} (2006) 181--301},
  [\href{http://arxiv.org/abs/astro-ph/0608032}{{\tt astro-ph/0608032}}].

\bibitem{Lewis2007b}
A.~Lewis and A.~Challinor, \emph{{The 21cm angular-power spectrum from the dark
  ages}}, \href{http://dx.doi.org/10.1103/PhysRevD.76.083005}{\emph{Phys. Rev.
  D} {\bf 76} (2007) 083005},
  [\href{http://arxiv.org/abs/astro-ph/0702600}{{\tt astro-ph/0702600}}].

\bibitem{Ali-Haimoud2014}
Y.~Ali-Ha\"\i{}moud, P.~D. Meerburg and S.~Yuan, \emph{{New light on 21 cm
  intensity fluctuations from the dark ages}},
  \href{http://dx.doi.org/10.1103/PhysRevD.89.083506}{\emph{Phys. Rev. D} {\bf
  89} (2014) 083506}, [\href{http://arxiv.org/abs/1312.4948}{{\tt 1312.4948}}].

\bibitem{Lewis2007}
A.~Lewis, \emph{{Linear effects of perturbed recombination}},
  \href{http://dx.doi.org/10.1103/PhysRevD.76.063001}{\emph{Physical Review D -
  Particles, Fields, Gravitation and Cosmology} {\bf 76} (2007) 1--5},
  [\href{http://arxiv.org/abs/0707.2727}{{\tt 0707.2727}}].

\bibitem{Peebles1968}
P.~J.~E. Peebles, \emph{{Recombination of the Primeval Plasma}},
  \href{http://dx.doi.org/10.1086/149628}{\emph{Astrophys. J.} {\bf 153} (1968)
  1}.

\bibitem{Seager:1999km}
S.~Seager, D.~D. Sasselov and D.~Scott, \emph{{How exactly did the universe
  become neutral?}}, \href{http://dx.doi.org/10.1086/313388}{\emph{Astrophys.
  J. Suppl.} {\bf 128} (2000) 407--430},
  [\href{http://arxiv.org/abs/astro-ph/9912182}{{\tt astro-ph/9912182}}].

\bibitem{Cheung2007}
C.~Cheung, P.~Creminelli, A.~L. Fitzpatrick, J.~Kaplan and L.~Senatore,
  \emph{{The Effective Field Theory of Inflation}},
  \href{http://dx.doi.org/10.1088/1126-6708/2008/03/014}{\emph{JHEP} {\bf 03}
  (2008) 014}, [\href{http://arxiv.org/abs/0709.0293}{{\tt 0709.0293}}].

\bibitem{Chen:2012ja}
X.~Chen and C.~Ringeval, \emph{{Searching for Standard Clocks in the Primordial
  Universe}},
  \href{http://dx.doi.org/10.1088/1475-7516/2012/08/014}{\emph{JCAP} {\bf 08}
  (2012) 014}, [\href{http://arxiv.org/abs/1205.6085}{{\tt 1205.6085}}].

\bibitem{Chen:2014cwa}
X.~Chen, M.~H. Namjoo and Y.~Wang, \emph{{Models of the Primordial Standard
  Clock}}, \href{http://dx.doi.org/10.1088/1475-7516/2015/02/027}{\emph{JCAP}
  {\bf 02} (2015) 027}, [\href{http://arxiv.org/abs/1411.2349}{{\tt
  1411.2349}}].

\bibitem{Chen:2014joa}
X.~Chen and M.~H. Namjoo, \emph{{Standard Clock in Primordial Density
  Perturbations and Cosmic Microwave Background}},
  \href{http://dx.doi.org/10.1016/j.physletb.2014.11.002}{\emph{Phys. Lett. B}
  {\bf 739} (2014) 285--292}, [\href{http://arxiv.org/abs/1404.1536}{{\tt
  1404.1536}}].

\bibitem{Chen:2015lza}
X.~Chen, M.~H. Namjoo and Y.~Wang, \emph{{Quantum Primordial Standard Clocks}},
  \href{http://dx.doi.org/10.1088/1475-7516/2016/02/013}{\emph{JCAP} {\bf 02}
  (2016) 013}, [\href{http://arxiv.org/abs/1509.03930}{{\tt 1509.03930}}].

\bibitem{Chen2009}
X.~Chen and Y.~Wang, \emph{{Large non-Gaussianities with Intermediate Shapes
  from Quasi-Single Field Inflation}},
  \href{http://dx.doi.org/10.1103/PhysRevD.81.063511}{\emph{Phys. Rev. D} {\bf
  81} (2010) 063511}, [\href{http://arxiv.org/abs/0909.0496}{{\tt 0909.0496}}].

\bibitem{Suyama2007}
T.~Suyama and M.~Yamaguchi, \emph{{Non-Gaussianity in the modulated reheating
  scenario}}, \href{http://dx.doi.org/10.1103/PhysRevD.77.023505}{\emph{Phys.
  Rev. D} {\bf 77} (2008) 023505}, [\href{http://arxiv.org/abs/0709.2545}{{\tt
  0709.2545}}].

\bibitem{Smith2015}
K.~M. Smith, L.~Senatore and M.~Zaldarriaga, \emph{{Optimal analysis of the CMB
  trispectrum}},  \href{http://arxiv.org/abs/1502.00635}{{\tt 1502.00635}}.

\bibitem{Senatore2010}
L.~Senatore and M.~Zaldarriaga, \emph{{The Effective Field Theory of Multifield
  Inflation}}, \href{http://dx.doi.org/10.1007/JHEP04(2012)024}{\emph{JHEP}
  {\bf 04} (2012) 024}, [\href{http://arxiv.org/abs/1009.2093}{{\tt
  1009.2093}}].

\bibitem{Senatore:2010jy}
L.~Senatore and M.~Zaldarriaga, \emph{{A Naturally Large Four-Point Function in
  Single Field Inflation}},
  \href{http://dx.doi.org/10.1088/1475-7516/2011/01/003}{\emph{JCAP} {\bf 01}
  (2011) 003}, [\href{http://arxiv.org/abs/1004.1201}{{\tt 1004.1201}}].

\bibitem{Assassi2012}
V.~Assassi, D.~Baumann and D.~Green, \emph{{On Soft Limits of Inflationary
  Correlation Functions}},
  \href{http://dx.doi.org/10.1088/1475-7516/2012/11/047}{\emph{JCAP} {\bf 11}
  (2012) 047}, [\href{http://arxiv.org/abs/1204.4207}{{\tt 1204.4207}}].

\bibitem{Bernardeau2001}
F.~Bernardeau, S.~Colombi, E.~Gaztanaga and R.~Scoccimarro, \emph{{Large scale
  structure of the universe and cosmological perturbation theory}},
  \href{http://dx.doi.org/10.1016/S0370-1573(02)00135-7}{\emph{Phys. Rept.}
  {\bf 367} (2002) 1--248}, [\href{http://arxiv.org/abs/astro-ph/0112551}{{\tt
  astro-ph/0112551}}].

\bibitem{Shoji2009}
M.~Shoji and E.~Komatsu, \emph{{Third-order Perturbation Theory With Non-linear
  Pressure}},
  \href{http://dx.doi.org/10.1088/0004-637X/700/1/705}{\emph{Astrophys. J.}
  {\bf 700} (2009) 705--719}, [\href{http://arxiv.org/abs/0903.2669}{{\tt
  0903.2669}}].

\bibitem{Chen2016a}
X.~Chen, P.~D. Meerburg and M.~M\"unchmeyer, \emph{{The Future of Primordial
  Features with 21 cm Tomography}},
  \href{http://dx.doi.org/10.1088/1475-7516/2016/09/023}{\emph{JCAP} {\bf 09}
  (2016) 023}, [\href{http://arxiv.org/abs/1605.09364}{{\tt 1605.09364}}].

\bibitem{Bertolini:2015fya}
D.~Bertolini, K.~Schutz, M.~P. Solon, J.~R. Walsh and K.~M. Zurek,
  \emph{{Non-Gaussian Covariance of the Matter Power Spectrum in the Effective
  Field Theory of Large Scale Structure}},
  \href{http://dx.doi.org/10.1103/PhysRevD.93.123505}{\emph{Phys. Rev. D} {\bf
  93} (2016) 123505}, [\href{http://arxiv.org/abs/1512.07630}{{\tt
  1512.07630}}].

\bibitem{SKA}
\emph{{Square Kilometer Array}}.
\newblock https://www.skatelescope.org/.

\bibitem{HERA}
J.~C. Pober et~al., \emph{{What Next-Generation 21 cm Power Spectrum
  Measurements Can Teach Us About the Epoch of Reionization}},
  \href{http://dx.doi.org/10.1088/0004-637X/782/2/66}{\emph{Astrophys. J.} {\bf
  782} (2014) 66}, [\href{http://arxiv.org/abs/1310.7031}{{\tt 1310.7031}}].

\bibitem{DAPPER}
J.~O. Burns, \emph{Transformative science from the lunar farside: observations
  of the dark ages and exoplanetary systems at low radio frequencies},
  \href{http://dx.doi.org/10.1098/rsta.2019.0564}{\emph{Philosophical
  Transactions of the Royal Society A: Mathematical, Physical and Engineering
  Sciences} {\bf 379} (Nov, 2020) 20190564}.

\bibitem{KoopmansPeering}
L.~V.~E. Koopmans et~al., \emph{{Peering into the dark (ages) with
  low-frequency space interferometers: Using the 21-cm signal of neutral
  hydrogen from the infant universe to probe fundamental (Astro)physics}},
  \href{http://dx.doi.org/10.1007/s10686-021-09743-7}{\emph{Exper. Astron.}
  {\bf 51} (2021) 1641--1676}, [\href{http://arxiv.org/abs/1908.04296}{{\tt
  1908.04296}}].

\bibitem{Parsons:2012qh}
A.~R. Parsons, J.~C. Pober, J.~E. Aguirre, C.~L. Carilli, D.~C. Jacobs and
  D.~F. Moore, \emph{{A Per-Baseline, Delay-Spectrum Technique for Accessing
  the 21cm Cosmic Reionization Signature}},
  \href{http://dx.doi.org/10.1088/0004-637X/756/2/165}{\emph{Astrophys. J.}
  {\bf 756} (2012) 165}, [\href{http://arxiv.org/abs/1204.4749}{{\tt
  1204.4749}}].

\bibitem{Liu:2014bba}
A.~Liu, A.~R. Parsons and C.~M. Trott, \emph{{Epoch of reionization window. I.
  Mathematical formalism}},
  \href{http://dx.doi.org/10.1103/PhysRevD.90.023018}{\emph{Phys. Rev. D} {\bf
  90} (2014) 023018}, [\href{http://arxiv.org/abs/1404.2596}{{\tt 1404.2596}}].

\bibitem{Liu:2014yxa}
A.~Liu, A.~R. Parsons and C.~M. Trott, \emph{{Epoch of reionization window. II.
  Statistical methods for foreground wedge reduction}},
  \href{http://dx.doi.org/10.1103/PhysRevD.90.023019}{\emph{Phys. Rev. D} {\bf
  90} (2014) 023019}, [\href{http://arxiv.org/abs/1404.4372}{{\tt 1404.4372}}].

\bibitem{Ghosh:2017woo}
A.~Ghosh, F.~Mertens and L.~V.~E. Koopmans, \emph{{Deconvolving the wedge:
  maximum-likelihood power spectra via spherical-wave visibility modelling}},
  \href{http://dx.doi.org/10.1093/mnras/stx2959}{\emph{Mon. Not. Roy. Astron.
  Soc.} {\bf 474} (2018) 4552--4563},
  [\href{http://arxiv.org/abs/1709.06752}{{\tt 1709.06752}}].

\bibitem{Modi2019}
C.~Modi, M.~White, A.~Slosar and E.~Castorina, \emph{{Reconstructing
  large-scale structure with neutral hydrogen surveys}},
  \href{http://dx.doi.org/10.1088/1475-7516/2019/11/023}{\emph{JCAP} {\bf 11}
  (2019) 023}, [\href{http://arxiv.org/abs/1907.02330}{{\tt 1907.02330}}].

\bibitem{CHIME}
J.~R. Shaw, K.~Sigurdson, U.-L. Pen, A.~Stebbins and M.~Sitwell, \emph{All-sky
  interferometry with spherical harmonic transit telescopes},
  \href{http://dx.doi.org/10.1088/0004-637x/781/2/57}{\emph{The Astrophysical
  Journal} {\bf 781} (Jan, 2014) 57}.

\bibitem{TIANLAI}
X.~{Chen}, \emph{{Radio detection of dark energy{\textemdash}the Tianlai
  project}}, \href{http://dx.doi.org/10.1360/132011-972}{\emph{Scientia Sinica
  Physica, Mechanica \& Astronomica} {\bf 41} (Jan., 2011) 1358}.

\bibitem{HIRAX}
L.~B. Newburgh, K.~Bandura, M.~A. Bucher, T.-C. Chang, H.~C. Chiang, J.~Cliche
  et~al., \emph{Hirax: a probe of dark energy and radio transients},
  \href{http://dx.doi.org/10.1117/12.2234286}{\emph{Ground-based and Airborne
  Telescopes VI} (Aug, 2016) }.

\bibitem{PUMA}
{\scshape PUMA} collaboration, A.~Slosar et~al., \emph{{Packed Ultra-wideband
  Mapping Array (PUMA): A Radio Telescope for Cosmology and Transients}},
  {\emph{Bull. Am. Astron. Soc.} {\bf 51} (2019) 53},
  [\href{http://arxiv.org/abs/1907.12559}{{\tt 1907.12559}}].

\bibitem{LOFAR}
{LOFAR}, \emph{{Low-Frequency Array}}.
\newblock https://www.astro.rug.nl/~lofareor/scibgd/scibgd.html.

\bibitem{LWARe}
{LWA}, \emph{{Long Wavelength Array}}.
\newblock http://lwa.phys.unm.edu/science.html.

\bibitem{MWA}
{MWA}, \emph{{Murchison Widefield Array}}.
\newblock https://www.mwatelescope.org/science/epoch-of-reionization-eor.

\bibitem{NenuFAR}
F.~G. Mertens, B.~Semelin and L.~V.~E. Koopmans, \emph{{Exploring the Cosmic
  Dawn with NenuFAR}},  in \emph{{Semaine de l'astrophysique fran\c{c}aise
  2021}}, 9, 2021.
\newblock \href{http://arxiv.org/abs/2109.10055}{{\tt 2109.10055}}.

\bibitem{LWADawn}
C.~DiLullo, G.~B. Taylor and J.~Dowell, \emph{{Using the Long Wavelength Array
  to Search for Cosmic Dawn}},
  \href{http://dx.doi.org/10.1142/S2251171720500087}{\emph{J. Astron. Inst.}
  {\bf 09} (2020) 2050008}, [\href{http://arxiv.org/abs/2005.10669}{{\tt
  2005.10669}}].

\bibitem{Scoccimarro2003}
R.~Scoccimarro, E.~Sefusatti and M.~Zaldarriaga, \emph{{Probing primordial
  non-Gaussianity with large - scale structure}},
  \href{http://dx.doi.org/10.1103/PhysRevD.69.103513}{\emph{Phys. Rev. D} {\bf
  69} (2004) 103513}, [\href{http://arxiv.org/abs/astro-ph/0312286}{{\tt
  astro-ph/0312286}}].

\bibitem{Baldauf2016}
T.~Baldauf, M.~Mirbabayi, M.~Simonovi\'c and M.~Zaldarriaga, \emph{{LSS
  constraints with controlled theoretical uncertainties}},
  \href{http://arxiv.org/abs/1602.00674}{{\tt 1602.00674}}.

\bibitem{Fergusson2012}
J.~R. Fergusson, D.~M. Regan and E.~P.~S. Shellard, \emph{{Rapid Separable
  Analysis of Higher Order Correlators in Large Scale Structure}},
  \href{http://dx.doi.org/10.1103/PhysRevD.86.063511}{\emph{Phys. Rev. D} {\bf
  86} (2012) 063511}, [\href{http://arxiv.org/abs/1008.1730}{{\tt 1008.1730}}].

\bibitem{Chen2009tri}
X.~Chen, B.~Hu, M.~X. Huang, G.~Shiu and Y.~Wang, \emph{Large primordial
  trispectra in general single field inflation},
  \href{http://dx.doi.org/10.1088/1475-7516/2009/08/008}{\emph{Journal of
  Cosmology and Astroparticle Physics} {\bf 2009} (5, 2009) }.

\bibitem{Lepage2020}
G.~P. Lepage, \emph{{Adaptive multidimensional integration: VEGAS enhanced}},
  \href{http://dx.doi.org/10.1016/j.jcp.2021.110386}{\emph{J. Comput. Phys.}
  {\bf 439} (2021) 110386}, [\href{http://arxiv.org/abs/2009.05112}{{\tt
  2009.05112}}].

\end{thebibliography}\endgroup
\end{document}